\documentclass{ar-1col}

\jname{Xxxx. Xxx. Xxx. Xxx.}
\jvol{55}
\jyear{2016}
\doi{10.1146/((please add article doi))}

\newcommand{\msun}{\rm{M}_\odot}
\newcommand{\lsun}{\rm{L}_\odot}
\newcommand{\hii}{{H{\scriptsize II} }}

\usepackage{natbib}
\usepackage{journaux}
\usepackage{color}

\begin{document}

\markboth{Motte et al.}{High-mass star and massive cluster formation in the Milky Way}

\title{High-mass star and massive cluster formation in the Milky Way}

\author{Fr\'ed\'erique Motte,$^{1,2}$, Sylvain Bontemps,$^3$, and Fabien Louvet,$^4$
\affil{$^1$University Grenoble Alpes, CNRS, Institut de Plan\'etologie et d'Astrophysique de Grenoble, F-38000 Grenoble, France; email: frederique.motte@univ-grenoble-alpes.fr}
\affil{$^2$AIM Paris-Saclay/Service d'Astrophysique, CEA/IRFU -- CNRS/INSU -- Univ. Paris Diderot, CEA-Saclay, F-91191 Gif-sur-Yvette Cedex, France}
\affil{$^3$OASU/Laboratoire d'Astrophysique de Bordeaux, Univ. de Bordeaux -- CNRS/INSU, BP 89, F-33271 Floirac Cedex, France}
\affil{$^3$ Department of Astronomy, Universidad de Chile, Las Condes, Santiago, Chile}}

\begin{abstract}
This review examines the state-of-the-art knowledge of high-mass star and massive cluster formation, gained from ambitious observational surveys, which acknowledge the multi-scale characteristics of these processes. After a brief overview of theoretical models and main open issues, we present observational searches for the evolutionary phases of high-mass star formation, first among high-luminosity sources and more recently among young massive protostars and the elusive high-mass prestellar cores.
We then introduce the most likely evolutionary scenario for high-mass star formation, which emphasizes the link of high-mass star formation to massive cloud and cluster formation.
Finally, we introduce the first attempts to search for variations of the star formation activity and cluster formation in molecular cloud complexes, in the most extreme star-forming sites, and across the Milky Way. The combination of Galactic plane surveys and high-angular resolution images with submillimeter facilities such as Atacama Large Millimeter Array (ALMA) are prerequisites to make significant progresses in the forthcoming decade.

\end{abstract}

\maketitle

\tableofcontents

\section{INTRODUCTION}
\label{s:intro}

High-mass stars, also called OB stars, have luminosities larger than $10^3~\lsun$, spectral types of B3 or earlier, and stellar masses from $8~\msun$ up to possibly more than $150~\msun$ \citep{martins08}. From their births to their deaths, high-mass stars are known to play a major role in the energy budget of galaxies via their radiation, wind, and supernovae events. Despite that, the formation of high-mass stars remains an enigmatic process, being far less understood than it is for their low-mass (solar-type) counterparts. Solving the mystery of the high-mass star-formation process is important for itself but it is also fundamental to fully constrain the origin of the initial mass function (IMF) and the formation of massive star clusters and to provide accurate star-formation recipes such as star-formation rate (SFR) and IMF for extragalactic studies and numerical simulations.

Theoretical models proposed for the formation of high-mass stars tried to solve the UV radiation pressure problem \citep{WoCa87}. Stars reaching a few $10~\msun$ masses and a few $10^3~\lsun$ luminosities were indeed supposed to develop a pressure barrier halting further accretion. Most recent 3D modeling mostly solved this problem by showing that equatorial accretion can continue for ionizing protostar embryos \citep[e.g.,][]{krumholz09b, kuiper11}. Competing concepts for high-mass star formation currently are (a) monolithic collapse of a turbulent, pre-assembled core in Virial equilibrium \citep[e.g.,][]{MKT02, MKT03, HoOm09}, (b) protostar collision and coalescence in very dense systems \citep[e.g.,][]{bonnell98, BoBa02}, and (c) competitive accretion in a protocluster environment through Bondi-Hoyle accretion \citep[e.g.,][]{bonnell01,MuCh12} and/or gravitationally-driven cloud inflow \citep{smith09, hartmann12}. Numerical simulations are now able to form stars with masses of up to $40-140~\msun$ stars thanks to non spherical accretion, improved radiation transfer, and feedback effects such as heating and ionization \citep[e.g.,][]{YoSo02, krumholz09b, kuiper10, kuiper11}. Modeling the formation of higher mass stars may remain a challenge \citep[see][]{krumholz15}. For a complete description of high-mass star-formation theories, readers are directed to reviews by, e.g., \cite{ZiYo07}, \cite{beuther07a}, \cite{tan14}, and \cite{krumholz15}.

The main open issues on high-mass and massive cluster formation include the following: How different are the processes, that form high-mass stars and massive clusters with respect to their low-mass analogs? How is high-mass star formation linked to the formation of their parental clouds and descendant clusters? Does it vary across the Milky Way?
Observational constraints take time to gather because understanding star formation and especially high-mass star formation requires studies over several decades of spatial scales and densities. Furthermore, studying the formation of high-mass stars and their companion low-mass stars implies dissecting their parental protoclusters. The latter are complex structures composed of molecular gas and stars in the making, generally located at more than 1~kpc from the Sun, and largely embedded within high-density clouds.
Investigating high-mass star and massive cluster formation thus requires high angular resolution imaging at far-IR to (sub)millimeter wavelengths over large areas of the Milky Way.
Because we suspect high-mass star-forming regions to be exposed to shock waves, powered by cloud collision, infall motions, OB stellar winds, and ionization fronts, studying both cloud structures and kinematics is mandatory.

This observational review follows those done by, e.g., \cite{churchwell02}, \cite{ZiYo07} and \cite{beuther07a}. We intentionally refrain from discussing detailed characteristics of high-mass precursors, such as disk and binary formation or chemistry evolution, because observational constraints remain sparse and are based on studies of a few very luminous objects.
In the remainder of this review, we characterize the evolutionary phases of high-mass star formation as defined from large surveys of infrared-bright (IR-bright) to infrared-quiet (IR-quiet) objects (see Sect.~\ref{s:HMSF}), pointing out their strengths and biases. We end up proposing the most probable evolutionary scenario for the formation of high-mass stars in relation to source statistics and cloud kinematics.
We then investigate the importance of cloud characteristics to form high-mass stars and massive stellar clusters and present initial searches for variations across the Milky Way (see Sects.~\ref{s:cloud+cluster}-\ref{s:prospects}). Finally, we point out directions of improvement for the coming decade (see Sect.~\ref{s:conc}.)

\section{HIGH-MASS STAR FORMATION}
\label{s:HMSF}

Unlike the case for low-mass stars \citep[see, e.g.,][]{shu87,andre00}, there is no observational evolutionary sequence that is firmly established for high-mass star formation. One of the main differences between high-mass and low-mass stars is that the radiation field of a massive star plays a more important role during its whole life and already in its formation phase. Theoretically, a massive protostellar embryo heats and eventually ionizes the gas of its surrounding envelope, creating an \hii region that develops by expanding within the cloud \citep[see the Str\"omgren theory in][]{spitzer78}. 

Despite the lack of an evolutionary sequence, a nomenclature of high-mass star precursors exists. Following that of low-mass stars, objects associated with the first phase of high-mass star formation have been called massive starless clumps, high-mass prestellar cores, massive cold molecular cores, or even IR-dark clouds (IRDCs). High-mass prestellar cores would be pre-assembled, gravitationally bound cores that will form individual high-mass stars or binaries. The nature of larger-scale cloud structures remains unclear.
In the subsequent phase, high-mass star precursors have been named massive protostars, high-mass protostellar objects (HMPOs), protostellar massive dense cores (MDCs), or hot molecular cores (HMCs). These collapsing cloud fragments qualify as high-mass protostars when they have the ability to form a high-mass star binary but not a full cluster.
The final phase corresponds to \hii regions being from hyper-compact to classical. Below, following the chronological order of the bibliography and thus going backward in time for high-mass star formation, we present surveys, th	at have discovered and characterized such objects (see Sects.~\ref{s:irb}-\ref{s:irq}).
For a meaningful comparison of the precursors of high-mass stars identified by these studies (see \textbf{Table}~\ref{tab:cloudstr}), we choose to use and extend the terminology recommended by \cite{williams00}: 
cloud complexes have  $\sim$100~pc sizes, clouds $\sim$10~pc, clumps are  $\sim$1~pc cloud structures, dense cores have $\sim$0.1~pc, and individual cores $\sim$0.01~pc sizes. Interestingly, individual protostars are observed with $\sim$0.02~pc or $3\,000-5\,000$~AU sizes and may not further subfragment \cite{bontemps10, palau13, beuther15}. In the following, the term protostar is used to refer to a protostellar embryo surrounded by a protostellar envelope/core, not the protostellar embryo alone.

\begin{table}[h]
\caption[]{Cloud structures of a few reference studies of high-mass (left) and low-mass (right) star formation}
\label{tab:cloudstr}
\begin{center}
\begin{tabular}{|l|ccc|cc|}
\hline
Source & HMPOs	& IRDCs & MDCs,	& Isolated		& Clustered\\
& 	& fragments,	& 		& prestellar & pre- or protostellar\\
Nature & clumps	& clumps	& dense cores	&  cores & cores\\
\hline
\hline
FWHM [pc]		& $\sim$0.5		& $\sim$0.5		& $0.1-0.2$	& $\sim$0.08 	& $\sim$0.007\\
Mass [$\msun$]	& $\sim$290	& $\sim$150	& $\sim 150$ 	& $\sim$5		& $\sim$0.15 \\

$<n_{\mbox{\tiny H$_2$}}>$ [cm$^{-3}$]	& $\sim$$6\times 10^4$ 	& $\sim$$5\times 10^4$ 	 & $\sim$$2\times 10^6$ & $\sim$$2\times 10^5$	& $\sim$$2\times 10^7$ \\
$d_{\mbox{\scriptsize Sun}}$ [kpc] & $0.3-14$  & $1.8-7.1$ & 1.4 & $0.14-0.44$ & 0.14\\
References  & (1) & (2), (3), (4) & (5), (6) & (7), (8) & (8)\\
\hline
\end{tabular}
\end{center}
\begin{tabnote}
References: (1) \cite{beuther02a}; (2) \cite{rathborne06}; (3)  \cite{BuTa09}; (4) \cite{PeFu10};
(5) \cite{motte07}; (6)  \cite{russeil10}; (7) \cite{ward99}; (8) \cite{motte98}.
\end{tabnote}
\end{table}

Because high-mass stars represent less than 1\% of the stars \citep[when integrating the IMF by][]{kroupa01}, observing a statistically significant sample of high-mass star precursors requires probing cloud complexes more massive than any of the Gould Belt clouds, including Orion. The present knowledge of high-mass star formation is mainly based on surveys of the most nearby, massive cloud complexes that are more massive than Orion (see Sect.~\ref{s:HOBYSclouds}).

\subsection{Evolution from \hii regions back to IR-bright protostars}
\label{s:irb}

Because high-mass stars are luminous (above $10^3~\lsun$) on the main sequence, from the 1980s to the end of the 1990s their precursors have been searched for among sources strongly emitting UV or IR radiation. 
Massive young stellar objects that have developed an \hii region are strong free-free emitters at centimeter wavelengths and have thus been studied in great details for several decades \citep[see, e.g.,][]{churchwell02}. In his ARAA review,  \cite{churchwell02} proposed an empirical evolutionary scenario, based on ionization expansion, leading from ultra-compact \hii (UCH\mbox{\sc ii}) regions to compact \hii regions, and then classical/developed \hii regions. It has been completed with a new class of objects qualified as hyper-compact H\mbox{\sc ii}s \citep[HCH\mbox{\sc ii}s,][]{hoare07}. The physical size, density, line profile, and spectral index of \hii regions detected in the radio centimeter and recombination line surveys are the main characteristics used in this empirical classification.
The smallest \hii regions, UCH\mbox{\sc ii}s and HCH\mbox{\sc ii}s, that have more to tell about the process of high-mass star formation have $\sim$0.1~pc and $<$0.05~pc sizes, respectively, and $\sim$$10^4$~cm$^{-3}$ and $\sim$$10^6$~cm$^{-3}$ densities, respectively \citep{kurtz00, hoare07}. HCH\mbox{\sc ii} regions themselves could correspond to a very early phase of \hii regions, quenched by infalling gas, or to high-mass protostars, whose photo-evaporating disks and ionized accretion flows or jets are detected at centimeter wavelengths \citep[e.g.,][]{keto03, hoare07}.

In 1989, Wood \& Churchwell started searching for the youngest \hii regions by using the Galaxy-wide survey of high-luminosity IR sources provided by the \emph{IRAS} point source catalog.
They applied the Log$(F_{\rm 60\mu m}/F_{\rm 12\mu m}) > 1.3$ and Log$(F_{\rm 25\mu m}/F_{\rm 12\mu m}) > 0.57$ color-color criteria to select bright red \emph{IRAS} sources that could correspond to young stellar objects with a stellar embryo more massive than $8~\msun$. The resulting catalog contains 1646 sources spread near and far across the Galaxy. Most of these sources indeed are UC\hii regions, but some of them could even be in the earlier protostellar phase.
\begin{marginnote}[]
\entry{\emph{Infrared Astronomical Satellite (IRAS):}}{the first space telescope to perform a survey of the entire sky at 12, 25, 60, and $100~\mu$m; its compact source catalog contains over 250\,000 sources.}
\end{marginnote}

Many authors have searched for protostellar objects within the Wood \& Churchwell catalog of \emph{IRAS} sources \citep[e.g.,][]{bronfman96, plume97, walsh98, molinari00, sridharan02, mueller02, faundez04,hill05}. They postulated that protostellar objects were all high-luminosity IR sources embedded within massive envelopes that have not yet developed an \hii region. These authors have thus investigated the association of the Wood \& Churchwell sources with dense gas, detected through for instance CS molecular lines or millimeter continuum, with a hot core through detection of complex molecules, and/or with masers. They checked for the absence of any \hii region via no or weak emission at centimeter wavelengths. These sources, in the pre-UC\hii phase, have been named differently in each of the papers referenced above, but two of these names remained: HMCs \citep{GaLi99, kurtz00} or HMPOs \citep{beuther02a}. The main drawback of these large samples is their inhomogeneity in terms of distance and thus spatial scales. Following the terminology of \cite{williams00}, these \emph{IRAS}-selected sources spanning 0.1 to 10~pc scales are dense cores, clumps or even clouds, each of them associated sometimes loosely with at least one bright IR source.

One of the best-studied sample by \cite{beuther02a} contains 69 HMPOs, which are located at 300~pc up to 14~kpc from the Sun. The median HMPO (see \textbf{Table}~\ref{tab:cloudstr}) is thus a clump, i.e. a $\sim$1~pc cloud structure hosting several individual high-mass protostars with expected 0.02~pc sizes \citep[e.g.][]{beuther15}. HMPO clumps closely associated with  \emph{IRAS} sources are good candidates to contain IR-bright high-mass protostars.

Several attempts to derive an evolutionary sequence for high-mass star formation have been made in these surveys by using three types of diagnostics: hot core chemistry enrichment, maser types, and luminosity. Because the warm inner parts of high-mass protostellar envelopes evolve with time, the physical and chemical properties of a hot core (e.g., its size, temperature, molecular abundances, and associated masers) can in principle be used as a clock \citep[e.g.,][]{HevD97, GaLi99}. Both methanol and OH masers are associated with hot cores formed during the high-mass star formation process \citep[see catalogs by, e.g.,][]{pestalozzi05, walsh16}. A timeline based on masers has been proposed with OH masers generally associated with \hii regions and methanol masers at 6.7~GHz exclusively tracing the earliest protostellar phases \citep[e.g.,][]{minier05, breen10}. 
Besides, because a high-mass star is expected to grow in mass across its formation process, its luminosity should increase. The envelope mass to bolometric luminosity ratio, $M/L$, can thus be used to qualitatively separate the early or late state of evolution of a high-mass protostellar object \citep[e.g.,][]{sridharan02, elia17}.
Because the evolutionary sequences proposed for high-mass star formation before the decade 2000 are almost exclusively based on follow-up studies of bright sources found by \emph{IRAS}, they are biased against its earliest phases, which are expected to be colder and thus IR-quiet. To make progress, dedicated unbiased surveys of the IR-quiet phases of high-mass star formation were therefore required.

\subsection{IR-quiet high-mass protostars}
\label{s:irq}

The precursors of UC\hii regions and IR-bright protostars could be the high-mass analogs of low-mass prestellar cores and Class~0 protostars and thus massive cloud structures, cold enough not to be detected by near- to mid-IR surveys. For the past ten years, they have been searched for through mid-IR, far-IR, and (sub)millimeter surveys. In this section, we review the major studies that indeed found precursors of IR-bright protostars, with luminosity lower than $10^3-10^4~\lsun$ and size varying from 1~pc to 0.01~pc.

\subsubsection{Serendipitous discoveries}
\label{s:serendipitous}

The first good candidates for being IR-quiet precursors of high-mass stars have been found by two different observational methods. The first one uses high-density tracers, often submillimeter continuum, to map the surroundings of high-mass IR-bright objects associated with well-known \hii regions, H$_2$O or CH$_3$OH masers, or \emph{IRAS} sources. Many of these mappings have serendipitously revealed some dense and massive cloud fragments which remain undetected at mid-IR wavelengths \citep[e.g.,][]{motte03, garay04, hill05, klein05, sridharan05, beltran06, thompson06,BeSt07}. These studies are evidently plagued by very low-number statistics and large inhomogeneity because the cloud fragments identified this way have sizes ranging from 0.1~pc to more than 1~pc.
\begin{figure}[h]
\vskip -2.2cm
\centerline{\includegraphics[width=11.5cm,angle=270]{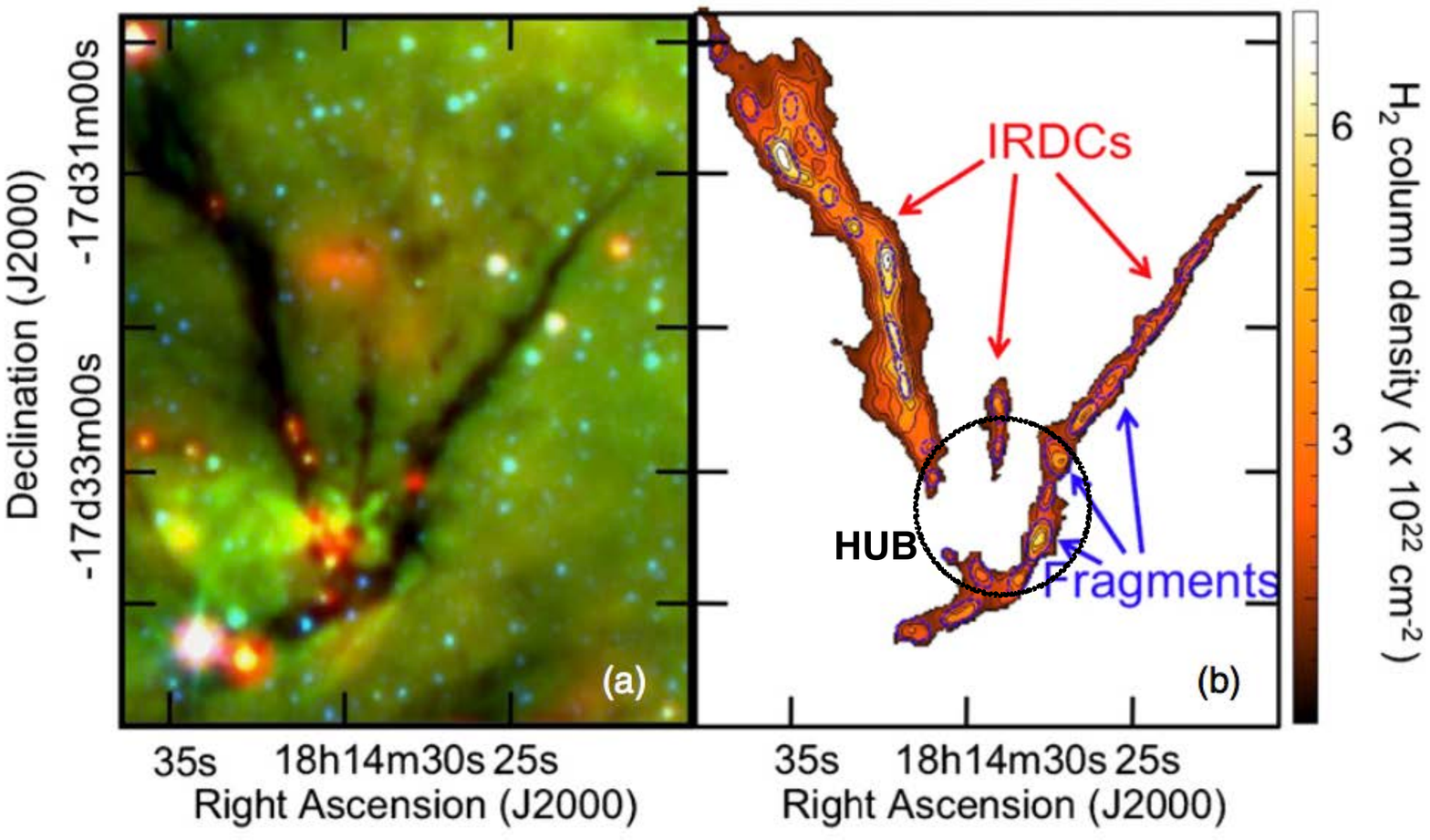}}
\vskip -1.5cm
\caption{IRDCs seen in {\bf (a):} extinction in the \emph{Spitzer} three-color image  (red=24~$\mu$m, green=8~$\mu$m, and blue=3.6~$\mu$m) and in {\bf (b):} H$_2$ column density in a map constructed from the 8~$\mu$m extinction. Fragments/MDCs are the $\sim$0.1~pc substructures seen within $\sim$1~pc IRDCs/ clumps connecting toward a hub. Adapted from \cite{PeFu10} with permission.}
\label{f:irdc}
\end{figure}

A second method is to search for compact sources within cold clouds seen in absorption against the diffuse mid-IR background of square degrees images taken by the \emph{ISO}, \emph{MSX}, \emph{Spitzer}, and \emph{Herschel} space observatories. Indeed, these absorption features, referred as IRDCs, could be the footprints of cold cloud structures (see, e.g., \textbf{Figure}~\ref{f:irdc}). These IRDCs surveys provide large samples of IR-quiet sources generally located at large and inhomogeneous distances from the Sun \citep[e.g.,][]{perault96, egan98, simon06a, BuTa09, PeFu09}. Their existence and gas content generally are confirmed by maps of high-density cloud tracers \citep[e.g.,][]{carey00, teyssier02, simon06b, rathborne06, ragan06, sakai08}. Even in the most recent studies by \cite{BuTa09} and \cite{PeFu10}, the selected sources,  called in their paper IRDC cores or fragments, have the $0.1-1$~pc sizes and should harbor several collapsing protostars and/or prestellar cores (see \textbf{Table}~\ref{tab:cloudstr}). Their large-scale structure often resembles filament hubs (see \textbf{Figure}~\ref{f:irdc}), as initially proposed by \cite{myers09}.
\begin{marginnote}[]
\entry{\emph{ISO}}{The IR Space Observatory was designed to image selected areas at 2.5 to $240~\mu$m.}
\entry{\emph{MSX}}{The Midcourse Space Experiment is a military satellite experiment, which mapped the Galactic plane at 4 to $21~\mu$m.}
\entry{\emph{Spitzer}}{This space observatory performed continuum imaging and spectroscopy at $3.6-160~\mu$m. One of its three instruments is still partly operable.}
\entry{\emph{Herschel}}{This space observatory was equipped with the largest IR telescope ever launched and three instruments (SPIRE, PACS, HIFI) sensitive in photometry and spectroscopy to the far IR and submillimeter wavebands ($70-500~\mu$m).}
\end{marginnote}

The sources identified above are definitively colder and less luminous than the high-mass IR-bright sources discussed in Sect.~\ref{s:irb}: $10-20$~K versus $30-100$~K and $10^2-10^3~\lsun$ versus $>$$10^4~\lsun$. They could be either starless clumps or clumps hosting IR-quiet protostars, depending on the existence of protostellar activity signatures such as outflows, hot cores, or masers. Given their large size and moderate mass (see \textbf{Table}~\ref{tab:cloudstr}), many of them are however probably not dense enough to form high-mass stars in the near future. This statement is confirmed by recent observations \citep{ragan06, rathborne09} as well as statistical arguments provided by the complete catalog of \emph{Spitzer} IRDCs in the Milky Way \citep[see][]{PeFu09}. Inspired by the sequence from Class~0 to Class~I observed for low-mass protostars \cite[e.g.,][]{andre00}, the ratio of submillimeter to bolometric luminosity has also been employed for OB-type protostellar objects separating massive Class~0-like from high-luminosity protostellar objects \citep[with $L_{\rm submm}/L_{\rm bol} \ge 1\%$, see e.g.,][]{molinari98, motte03, molinari08}. Only a dozen of sources identified by the above methods were studied with enough spatial resolution, spectral energy distribution (SED) coverage, and follow-up studies to qualify as high-mass equivalent of Class~0 protostars \citep{hunter98, molinari98, sandell00, garay02, SaSi04}.

\begin{textbox}[h]
\section{\emph{Herschel} surveys of nearby, massive molecular cloud complexes}
HOBYS, the \emph{Herschel} imaging survey of OB Young Stellar objects \citep[][see http:\/\/hobys-herschel.cea.fr]{motte10}, aims at making the census of MDCs in essentially all the molecular cloud complexes at less than 3~kpc (7 out of the 10 molecular complexes of \textbf{Table}~\ref{tab:hobysC} in Sect.~\ref{s:HOBYSclouds}). Its wide-field photometry part with both the SPIRE and PACS cameras along with the necessary interferometric follow-ups is expected to multiply by large factors the number of high-mass analogs of Class~0 protostars known before 2010. 

Among the three most nearby, massive molecular cloud complexes not targeted by HOBYS, Carina was imaged with \emph{Herschel} by \cite{preibisch12} and G345 and Vulpecula  were covered by the \emph{Herschel} imaging of the Galactic Plane survey \citep[Hi-GAL,][]{molinari10}. 
Completing the imaging of entire molecular complexes, \emph{Herschel} focused on several clumps forming high-mass stars \citep[e.g.,][]{zavagno10SI, ragan12}. 

The three-color  \emph{Herschel} images (red\,=\,250~$\mu$m, green\,=\,160~$\mu$m, and blue\,=\,70~$\mu$m) obtained for the ten most nearby, massive molecular complexes are given in Appendix \citep[see also][]{motte10, molinari10, nguyen11a, hill12, preibisch12, fallscheer13, rivera13, schneider16b}.
\end{textbox}

\subsubsection{Surveys within entire molecular cloud complexes}
\label{s:survey}
\begin{figure}[h]
\hskip 0.8cm\includegraphics[angle=0,width=5.5cm]{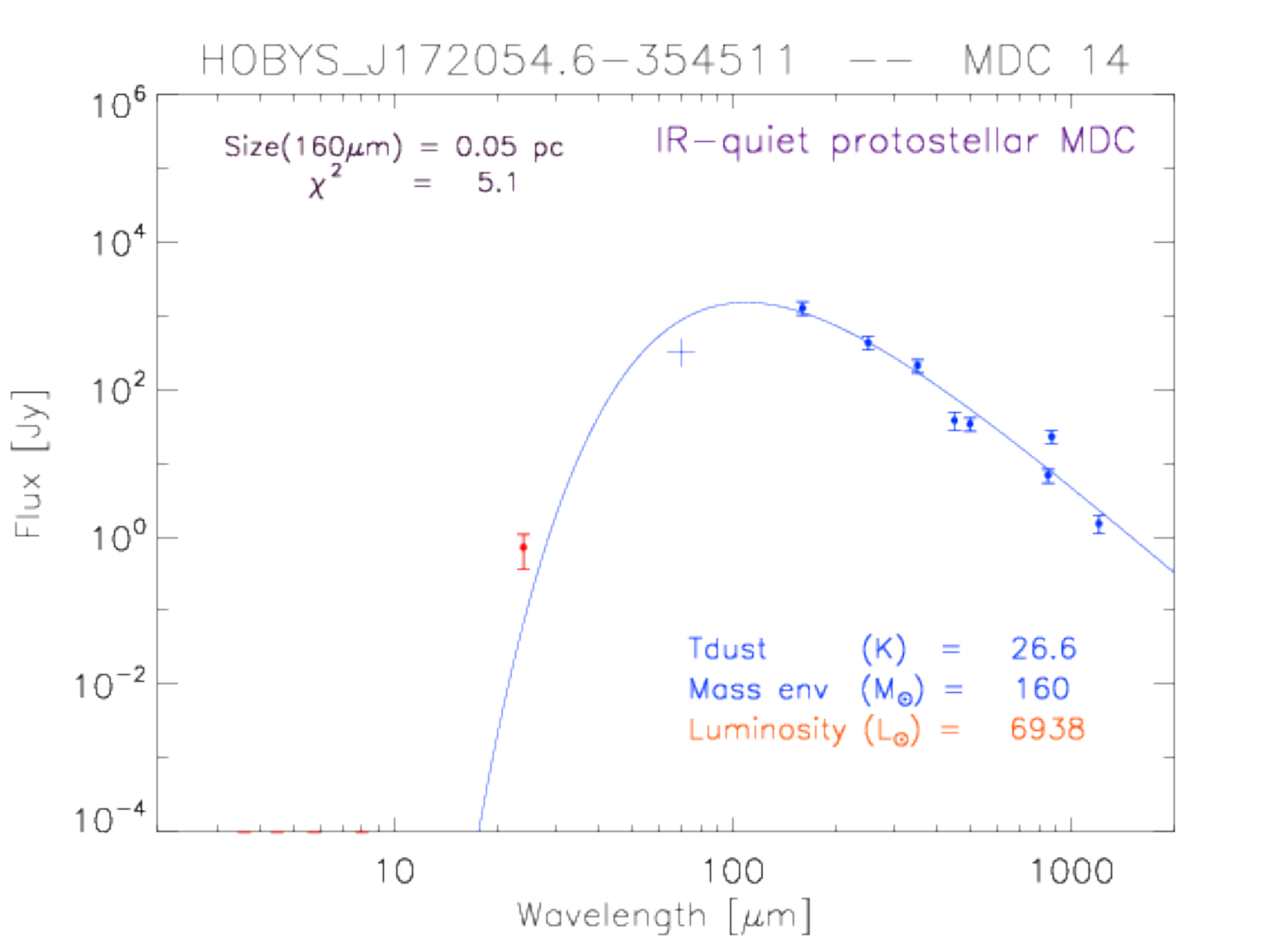}\hskip 3.1cm\includegraphics[angle=0,width=4cm]{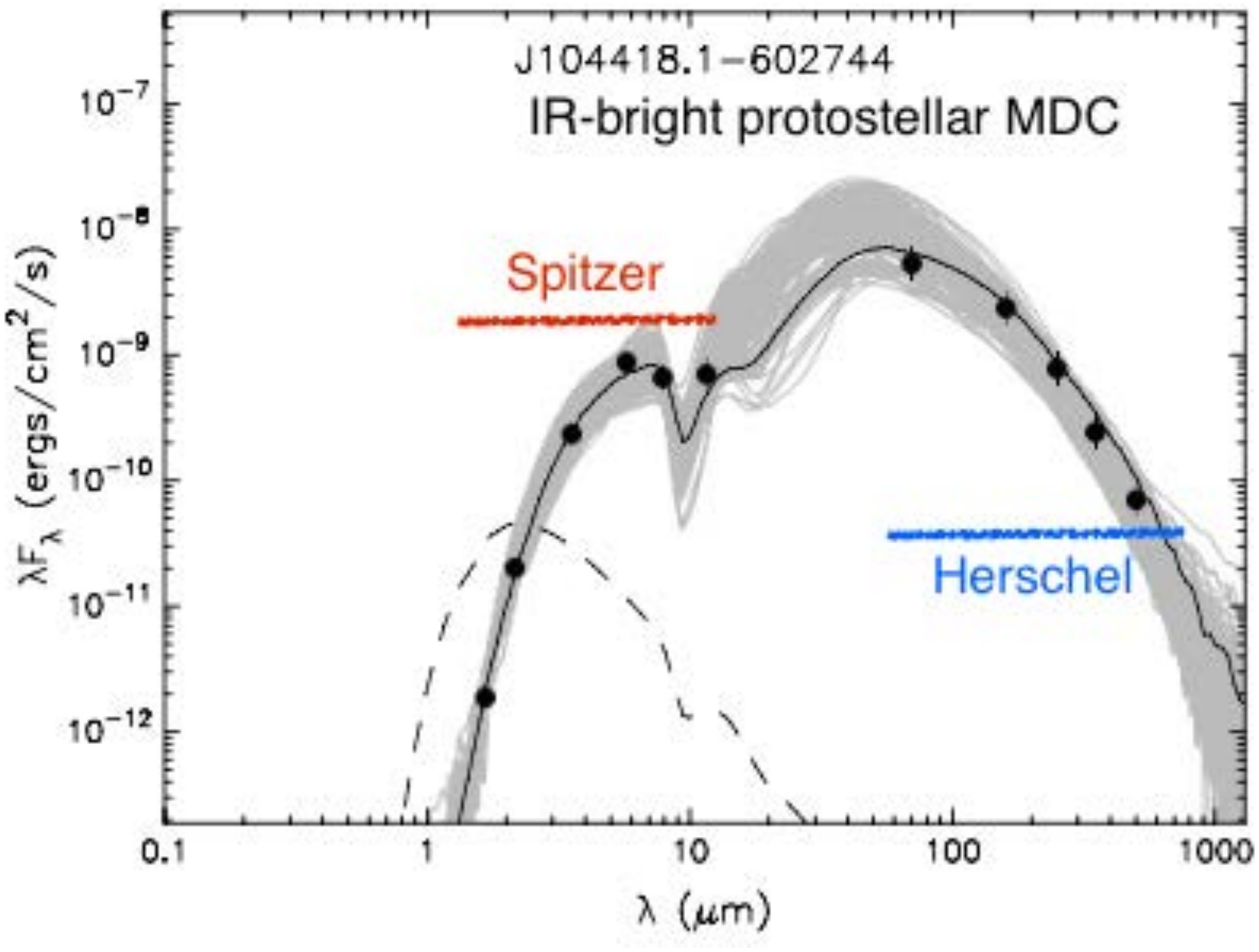}
\vskip 0.5cm
\caption{SEDs built from \emph{Herschel} and \emph{Spitzer} fluxes of in {\bf (Left):} the IR-quiet NGC6334-I(N) MDC and in {\bf (Right):} the IR-bright J104418.1-602744 MDC in Carina. When compared to modified blackbody models, the far-IR to submillimeter fluxes of NGC6334-I(N) and J104418.1\-602744 suggest $26$~K and $\sim$50~K envelopes. More complex models by, e.g., \cite{robitaille07} are necessary to fit the complete SED of IR-bright MDCs. Adapted from \cite{tige17} and \cite{gaczkowski13} with permission.}
\label{f:sed}
\end{figure}

To go beyond, one needed to search, in a systematic and unbiased way, for high-mass analogues of prestellar cores, Class~0 and Class~I protostars. If they exist and are somewhat similar to their low-mass counterparts, one should look for small-scale cloud fragments: $\sim$0.02~pc for protostars \citep{bontemps10} and $0.02-0.1$~pc for prestellar cores. They should also be massive enough to allow the formation of a couple of high-mass star, leading to huge volume-averaged densities, $n_{\mbox{\tiny H$_2$}}=10^6-10^8$~cm$^{-3}$. High-mass star progenitors should therefore be best detected via (sub)millimeter or far-IR dust continuum (see \textbf{Figure}~\ref{f:sed}). This happens to be the wavelengths domain of ground-based (sub)millimeter telescopes like the IRAM 30~m and APEX, of the \emph{Herschel} space observatory, and of submillimeter interferometers such as NOEMA and ALMA. To achieve sufficient spatial resolution and statistics, it is judicious to focus on the closest molecular cloud complexes which are actively forming OB stars. Ten such complexes were identified at intermediate distances, 1.4 to 3~kpc (see Sect.~\ref{s:HOBYSclouds}), ensuring reasonable $\sim$0.1~pc resolution with past and present single-dish submillimeter facilities:  \emph{HPBW}$=8''-19''$ with IRAM 30~m, APEX, and JCMT, $33''$ with CSO and \emph{HPBW}$=6''-36''$ with \emph{Herschel}. With $\sim$0.1~pc typical sizes, these so-called MDCs are intermediate cloud structures between clumps like IRDCs or HMPOs and individual cores forming single stars or binaries (see \textbf{Table}~\ref{tab:cloudstr}). See the sidebar titled \emph{Herschel surveys of nearby, massive molecular cloud complexes}. As discussed in Sect.~\ref{s:HOBYSclouds}, the amount of molecular gas contained in these ten most nearby, massive molecular cloud complexes should statistically permit studying the precursors of OB stars with masses up to $20~\msun$. Multi-tracer studies of such complexes are thus expected to provide more statistically significant and more homogeneous samples of precursors of high-mass stars than any of the studies discussed in Sect.~\ref{s:serendipitous}. 
\begin{marginnote}[]
\entry{\emph{IRAM}}{The Institut de RadioAstronomie Millim\'etrique operates, at $850~\mu$m to 3~mm, a 30~m radiotelescope and the NOEMA interferometer, soon consisting of twelve 15~m antennas.}
\entry{\emph{APEX}}{The Atacama Pathfinder Experiment is a 12~m telescope optimized at $350~\mu$m to 1~mm.}
\entry{\emph{ALMA}}{The Atacama Large Millimeter Array is the largest submillimeter interferometer with fifty 12~m and twelve 7~m antennas, optimized at $450~\mu$m to 3~mm.}
\entry{\emph{CSO and JCMT}}{The 10~m Caltech Submillimeter Observatory and 15~m James Clerk Maxwell Telescope were/are working from $450~\mu$m to 1-3~mm.}
\end{marginnote}

\begin{figure}[h]
\vskip -0.2cm
\hskip -1.7cm \includegraphics[angle=0,width=5.9cm]{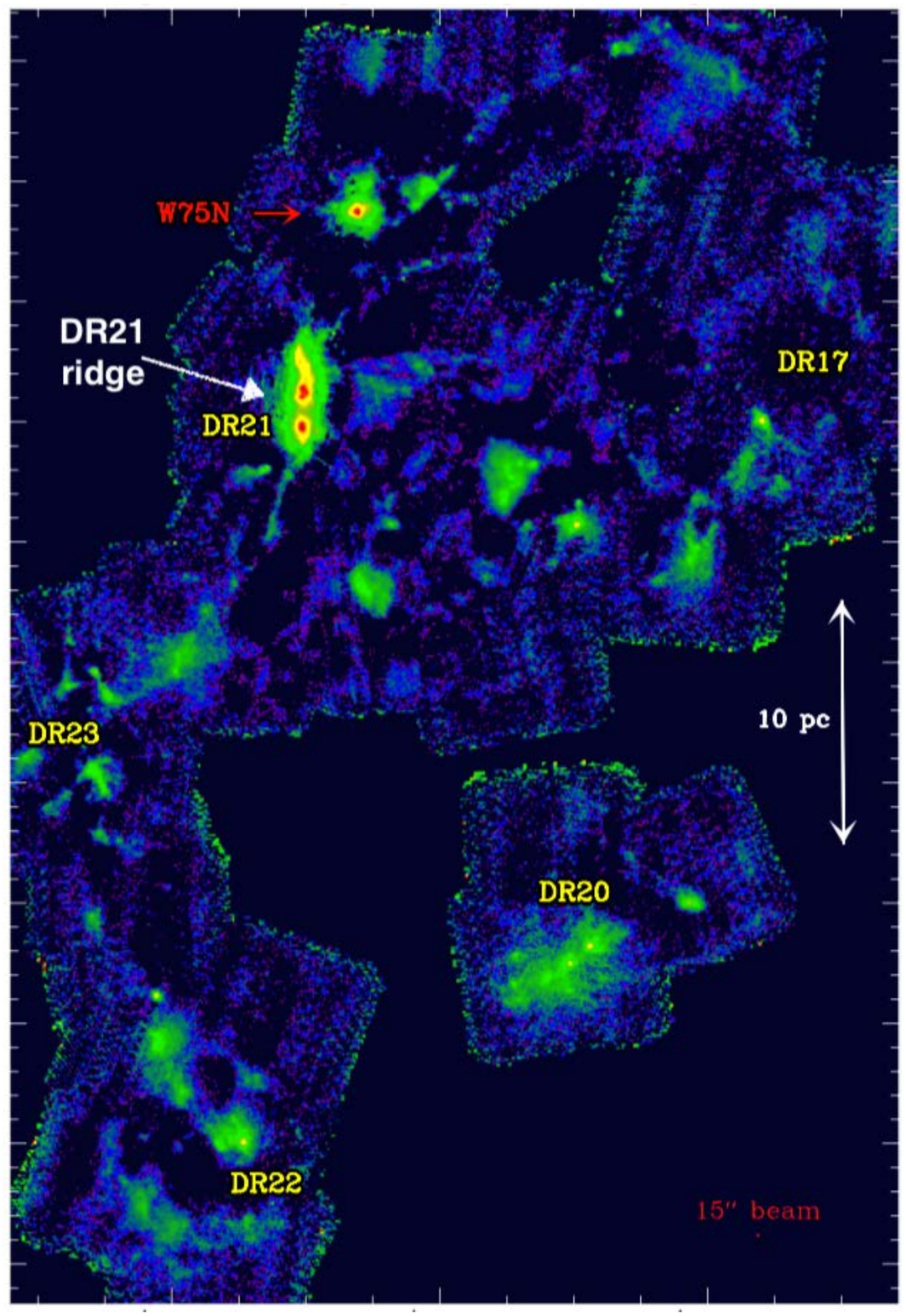} \vskip -13.3cm \hskip 2.6cm \includegraphics[width=0.72\textwidth]{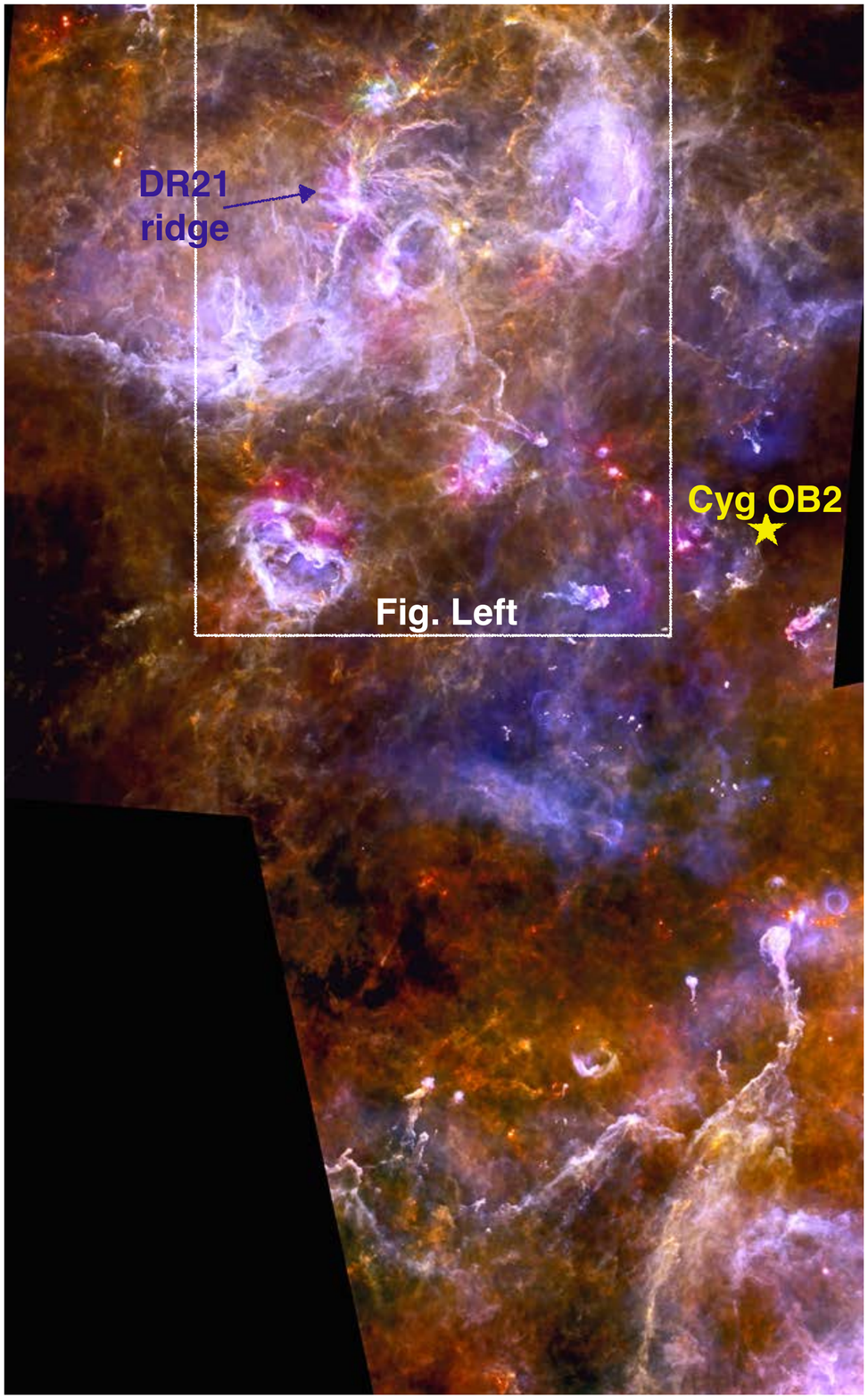}
\vskip -0.2cm
  \caption{The Cygnus~X molecular cloud complex imaged in \textbf{(Left):} at 1.2~mm with the IRAM 30~m and in \textbf{(Right):} at 250\,$\mu$m, 160\,$\mu$m, and 70\,$\mu$m (RGB) with \emph{Herschel}. The mosaics   approximately cover the northern part of Cygnus~X \textbf{(Left)} and the $5^\circ\times2.5^\circ$ (or 120\,pc\,$\times$\,60\,pc) area of the complete Cygnus~X complex \textbf{(Right)}. 
  \textbf{Left:} At the center of the CygX-North complex, one finds a  5~pc-long dominating filament called the DR21 ridge. It contains half of the Cygnus~X MDCs. 
  \textbf{Right:} The blue diffuse emission corresponds to the photo-dissociation region associated with the massive Cyg~OB2 cluster. Earlier stage star-forming sites are themselves seen as pink filaments and MDCs.
Adapted from \cite{motte07} and \cite{schneider16b}  with permission.}
  \label{f:cygnus}
\end{figure}

Among the most active star-forming complexes at less than 3~kpc, Cygnus~X is the one that has caught most of the attention. According to \cite{schneider06}, this region is dominated by a massive ($3.4\times 10^6~\msun$) molecular complex, tightly associated with several OB associations, the largest being Cyg~OB2. It is located at only 1.4~kpc from the Sun \citep{rygl12}. The high-density clouds of Cygnus~X have been completely imaged in millimeter continuum with the IRAM 30~m and as part of \emph{Herschel}/HOBYS \citep[][see \textbf{Figures}~\ref{f:cygnus}]{motte07, hennemann12, schneider16b}. The millimeter imaging survey of the entire Cygnus~X molecular complex has revealed hundreds of 0.1~pc dense cores, among which $\sim$42 proposed to be massive enough, $>$$40~\msun$, to be MDCs (see \textbf{Table}~\ref{tab:cloudstr}).

Similar studies have been done for the NGC~6334-6357 and Carina MDCs using a millimeter imaging of the complex \citep{russeil10} and \emph{Herschel} images \citep{tige17, gaczkowski13}. As for the most remote star-forming regions, they have right-away been imaged with submillimeter interferometers to pinpoint 0.1~pc MDCs \citep[e.g.,][]{beuther07b, zhang09, wang11, louvet14}.

Prototellar MDCs distinguish from starless MDC candidates discussed in Sect.~\ref{s:pcore} by the fact that they drive outflows, power hot cores and masers, and/or are associated with mid-IR \emph{Spitzer} emission. Outflows are traced by high-velocity wings of, e.g. CO or SiO, molecular lines or suggested by extremely green \emph{Spitzer} objects (EGOs) while hot cores are detected through line forests \citep[e.g.,][]{motte07,cyganowski11, wang11, louvet16}.
Before the advent of \emph{Herschel}, \cite{motte07} proposed to use the mid-IR fluxes detected toward MDCs to identify high-mass IR-bright MDCs, with a luminosity larger than $10^3~\lsun$ and thus a protostellar embryo larger than 8~$\msun$. In this scheme, IR-quiet prototellar MDCs would themselves consist of a couple of stellar embryos of mass smaller than 8~$\msun$ buried in a 0.1~pc MDC of mass larger than $40~\msun$. IR-quiet protostellar MDCs were thus recognized as small scale, 0.1~pc, and dense, $10^6$~cm$^{-3}$, cloud fragments with no or weak mid-IR emission \citep[$F_{\rm 21\mu m}<$10~Jy,][]{motte07} but clear signposts of high-mass protostellar activity. 
With \emph{Herschel}, a much more direct classification arises from the complete SED one can build for MDCs. While IR-quiet protostellar MDCs can generally be described by simple modified blackbody models, IR-bright protostellar MDCs display clear mid-IR, $4-70~\mu$m, excesses (compare \textbf{Figures}~\ref{f:sed}). Interestingly, \cite{tige17} showed that the $M_{\rm env}/L_{\rm bol}$ ratio one can derive from such well-constrained SEDs is consistent with the classification made by solely using the mid-IR flux threshold proposed by \cite{motte07}.
In Cygnus~X, 17 MDCs qualify as good candidates for hosting IR-quiet high-mass protostars, i.e. protostellar embryos of masses smaller than $8~\msun$ surrounded by 0.02~pc envelopes massive enough to form at least one high-mass star \citep{motte07}.
 As a matter of fact, the five most massive IR-quiet MDCs have been confirmed to host nine individual high-mass protostars, driving outflows \citep{bontemps10}.

The velocity dispersion of IR-quiet MDCs, estimated from the width of molecular lines such as NH$_3$, N$_2$H$^+$, or N$_2$D$^+$, are $\sigma_{\rm MDCs} \sim 1-2$~ km\,s$^{-1}$ \cite[e.g.,][]{ragan06,ragan12b, bontemps10, csengeri11b, wienen12, kauffmann13b, tan13}.
Therefore, despite the high level of turbulence measured in IR-quiet MDCs, they are virially supercritical and should be collapsing. It recalls the global hierarchical collapse model of \cite{vazquez09, vazquez17} and \cite{ballesteros11}, where velocity dispersions caused primarily by infall motions decrease with physical scales.
The alternative interpretation is, as in the turbulent core model of \cite{MKT02}, that IR-quiet starless MDCs and prestellar cores are supported against collapse by a strong magnetic field, which unfortunately remains difficult to measure (see Sect.~\ref{s:bfield}).

\subsection{Lifetime of high-mass star precursors and protostellar accretion rate}
\label{s:prot}

\subsubsection{High-mass protostellar lifetime}
\label{s:lifetimeProt}

The surveys mentioned in Sects.~\ref{s:irb} and \ref{s:survey} allow, for the first time, getting statistical lifetime estimates for each of the embedded phases of high-mass star formation (see \textbf{Table}~\ref{tab:lifetime}). Entries in \textbf{Table}~\ref{tab:lifetime} are ordered by spatial scales and evolutionary stages. High-mass star precursors, far from being mutually exclusive, form a Russian-doll structure, reflecting both the surveys resolution limitation and the hierarchical structure of clouds (see terminology in Sect.~\ref{s:HMSF}).

\begin{table}[h]
\caption[]{Characteristics and lifetime estimates of high-mass star precursors}
\label{tab:lifetime}
\centering
\begin{tabular}{|l|ccccc|}
\hline
& Median 	& Envelope	& Density 					& Statistical 		& References\\
& FWHM 	& Mass			& $<n_{\rm H_{2}}>$$^{\rm a}$	& Lifetime$^{\rm b, c}$ 	& \\
& [pc]	& [$\msun$]		& [cm$^{-3}$]				& [yr] 			& \\
\hline
Massive starless clumps & $\sim$0.5 & $100-10^4$	& $10^3-10^5$ & $<$$1-3 \times 10^4$	&(1),(2),(3),(4) \\
\hline
UC\hii regions	& $\sim$0.1	& $1-10^3$	& $10^3-10^5$ & $\sim$$3 \times 10^5$	&(5),(6) \\
IR-bright MDCs & $\sim$0.1	& $40-10^3$	& $10^5-10^7$ & $0.6-0.9 \times 10^5$	&(1),(2),(7) \\
IR-quiet MDCs 	& $\sim$0.1	& $40-10^3$	& $10^5-10^7$ & $0.5-1 \times 10^5$	& (1),(2),(7) \\ 
Starless MDCs	& $\sim$0.1 & $30-80$		& $\sim$$10^6$ & $<$$1\times 10^4$
	& (1),(8),(9),(10) \\
\hline
IR-bright high-mass protostars & $\sim$0.02$^{\rm d}$ & &
& $\sim$$1.2\times 10^5$ & (7) \\
IR-quiet high-mass protostars & $\sim$0.02 & $10-100$	& $10^6-10^8$ 	& $\sim$$2 \times 10^5$	& (11),(12),(7) \\
\emph{All} high-mass protostars & $\sim$0.02 & $>$10	& $\sim$$10^7$ & $\sim$$3 \times 10^5$	& (13),(7)\\
High-mass prestellar cores & $0.01-0.1$$^{\rm d}$	& $>$30$^{\rm d}$	& $10^5-10^7~^{\rm d}$ & $<$$1-7 \times 10^4$	& (13),(7)\\
\hline
\end{tabular}
\begin{tabnote}
$^{\rm a}$ Median value of the volume-averaged density, averaged over the a sphere with a FWHM diameter:
$<n_{\rm H_{2}}>~=  \frac{\rm\it Mass}{\frac{4}{3} \pi \mu m_{\rm H} ({\rm\it FWHM}/2)^{3}}$, where $\mu = 2.8$ is the mean molecular weight \citep{kauffmann08}, and $m_{\rm H}$ is the hydrogen mass. \\
$^{\rm b}$ The numbers of OB3 stars in Cygnus~X and thus the statistical lifetimes of MDCs have been corrected from the values given in \cite{motte07} \citep[see Table~5 of][]{russeil10}.\\
$^{\rm c}$ To estimate the lifetime of individual protostars and prestellar cores, protostellar MDCs are assumed to host $\sim$2 protostars (see Sect.~\ref{s:lifetimeProt}, \citealt{bontemps10}), while starless MDCs could host at most one high-mass prestellar core (see Sect.~\ref{s:lifetimePcore}, \citealt{tige17}). \\
$^{\rm d}$ Characteristics, which is postulated and thus not (yet) measured.\\
References: (1) \cite{motte07}; (2) \cite{russeil10}; (3) \cite{svoboda16}; (4) \cite{csengeri14}; (5) \cite{WoCh89}; (6) \cite{mottram11}; (7) \cite{tige17}; (8) \cite{BuTa12}; (9) \cite{tan13}; (10) \cite{peretto14}; (11) \cite{bontemps10}; (12) \cite{louvet14}; (13) \cite{duarte13}.
\end{tabnote}
\end{table}
 
From a Galactic plane survey of bright and red \emph{IRAS} sources, \cite{WoCh89} estimated that the typical lifetime of the UC\hii phase was about $10^5$~yr. This has recently been confirmed by statistical lifetimes determined by \cite{mottram11} for compact \hii regions carefully identified by the Red \emph{MSX} Source (RMS) survey \citep{lumsden13}, $\sim$$3\times 10^5$~yr  (see \textbf{Table}~\ref{tab:lifetime}). These lifetimes are substantially longer than the  timescale predicted by the Str\"omgren theory, initiating the so-called Ôlifetime problemÕ of UCH\mbox{\sc ii}s \citep[e.g.,][]{churchwell02}. The latter has recently been solved by simulations showing a fast decrease of the expansion velocity from its initial sound speed value in ionized gas and thus longer expansion times \citep[see][and references therein]{didelon15}. 

When ground- or space-based submillimeter surveys make a complete census in a single molecular cloud complex, their sample covers every embedded phase of high-mass star formation, the IR-bright and IR-quiet protostellar phases, the starless/prestellar phase if it exists (see Sect.~\ref{s:pcore}), and to a lesser extent the UC\hii phase. The evolutionary stage of MDCs/clumps within these samples is estimated thanks to searches of protostellar activity signatures such as mid-IR, outflow, maser and hot core emission and searches of \hii signatures like free-free centimeter emission. These complete studies provide the statistical base to derive the lifetimes of precursors of UC\hii regions and consequently the first estimate of the high-mass protostellar lifetime. 
Estimated relatively to the known age and numbers of OB stars in the molecular complexes surveyed, typically hundreds of sources of a few $10^6$~yr, the lifetimes of IR-quiet and IR-bright MDCs in Cygnus~X and NGC~6334 are $t_{\rm IRquiet-MDC}\sim5-6\times 10^4$~yr and $t_{\rm IRbright-MDC}\sim8-9\times 10^4$~yr, respectively \citep[][see \textbf{Table}~\ref{tab:lifetime}]{motte07, russeil10}.
Interestingly, using a much larger but less homogeneous sample of IR-bright massive protostellar objects (Galaxy-wide, with $0.1-1$~pc sizes), a similar lifetime value was found for the IR-bright phase of high-mass protostars \citep{mottram11}.
 The protostellar MDC lifetime, derived from the Cygnus~X and NGC~6334 samples, thus is $t_{\rm MDC}\sim1.5\times 10^5$~yr, which roughly corresponds to twice the estimated free-fall time for MDCs, $\tau_{\rm ff-MDC}\sim 9\times 10^4$~yr.
This result is consistent with the idea that MDCs collapse in a few free-fall times, as suggested by \cite{wyrowski16}. But we mention a word of caution anyway: Statistical lifetimes are subject to large observational uncertainties and free-fall times depend on the unconstrained gas density before collapse. Moreover in the scenario of Sects.~\ref{s:EvolSeq} and \ref{s:starburst}, cores feed from their surrounding during bursts of star formation, making such estimates loose all their significance.

\begin{figure}[h]
\vskip 1.8cm
\hskip -1.5cm\includegraphics[angle=270,width=11.5cm]{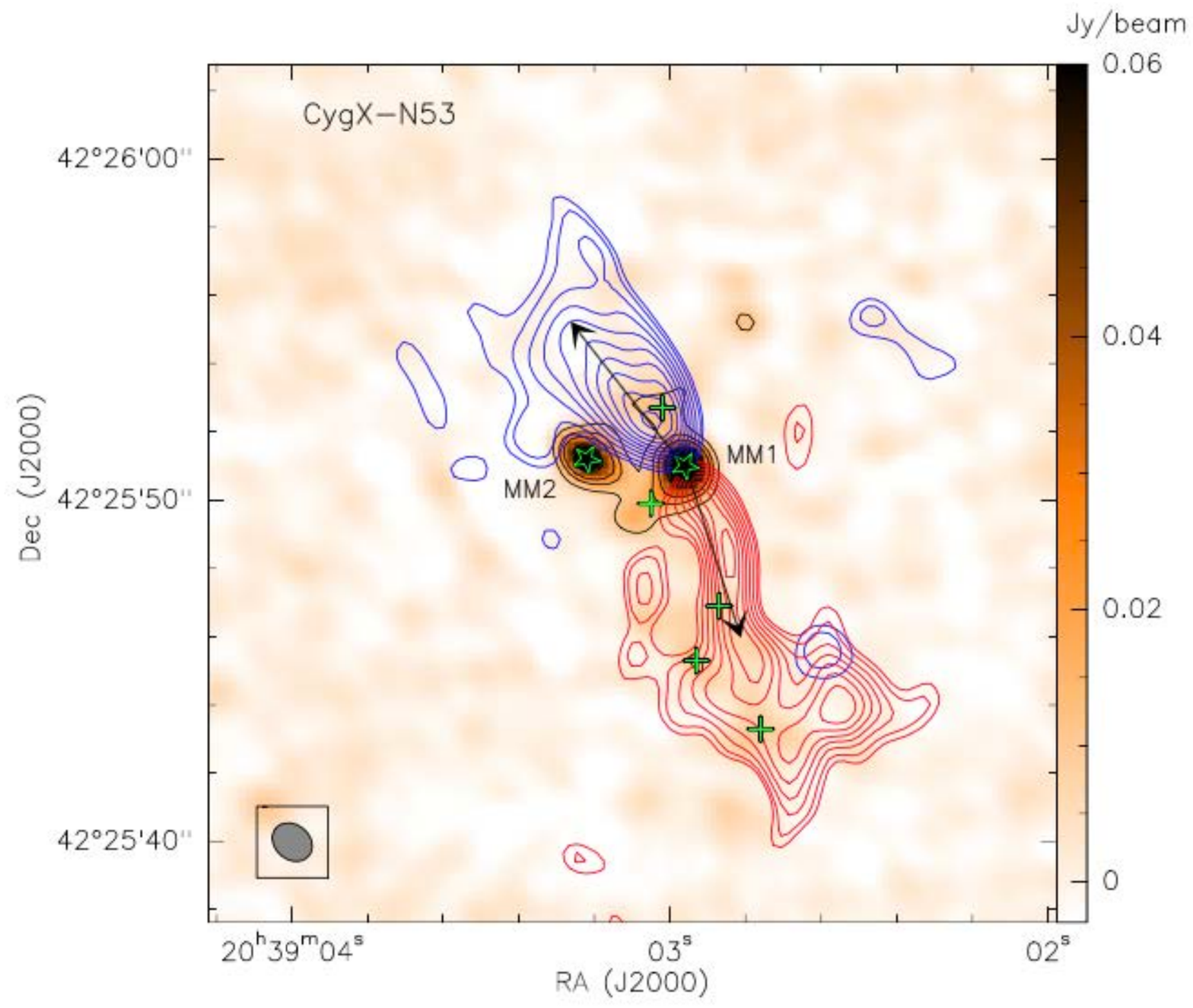}\vskip -7.5cm \hskip 10.5cm \includegraphics[angle=0,width=4.2cm]{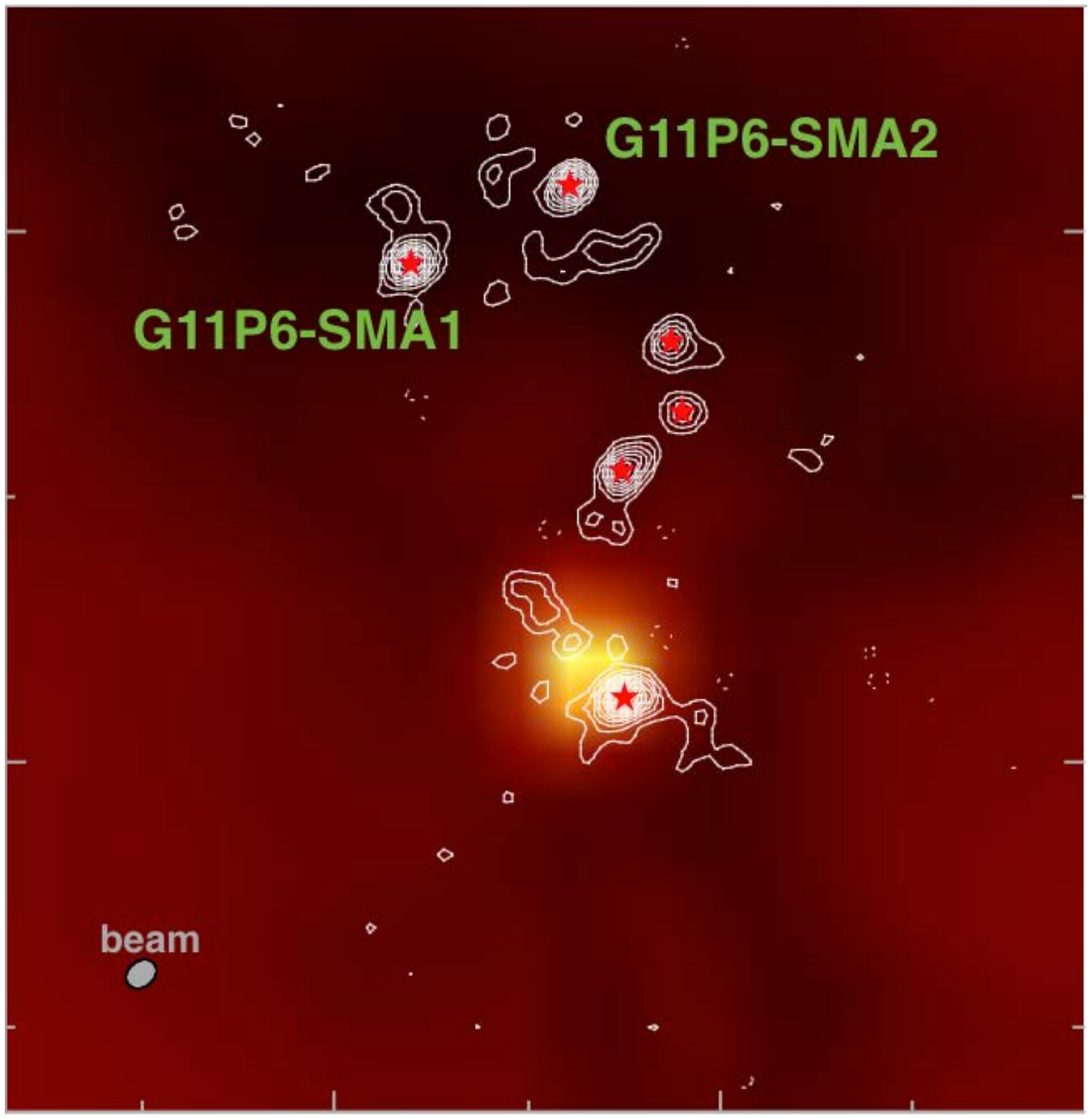}
\caption{The only two high-mass prestellar core candidates known to date, in {\bf (Left):} CygXN53-MM2 discovered in Cygnus~X and in {\bf (Right):} G11P6-SMA1 in IRDC~G11.11-0.12. Their neighboring high-mass IR-quiet protostars CygXN53-MM1 and G11P6-SMA2 drive outflows.
 Cores and envelopes are traced by submillimeter continuum from \cite{bontemps10} (1~mm, $1\,500$~AU resolution, color scale and contours in {\bf Left}) and from \cite{wang14} (880~$\mu$m, $\sim$$2\,500$~AU resolution, contours in {\bf Right}). 
 {\bf Left:} $^{12}$CO high and low-velocity line wings locate the red and blue lobes (contours) of the protostellar outflow. 
{\bf Right:} \emph{Spitzer} $8~\mu$m color-scale image, which separates IR-quiet high-mass cores (G11P6-SMA1 and SMA2) from IR-bright protostars.
Adapted from \cite{duarte13} and \cite{wang14} with permission.}
\label{f:N53}
\end{figure}

The relevance of single-dish (sub)millimeter and far-IR surveys for discussing the statistical protostellar (and prestellar) lifetime of high-mass stars is limited by the fact that they do not have the necessary angular resolution to pinpoint individual cores forming high-mass stars in high-pressure environments. To alleviate this difficulty one can observe MDCs at higher-angular resolution with (sub)millimeter interferometers. A handful of IR-quiet MDCs in Cygnus~X and several massive fragments of IRDCs were studied using IRAM/NOEMA \citep{bontemps10, beuther15}, SMA \citep{zhang09, rathborne11,wang11, wang14}, and ALMA \citep[e.g.,][]{peretto13, tan13}. 
\begin{marginnote}[]
\entry{\emph{SMA}}{The SubMillimeter Array consists of eight 6~m antennas, functioning from $850~\mu$m to 1~mm.}
\end{marginnote}
Unfortunately, only a few of these studies reach the $\sim$0.02~pc size of individual protostars \citep{bontemps10, wang11, wang14}. At this scale, interferometric observations revealed a large concentration of the gas into a few massive cores and much fewer low-mass cores than predicted by the IMF (see \textbf{Figure}~\ref{f:N53} Left). In contrast to IR-quiet objects, IR-bright MDCs and HMPOs fragment into a large number of low-mass fragments as expected in the thermal Jeans process \citep{palau15, cyganowski17}. However, protostellar feedback (heating, outflow, etc...) distorts the envelope structure and makes fragmentation level more difficult to study without a  good dust temperature model or a complete modeling of the radiative transfer \citep{beuther07b, leurini07b, wang13}. In \textbf{Table}~\ref{tab:lifetime}, lifetime estimates of IR-bright MDCs therefore assume the same fragmentation level as for the better-constrained IR-quiet MDCs.

In Cygnus~X, nine high-mass IR-quiet protostars have been identified within the 5 surveyed IR-quiet MDCs, making an average of $\sim$2 high-mass protostars of $\sim$0.02~pc typical size per $\sim$0.1~pc MDCs \citep{bontemps10}. This high concentration of mass into a small number of high-mass protostars makes the statistical protostostar lifetime a factor of only two longer than the lifetimes estimated for MDCs, $t_{\rm HMprotostar}\sim 3\times 10^5$~yr (see \textbf{Table}~\ref{tab:lifetime}). This exact same value was proposed by \cite{duarte13} based on the observed power of outflows driven by Cygnus~X high-mass protostars. The lifetime of high-mass protostars also corresponds to ten times their free-fall 
time\footnote{Free-fall time is measured from the density averaged over the full MDC volume, which is eight times smaller than $<n_{\rm H_{2}}>$ in \textbf{Table}~\ref{tab:lifetime}.},
which for cloud structures with full-volume averaged densities of $\sim$$1.3 \times 10^6$~cm$^{-3}$ is $\tau_{\rm ff-HMprotostar}\sim 3\times 10^4$~yr. As pointed out by \cite{duarte13}, because high-mass protostars should be free-falling cloud structures, their long lifetime may imply that the initial prestellar core had hundreds times smaller density than protostellar envelopes and that the latter grow in mass and density as they collapse. The following sections will give some arguments in favor of this scenario. Indeed, high-mass prestellar cores are still to be found (see Sect.~\ref{s:pcore}) and the gas surrounding high-mass protostars and MDCs is observed to be flowing toward protostars (see Sect.~\ref{s:ridge}).

The statistical lifetime of high-mass protostars is in rough agreement with what is found in nearby, low-mass star-forming regions \citep[$\sim2-5\times 10^5$~yr,][]{KeHa95, evans09}. This is in marked contrast with the general belief that high-mass stars are living an accelerated life at all phases. While the $\sim$$3\times 10^5$~yr-long UC\hii phase \citep{mottram11} is a factor of ten shorter than the pre-main sequence star phase of low-mass stars \citep[$\sim$$2\times 10^6$~yr,][]{KeHa95}, the protostellar phase of high- and low-mass stars seems to last a rather similar span of time. This fact may permit star formation events to simultaneously develop both low- and high-mass stars. 

\subsubsection{Protostellar accretion}
\label{s:accretion}
A concerted hunt for sources in the cold pre-UC\hii phase and high-resolution follow-ups are necessary to make definitive progress in building a complete evolutionary scenario and providing empirical classifications of high-mass star precursors. \emph{Herschel} surveys provided robust measurements of the basic properties (bolometric luminosity and mass) of MDCs thanks to the unprecedented wavelength coverage by SPIRE and PACS. This is crucial to building quantitative evolutionary diagrams such as the mass versus luminosity and outflow momentum versus luminosity, $M_\mathrm{env}-L_\mathrm{bol}$ and $F_\mathrm{outflow}-L_\mathrm{bol}$, diagrams.

\begin{figure}[h]
\vskip -0.5cm
 \hskip 0.5cm \includegraphics[angle=0,width=0.33\textwidth]{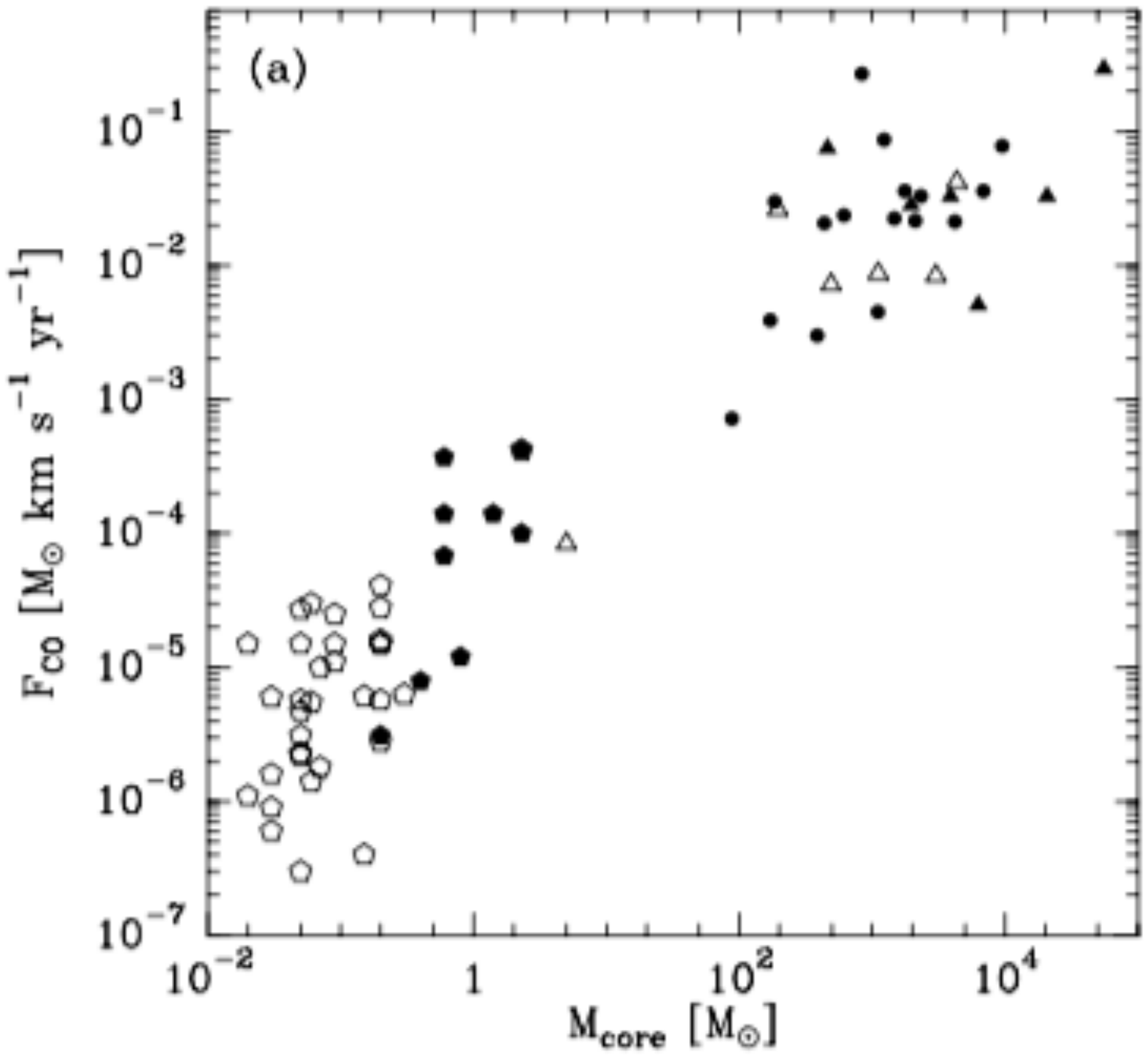} \vskip -9.5cm \hskip 7.8cm\includegraphics[angle=0,width=0.49\textwidth]{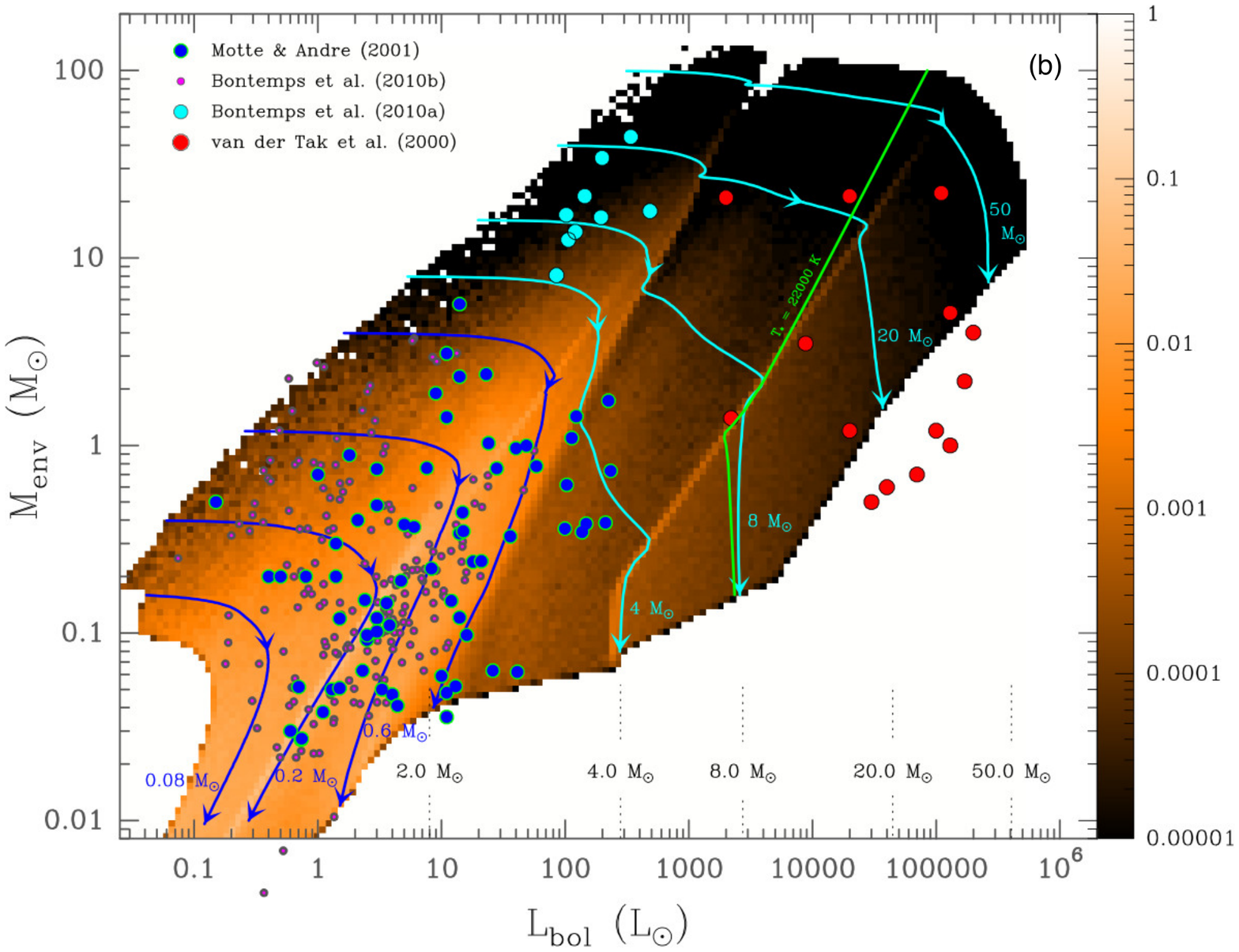}
\vskip -1cm
\caption[]{
Envelope mass of HMPOs/protostars with respect in {\bf (a):} to their outflow momentum (Y axis) and in {\bf (b):} to their bolometric luminosity (X axis). 
The outflow momentum correlation found for low-mass protostars \citep[pentagons,][]{bontemps96} holds for HMPOs \citep[dots and triangles,][]{beuther02b}. This result suggests high-mass stars form through protostellar accretion like low-mass stars but with an enhanced accretion rate. 
In {\bf (a):} the location of high-mass protostars \cite[with 5\,000~AU envelopes, IR-quiet and IR-bright in cyan and red circles,][]{bontemps10,duarte13,vandertak00} favors a scenario with decreasing accretion rates and intermittent accretion. Violet and cyan curves are evolutionary tracks of \cite{duarte13}, with a rate multiplied by 10 during $10\%$ of the protostellar lifetime. The colored area represents the surface density predicted for protostars. The green curve separates high-mass protostars from sources developing an \hii region. 
Adapted from \cite{beuther02b} and \cite{duarte13} with permission.}
\label{f:menvlbol}
\end{figure}

HMPOs selected as bright \emph{IRAS} sources embedded within massive envelopes \citep{beuther02a} are now recognized as having the same star-formation potential as IR-quiet clumps but being more evolved. Because HMPOs are more luminous than $3\times 10^3~\lsun$, they should indeed host high-mass protostellar embryos with masses larger than $8~\msun$. 
Given that most of the high-mass star precursors observed at $1-3$~kpc should become $8-20~\msun$ stars on the main sequence, most HMPOs could contain high-mass protostars which have accreted more than half of their final mass, i.e. the high-mass analog of low-mass Class~I protostars. 
Outflow studies of HMPOs suggest that the high- and low-mass star formation processes are similarly based on protostellar accretion but with much higher rates for the high-mass case, $\sim$$10^{-4}~\msun~\rm yr^{-1}$ \citep[][see \textbf{Figure}~\ref{f:menvlbol}a]{beuther02b}. Outflow studies of samples containing HMPOs and younger MDCs suggested that the outflow strength tracing protostellar accretion decreases with time \citep[e.g.,][]{motte07, lopez11}.
This is consistent with the mass accretion rates derived from optically thick line profiles of MDCs and protostellar cores, $10^{-4}-10^{-2}~\msun~\rm yr^{-1}$ \citep{fuller05, qiu11, herpin12, herpin16, wang13, wyrowski16}. This is also in line with accretion rates measured from the global infalls mentioned in Sect.~\ref{s:ridge}: a few $10^{-3}~\msun~\rm yr^{-1}$ \citep[e.g.,][]{schneider10, peretto13}.
According to the inflow survey by \cite{wyrowski16}, gas surrounding MDCs collapses at small fractions, $3\%-30\%$, of the clump free-fall velocity. The physical processes that could slow the clump and MDC global collapse need to be investigated, but to name a few, one can think of magnetic field, rotation, or adiabatic heating \cite[see Sect.~\ref{s:bfield}, \textbf{Figure}~\ref{f:collapse}c][]{MuCh15}. 

The initial results of \cite{beuther02a} are confirmed by \cite{duarte13}, who studied the younger IR-quiet phase at the protostellar scale of 0.02~pc. They derived proper envelope mass, bolometric luminosity, and outflow momentum flux of individual protostars, and their results are directly comparable with those of low-mass studies \citep{bontemps96,bontemps10SI, MA01}.  \cite{duarte13} built the protostellar evolutionary diagrams of mass versus luminosity and outflow momentum versus luminosity and proposed evolutionary tracks for individual high-mass stars (see, e.g., \textbf{Figure}~\ref{f:menvlbol}b). The dispersion of high-mass protostars in the outflow momentum versus luminosity diagram supports a picture in which accretion is strong but sporadic \citep[][]{duarte13}: variations of a factor of 2 around the mean value of $5\times 10^{-5}~\msun~\rm yr^{-1}$.
Such sporadic accretion is expected when gas reaches the protostellar envelopes through gas inflows, such as those observed by \cite{csengeri11b} and modeled by \cite{smith09}. 

According to \cite{bontemps10}, high-mass protostellar envelopes are all more massive than the thermal Jeans mass of their parental MDC medium, $M_{\rm Jeans}\sim0.3~\msun$ \citep[for $10^5$~cm$^{-3}$, 15~K, and Mach$ \sim 3.5$, see][]{palau15}, and called super-Jeans \citep[see also][]{wang11, tan13, ragan12b, peretto14}. This result along with the low fragmentation level found by \cite{bontemps10} (see also Sect.~\ref{s:lifetimeProt}) tend to exclude the protostar collision model of \cite{bonnell98} and equally favor the competitive accretion/global collapse models of \cite{bonnell01} or \cite{hartmann12} and the turbulent core model of \cite{MKT03}. 

\subsection{First magnetic field measurements in high-mass star-forming regions}
\label{s:bfield} 
As mentioned in the Sect.~\ref{s:intro}, the turbulent core model by \cite{MKT02} proposes that supersonic micro-turbulence prevents the fragmentation of MDCs and favors the formation of high-mass stars. Strong magnetic fields provide a natural alternative to such high turbulence levels. Numerical simulations of magnetized cores indeed demonstrated that the number of fragments is reduced by a factor of $\sim$2 in cores which are moderately super-critical, (M/$\Phi_{\rm B}$)/(M/$\Phi_{\rm B}$)$_{\rm crit}\sim2$ \citep{commercon11, myers13}. These theoretical works, together with observational constraints of massive clumps globally collapsing at only 3\% to 30\% their free-fall velocity \citep{wyrowski16} bring magnetic fields to the forefront of future studies on the high-mass star formation process. Unfortunately, it has been notoriously difficult to measure magnetic fields and even more difficult to follow their topology and strength evolution from clouds to cores.

\begin{figure}[h]
\vskip 1cm
\hskip -3.7cm\includegraphics[angle=270,scale=0.4]{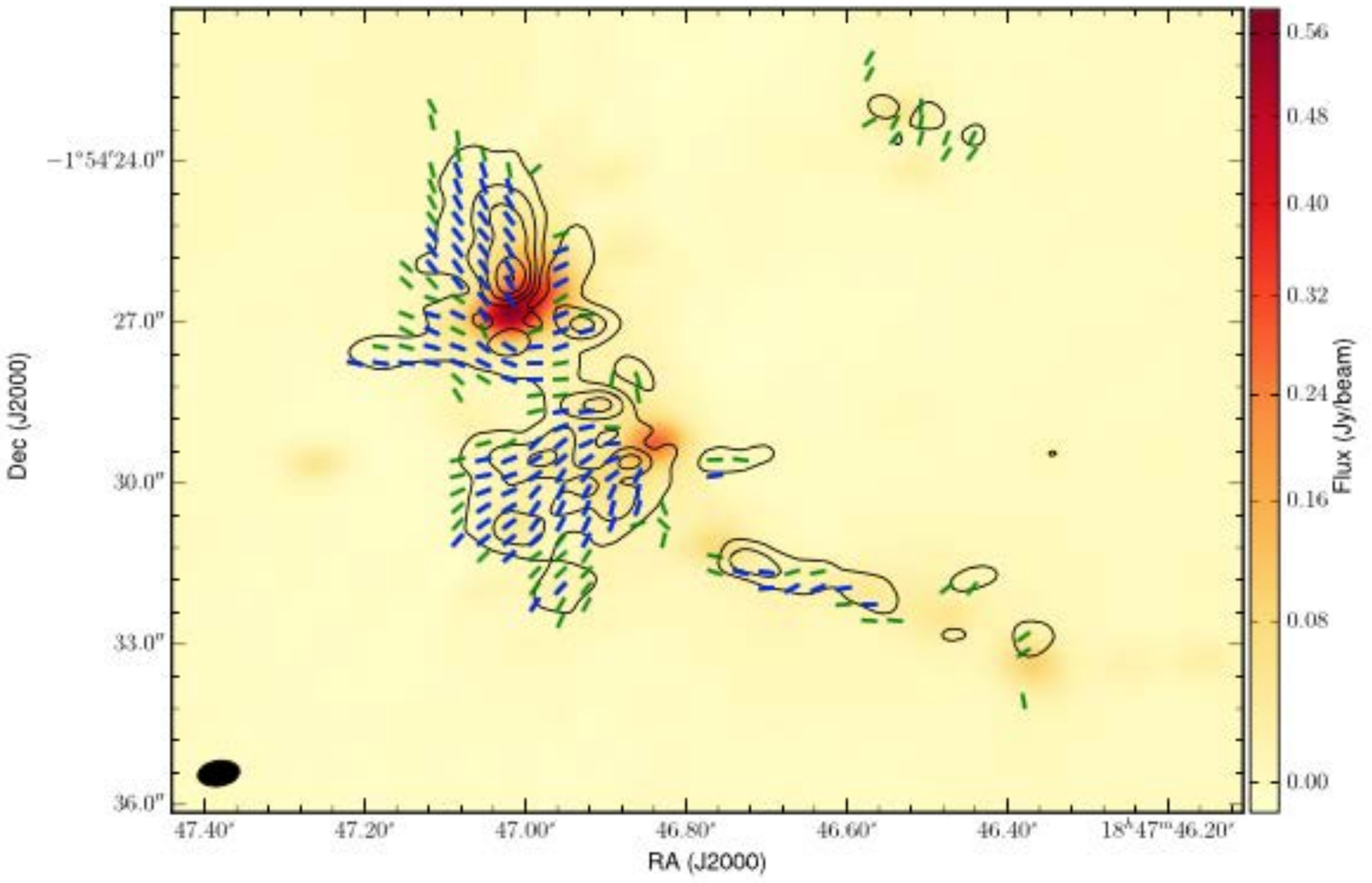}\vskip -4cm \hskip 7cm\includegraphics[angle=270,scale=0.22]{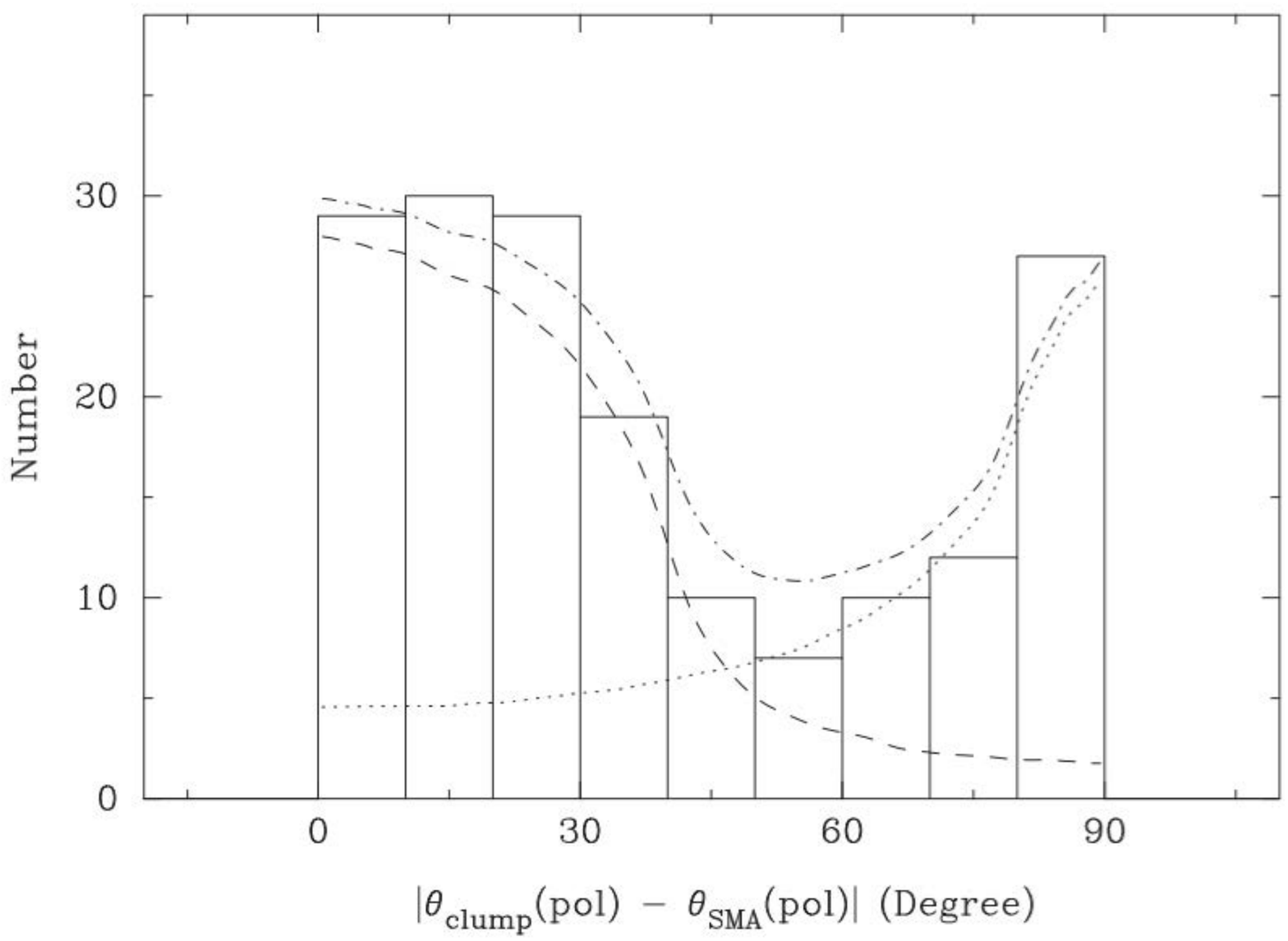}
\vskip 2.5cm
\caption{Dust polarization measurements displaying smooth and ordered polarization patterns.
{\bf Left:} Magnetic field morphology over part of the W43-MM1 ridge. The Stokes I emission is shown as colorscale, the polarized intensity as contours. The magnetic field morphology is represented, every half-beam, by normalized pseudo-vectors at a significance of 3$\sigma$ (green) and 5$\sigma$ (blue).
{\bf Right:} Variations of dust polarization angles from clump to dense core scales, i.e. from 1~pc to 0.1~pc. The dashed-dotted line is the combination of two polarization angle distributions, corresponding to the 0.1~pc MDC polarization being either parallel or perpendicular to the magnetic field lines in their parental clumps. 
Adapted from \cite{cortes16} and \cite{zhangQ14} with permission.}
\label{f:bfield}
\end{figure}

Magnetic field lines traced by optical polarimetry or dust polarization are observed, on 1-10~pc scales, to be perpendicular to long axes of ridges \citep[e.g.,][]{vafi06,li15}. At (sub)millimeter wavelengths, thermal emission is partially depolarized and magnetic field lines get pinched where MDCs are located \citep[see, e.g., \textbf{Figure}~\ref{f:bfield} Left,][]{cortes16}, as in the low-mass case \citep{girart06}. The hour-glass polarization configuration suggests that magnetic field lines are somewhat driven to smaller scales by the clump global collapse. The pioneer survey of  \cite{zhangQ14} investigated the magnetic field topology of 21 high-mass star-forming clumps, from clump to MDC, 1 to 0.03~pc, scales. It showed that sub-parsec magnetic fields are rather organized and either aligned or perpendicular to magnetic fields in their parental clumps (see \textbf{Figure}~\ref{f:bfield} Right). It may indicate that magnetic fields play an important role during the collapse and fragmentation of massive molecular clumps into high-mass protostars. This interpretation however relies on the assumption that the plane-of-sky magnetic field orientation derived from dust linear polarization traces dense regions, while there are indications that it could mostly trace polarization diffusion on clump outskirts \citep{crutcher12}. Plane-of-sky magnetic field measurements derived from the Goldreich-Kylafis effect of lines could help better constrain the evolution of magnetic fields down to protostellar core scales.

Several methods are used to estimate the magnetic field strength, the most well-known being the Zeeman effect, which directly measures the line-of-sight magnetic field and the Chandrasekhar-Fermi method  \citep{ChFe53}, which applies to plane-of-sky dust linear polarization measurements. Whereas in filaments the magnetic field energy dominates over turbulence, SMA observations of the DR21(OH) MDC showed that, at $\sim$0.1~pc scales, the magnetic field approximately is in equipartition with the turbulent energy  \citep{girart13}. Studies of high-mass star-forming clumps found slightly super-critical mass-to-magnetic flux ratios, (M/$\Phi_{\rm B}$)/(M/$\Phi_{\rm B}$)$_{\rm crit}$, of 1.5-2  \citep{falgarone08,li15, pillai16}.
If confirmed, such values show that magnetic fields of $0.1-10$~mG \citep[e.g.,][]{falgarone08, green12, cortes16} are too weak to sustain MDCs against gravity and thus do not favor  the turbulent core accretion model. Finally, the correlation of the magnetic field strength with density has a lower exponent than that for an isotropic gravitational contraction, $B\propto n^\alpha$ with $\alpha<2/3$ \citep{li15}. It therefore suggests that magnetic fields are, to some extent, still strong enough to channel the clump contraction.

The first magnetic field measurements performed in high-mass star-forming regions remain sparse and largely biased by the method used to constrain them. Larger surveys using complementary methods, like done by \cite{pillai16}, are necessary to confirm if magnetic fields slow down the clump global collapse, limit its fragmentation level, and possibly drive its clump gas inflows toward protostellar cores.

\subsection{High-mass prestellar cores, the current holy-grail}
\label{s:pcore} 

The controversy remains about the existence of high-mass analogs of low-mass prestellar cores. Indeed, while it is tempting to think that high-mass star formation goes through the same pathway as that of low-mass counterparts, observers tried, for the past ten years in vain, to identify good candidates for being high-mass prestellar cores. Extrapolating the prestellar core definition of, e.g., \cite{wsha94} and \cite{andre00}, high-mass prestellar cores should be small-scale, $0.01-0.1$~pc, gravitationally bound cores with high densities, $n_{\mbox{\tiny H$_2$}}=10^5-10^7$~cm$^{-3}$, and with no hydrostatic protostellar objects at their centers. 

The turbulent core model by \cite{MKT03} uses massive, gravitationally-bound starless cores in initial conditions. It invokes that these so-called high-mass prestellar cores are more massive than the thermal Jeans mass because they are supported against collapse and fragmentation by a large degree of turbulence and/or a strong magnetic field. High-mass virialized prestellar cores should also be in approximate pressure equilibrium with their surroundings and quasi-statically evolving toward higher degrees of central condensation, in marked contrast with protostars that are close to free-falling. As a direct consequence of these slow versus fast evolutions, one would expect, like in the low-mass case, to detect up to ten times more high-mass prestellar cores than high-mass protostars.

\subsubsection{The observational quest for high-mass prestellar cores} 
\label{s:quest}
This observational quest started with single-dish surveys, searching for the parental starless MDCs or starless clumps, whose $0.1-1$~pc sizes ensure them to be resolved at a distance of $1-3$~kpc. These surveys looked for these cloud structures, both though their far-IR to millimeter emission and their mid-IR extinction.

\paragraph{The quest within far-infrared to submillimeter massive dense core samples} Samples of cold MDCs were first built from ground-based millimeter surveys and later, interferometric high-resolution images of their internal structure permitted to look for prestellar cores at their centers.
In the pioneer study by \cite{motte07}, no starless object was found within their complete, unbiased, and homogeneous sample of MDCs built at 1.3~mm (see Sect.~\ref{s:survey}). This was rather surprising because one would expect to detect one to ten times more starless MDCs than protostellar MDCs and 42 protostellar MDCs were identified. The SiO emission used by \cite{motte07} to determine the protostellar nature of Cygnus~X MDCs was subsequently imaged at higher-angular resolution and showed not to be systemically associated with outflows \citep{duarte14}. These interferometric follow-up studies allowed revising the nature of Cygnus~X MDCs given by \cite{motte07} and recognizing CygX-N40 as a unique starless MDC candidate. It has a 0.16~pc size, a $\sim$100~$\msun$ mass and is not detected by \emph{Herschel}/PACS at 70~$\mu$m \citep{motte07, duarte13}.
 However when observed in continuum at the 0.02~pc scale of high-mass protostars, CygX-N40 gas mass is dispersed into diffuse cloud structures with only a low-mass, $<$$2~\msun$, core called CygXN40-MM1 \citep{bontemps10, duarte13}. 
Therefore, the initial result of \cite{motte07} arguing for starless MDC to be few in numbers or even missing remains valid.

Prestellar cores could also be hosted within IR-quiet (young) protostellar MDCs. An interferometric study of IR-quiet MDCs in Cygnus~X showed that all massive sub-fragments at 0.02~pc scales are associated with outflows, except CygXN53-MM2 \citep{duarte13}. With its $\sim$$25~\msun$ mass reservoir within $\sim$0.025~pc, this source is one of the best high-mass prestellar core candidate identified to date \citep[][see \textbf{Figure}~\ref{f:N53} Left]{bontemps10,duarte13}. Owing to the confusion caused by the neighbor CygXN53-MM1 protostar, it is however difficult to exclude that the N53-MM2 core is driving a weak outflow as a low- or intermediate-mass protostar would do.

In the much more extreme star-forming region called W43 (see Sect.~\ref{s:starburst}), \cite{motte03} identified tens of IR-quiet MDCs, among which is the W43-MM1 object having $3\,600~\msun$ and $2\times 10^4~\lsun$ within 0.23~pc \citep[see also][]{bally10}. Imaged at $\sim$0.05~pc resolution, W43 displays the extremely massive Class~0-like protostar W43-N1 ($500-1\,000~\msun$ within 0.03~pc) and five starless MDC candidates, without detected outflows: W43-N3, N8, N9, N10, and N11  \citep{louvet14,louvet16}. When observed with ALMA at 0.01~pc resolution, these $50-200~\msun$ 0.07~pc MDCs only host $0.5-1~\msun$ 0.02~pc cores of currently unknown nature (Nony et al. in prep.). Further investigation is therefore needed to convincingly identify a high-mass prestellar core in the W43-MM1 ridge.

At 0.1~pc scales, the similar lack of starless MDCs has been reported within the NGC~6334-6357 MDC sample through millimeter imaging of the complex \citep{russeil10}. Starless candidates are also found to be fewer in numbers, $5-30\%$, with respect to their protostellar counterparts in less homogeneous Galactic plane samples of larger-scale massive clumps/clouds surveyed at (sub)millimeter wavelengths \citep{ginsburg12, tackenberg12, csengeri14, traficante15}. 

Submillimeter surveys therefore show that high-mass prestellar cores are in an elusive phase. No final conclusion can however be derived from single-wavelength (sub)millimeter surveys alone because they remain partly biased against cold precursors of massive stars/clusters. Indeed, a single temperature, often $\sim$20~K, assumed for both cold starless objects and slightly warmer IR-quiet protostellar ones, underestimates the mass of starless objects and thus their number above a given mass threshold.
To solve this issue, the NGC~6334 molecular complex was investigated with \emph{Herschel}. Thanks to their careful dust temperature measurements, \cite{tige17} found as many starless as protostellar MDCs, 16 in numbers. The dust temperatures of starless MDCs have been measured to be $\sim$15~K, explaining by itself why \cite{russeil10} underestimated the number of starless MDC candidates, above their mass threshold of $75~\msun$ within 0.13~pc. Galactic plane surveys adjusting the temperature of each massive clumps/clouds have also found starless candidates with $\sim$15~K temperatures but, as for MDCs, in equal or smaller numbers ($20-50\%$) than their protostellar counterparts \citep{traficante15, svoboda16}. 

Starless MDC and starless clump candidates are also found to have smaller densities than the protostellar ones \citep{tige17, svoboda16}. So, many of the starless MDC candidates found in NGC~6334 are a factor of $3-10$ less dense, at a similar physical size, than protostellar MDCs and could as well form a cluster of intermediate-mass stars. A direct consequence is that no starless MDC from the sample could engender their neighbor protostellar MDCs assuming quasi-static compression. For their starless clump/cloud candidates (SCCs), \cite{svoboda16} themselves propose that their starless candidates will either not form any high-mass star or will further gain mass from their surroundings before reaching the protostellar state.

The complete and well-characterized MDC sample of \cite{tige17} is currently the more appropriate to evaluate the probability that high-mass prestellar cores do exist. The ability of NGC~6334 starless MDCs to form a high-mass star is in fact debatable. Most of them are located on top of filamentary structures, whose flux could have contaminated the \emph{Herschel} bands and thus overestimated the mass of MDCs. Confirming this suspicion, a handful of these MDCs were observed at higher-angular resolution and proved not to be centrally concentrated but to mostly contain diffuse gas (Ph. Andr\'e priv. com.). The complex structure and moderate density of starless MDC candidates put into doubt their ability to concentrate enough mass into a high-mass prestellar core. \cite{tige17} therefore estimate that the 16 starless MDCs in NGC~6334, should contain in total, at most, one to seven high-mass prestellar cores. MDC\#5, also called HOBYS\_J172053.0-354317, is their best candidate starless MDC with its unresolved size, $\le 0.08$~pc, and high mass, $\sim$210~$\msun$. Higher-resolution studies of NGC~6334 MDCs are necessary to finally prove that massive prestellar cores do or do not exist and some are actually ongoing.

\paragraph{The quest within infrared-dark clouds fragments}
IRDCs have, for long, been considered as the birth place of high-mass stars and could also reveal some high-mass prestellar cores. They however are too numerous and not massive and dense enough to all be forming high-mass stars \citep[e.g.,][]{PeFu10}. Moreover, the vast majority of massive IRDC fragments do harbor IR-quiet high-mass protostars \citep[e.g., ][]{pillai06a,rathborne10, PeFu10}. Fragmentation and follow-up studies of IRDCs have provided a few  samples of cold massive clumps, some of which are quiescent and could be starless \citep[e.g.,][]{rathborne10, traficante15}.

The best characterized sample was  selected among the 38 IRDCs studied by \cite{rathborne06} and consists of the ten IRDCs that are the closest and show the highest contrast against the Galactic mid-IR background \citep{BuTa09}. Among the IRDC fragments identified at 1~mm by \cite{rathborne06}, \cite{BuTa12} found 42 starless dense cores/clumps of $9-1700~\msun$ masses within $0.2-1.5$~pc. 
\cite{tan13} and \cite{kong17} observed with ALMA these clumps, which are more massive than $100~\msun$ and display high deuteration fractions -- these are generally taken as a youth indicators \citep{fontani11}. ALMA marginally resolved the size scale of individual cores ($0.03-0.09$~pc versus $0.02$~pc) and only revealed low- to intermediate-mass cores, with the notable exceptions of the G028C1-S and G028C9A MDCs: $\sim$$60~\msun$ within $\sim$0.09~pc and $\sim$$80~\msun$ within $\sim$0.05~pc. These MDCs are definitively super-Jeans and reminiscent of the starless MDCs identified by \cite{tige17}. Follow-up studies of G028C1-S however revealed it actually harbors two protostars driving outflows \citep{tan16} and G028C91 could consist of two lower-mass cores \citep{kong17}.
We recall that G028C1-S source was long considered as the prototype high-mass prestellar core, potentially representing the initial conditions necessary for the turbulent core model.

Other high-resolution studies toward IRDCs generally detected several starless MDC candidates, but only very few prestellar cores at the 0.02~pc scale of high-mass protostars. For instance, the starless MDC candidate S13-MM2, with $\sim$$80~\msun$ within 0.21~pc \citep{peretto14}, probably has a too-low density to be able to form a high-mass protostar in the near future. In parallel, interferometric studies of the G28.34+0.06 and G11.11-0.12 IRDCs, down to the protostellar scale, discovered several high-mass protostars and only a single high-mass prestellar core candidate, G11P6-SMA1, which has a $\sim$$30~\msun$ mass within 0.02~pc \citep[][see \textbf{Figure}~\ref{f:N53} Right]{wang11, wang14}.

For the sake of completeness, one should mention the strong submillimeter continuum source without molecular emission, G11.92-0.61-MM2, proposed by \cite{cyganowski14} to be a high-mass prestellar core with an estimated  mass of $\sim$$30~\msun$ within $\sim$0.005~pc. Because prestellar cores should probably be large and warm enough for molecular line emission to be detected, there is reasonable doubt that this object is a Galactic cloud structure.

\subsubsection{Gas mass concentration at starless stages} 
\label{s:gasConcent}

\begin{figure}[h]
\centerline{\includegraphics[angle=0,width=7cm]{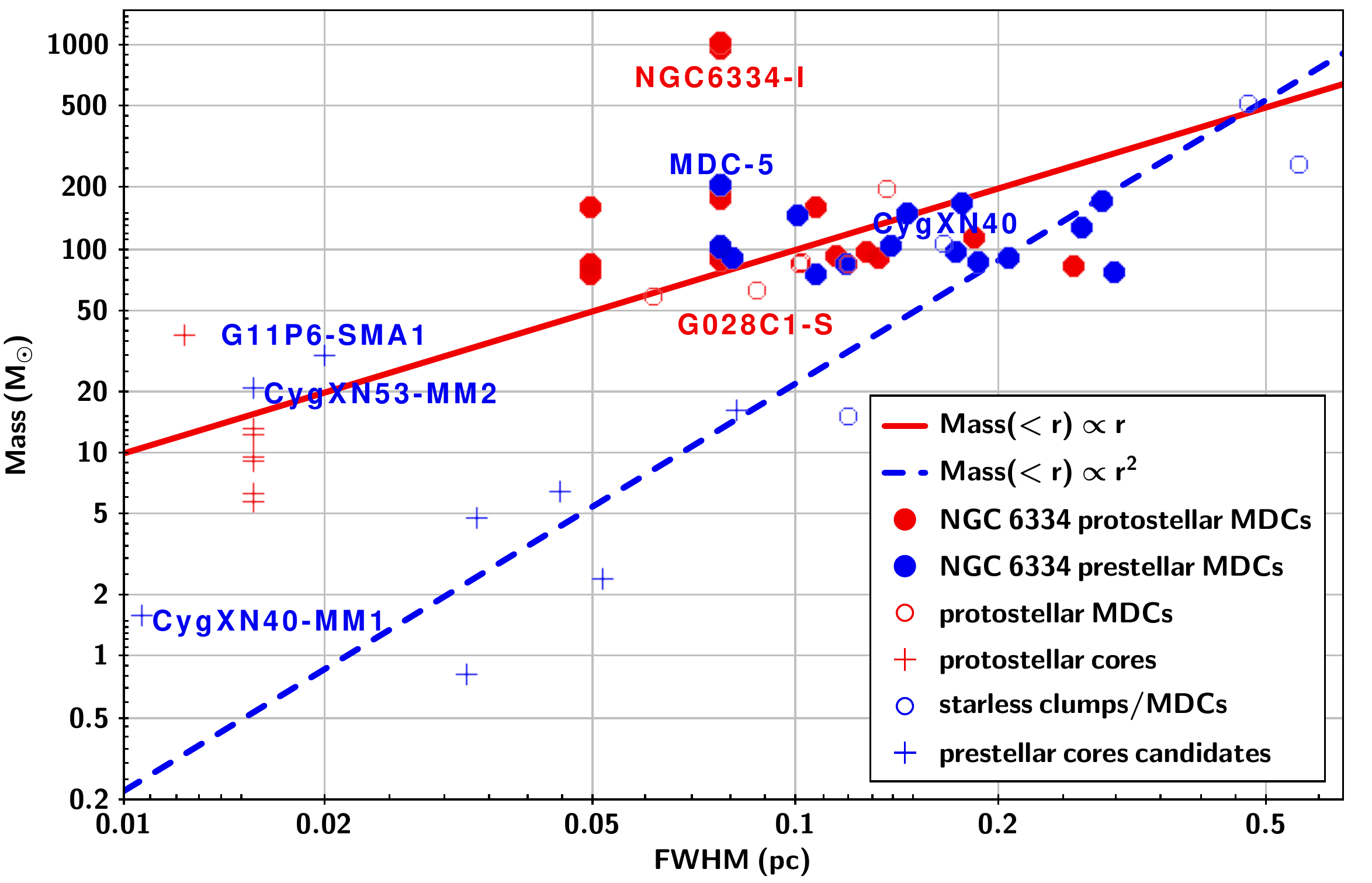}}
\vskip 0.3cm
\caption{Prediction of gas concentration within starless MDC candidates of NGC~6334 (filled blue circles).
If they should follow the $M(<r) \propto r^{2}$ relation (blue dashed line), linking turbulence-dominated starless clumps and their prestellar cores (open blue circles and crosses), most NGC~6334 starless MDCs should merely host low- to intermediate-mass prestellar cores. They could otherwise follow the $M(<r) \propto r$ relation (red line) of gravity-dominated MDCs/cores like CygX-N53 and CygXN53-MM2.
In NGC~6334, the best starless candidate MDC-5 (or HOBYS\-\_J172053.0-354317) has a $\sim$210~$\msun$ mass within $0.08$~pc. 
Adapted from \cite{tige17} with permission.}
\label{f:mass-radius}
\end{figure} 

High-resolution studies of starless MDC suggest that most of them may not contain fragments dense and massive enough to be the high-mass analogs of prestellar cores \citep{duarte13,tan13}. This could be related to the gas mass concentration of starless structures in a turbulent cloud. 

Indeed, when located in the mass versus radius diagram of \textbf{Figure}~\ref{f:mass-radius}, starless MDCs and their hosted fragments approximately follow a  $M(<r) \propto r^{2}$ relation \citep[][blue open circles and crosses, respectively]{BuTa12, tan13}. This mass concentration with spatial scale is close to the one found in non-centrally concentrated, turbulent clouds with large fragmentation level like CO fractal cloud structures: $M(<r) \propto r^{\gamma}$ with $\gamma=2.2-2.3$ \citep[Larson's law,][]{kramer96, heithausen98}.
This is in marked contrast with the $M(<r) \propto r$ relation predicted for Bonnor-Ebert spheres \citep{bonnor56, johnstone00} and the one found for samples of gravitationally-dominated low-mass prestellar cores \citep[e.g.,][]{motte01,konyves15}. We recall that a $M(<r) \propto r$ mass concentration is expected for cloud structures with a $\rho(r) \propto r^{-2}$ radial density structure, like observed for  protostellar envelopes and the outskirt of prestellar cores \citep{MA01, mueller02, ward99}. In agreement, the  IR-quiet protostellar MDCs found in Cygnus~X and their hosted high-mass Class~0-like protostars  \citep[][red open circles and crosses, respectively]{motte07, bontemps10} follow a $M(<r) \propto r$ relation in \textbf{Figure}~\ref{f:mass-radius}.

Interestingly, the CygXN53-MM2 and G11P6-SMA1 cores and their parental MDC/clump rather concentrate as $M(<r) \propto r$, arguing for their gravitationally-dominated  nature. In contrast, the CygXN40-MM1 core follows the $M(<r) \propto r^{2}$ relation, typical of turbulence-dominated starless sources.  According to this empirical mass versus size relation and assuming a 40\% star formation efficiency \citep[e.g.,][]{alves07,konyves15}, high-mass prestellar cores of $>$$30~\msun$ within 0.02~pc size should be found in turbulence-dominated, starless MDCs more massive than $100~\msun$, within a 0.1~pc size. In NGC~6334, only four of the starless MDC candidates are above this mass threshold \citep{tige17}, and HOBYS\_J172053.0-354317 is among them.

After ten years of research, only two high-mass prestellar core candidates have therefore been identified: CygXN53-MM2 and G11P6-SMA1 \citep[][see \textbf{Figures}~\ref{f:N53}]{bontemps10,wang14}. Interferometric studies toward large samples of starless MDC and IRDC fragments are ongoing. We are thus at the dawn of finally proving that massive prestellar cores do or do not exist.

\subsection{Evolutionary scenario of high-mass star formation}
\label{s:scenario}

\subsubsection{High-mass prestellar core lifetimes}
\label{s:lifetimePcore}
What do we know from the very few known high-mass prestellar cores (see Sect.~\ref{s:quest} and \textbf{Figures}~\ref{f:N53}) suggests that their lifetimes are, at most, very short.
Like for protostars described in Sect.~\ref{s:lifetimeProt}, the lifetime of a high-mass prestellar core in a given region is estimated relative to the known ages and numbers of OB stars. According to the detailed statistical study of \cite{tige17}, high-mass prestellar cores should live for less than $1-7\times 10^4$~yr (see \textbf{Table}~\ref{tab:lifetime}). This upper value agrees with extrapolations, from 0.1~pc to 0.02~pc scales, made of statistical MDC results of \cite{motte07} in Cygnus~X (see \textbf{Table}~\ref{tab:lifetime}). \cite{duarte13} also proposed a prestellar lifetime of $1\times 10^4$~yr, from the detection of a unique high-mass prestellar core candidate in Cygnus~X, CygXN53-MM2 (see \textbf{Figure}~\ref{f:N53} Left).
Because all studies only measured upper limits, the lifetime of the prestellar phase of high-mass star formation should be more than one order of magnitude smaller than what is found for low-mass stars in nearby star-forming regions \citep[$10-40 \times 10^4$~yr,][]{onishi02,kirk05}. 
As a matter of fact, a short prestellar phase is also suggested by the low level of deuteration of organics observed toward massive hot cores with respect to their low-mass counterparts \citep[e.g.,][]{faure15}. Interestingly the statistical lifetime of starless MDC and starless clump candidates is also estimated to be short:$1-3\times 10^4$~yr \citep[][see \textbf{Table}~\ref{tab:lifetime}]{motte07, russeil10, csengeri14, svoboda16}. These lifetimes therefore suggest that high-mass prestellar cores and starless MDCs/clumps should be in a highly dynamical state, as expected in a molecular cloud where turbulence and/or organized flow processes dominate.

In the intermediate-mass regime, \emph{Herschel} surveys of the Rosette molecular complex and the IRDC G035.39-00.33 have identified a few starless candidates, suggesting a lifetime of $\sim$$8 \times 10^4$~yr \citep{motte10,nguyen11a}. The lifetime of intermediate-mass prestellar cores may thus be a few times shorter than what is found in nearby low-mass star-forming regions and somewhat longer than that constrained in high-mass star-forming complexes. This result agrees with the statement by \cite{kirk05} that starless structures have lifetimes varying from 1 to 10 times their free-fall time, depending on their density. They proposed that the denser the starless structure or prestellar core the fewer free-fall times they live, with the minimum being close to a single free-fall time. This recalls the empirical correlation found by \cite{svoboda16} between the lifetime of starless and protostellar clumps/clouds and their mass.

In this framework, high-mass prestellar cores, postulated to be super-Jeans and thus very dense, should live for about one single free-fall time and be free-falling as soon as they form. The free-fall time of a putative high-mass prestellar core of full-volume averaged densities equivalent to that of high-mass protostars, $<$$n_{\rm H_{2}}$$>$$_{\rm full} \sim1.3 \times 10^6$~cm$^{-3}$, is $\tau_{\rm ff-prestellar}\sim 3\times 10^4$~yr. Therefore, current high-resolution studies, which remain limited in terms of statistics, could still have failed in detecting
 high-mass prestellar cores (see \textbf{Table}~\ref{tab:lifetime}). 

\subsubsection{Individual collapse of a turbulent core or global hierarchical collapse of a clump?}
\label{s:EvolSeq}
The evolutionary sequence found for high-mass star formation finally permits to start discussing the physical processes at work during the high-mass star formation.
With current lifetime constraints, it is statistically possible that high-mass stars form from high-mass pre-stellar cores, in a manner that can be considered a scaled-up version of low-mass star formation \citep[see, e.g.,][]{andre00}. Like proposed by the turbulent core model \citep{MKT02} and suggested by accretion and velocity dispersion constraints (see Sect.~\ref{s:accretion}), high-mass prestellar cores would form and remain un-fragmented, i.e. monolithic, thanks to turbulent and/or magnetic supports. 
Given that their lifetimes are as short as about one free-fall time (see Sect.~\ref{s:lifetimePcore}), high-mass prestellar cores cannot form quasi-statically over several free-fall times as was assumed by  \cite{MKT02}. High-mass prestellar cores must thus quickly assemble their mass and collapse as soon as they reach the necessary mass, which is qualified as super-Jeans (see Sect.~\ref{s:accretion}). In the turbulent core model, prestellar cores on the verge of their collapse should be static and isolated from their surroundings. As soon as they loose their supplementary turbulent and magnetic support, high-mass prestellar cores start to collapse and enter the protostellar phase. The process leading from the run-away global collapse observed at clump and MDC scales \citep[][see Sect.~\ref{s:ridge}]{csengeri11b, schneider10} to a quasi-static configuration at the prestellar core scale of $0.02-0.1$~pc still needs to be found.

The alternative interpretation of short lifetimes for the high-mass prestellar phase is that high-mass prestellar cores simply do not exist as small, $\sim$0.02~pc, condensations, isolated from their environment. Both the lifetime of high-mass protostars and the infalling gas observed down to the protostellar scale indeed invoke that high-mass stars form while still strongly interacting with their surroundings. First, the high-mass protostellar lifetime suggests that the collapse starts within a low-mass prestellar core and continues within a protostellar envelope, which grows from low to high mass \citep[e.g.,][see Sect.~\ref{s:lifetimeProt}]{tige17}. Moreover, high-mass stars form into infalling clumps at 1~pc scales, whose global collapse drives inflowing gas streams toward protostars at 0.01~pc scales \citep[e.g.,][see Sect.~\ref{s:cloud+cluster}]{schneider10, csengeri11b}. This evolutionary scenario corresponds to the global hierarchical collapse theory of, e.g., \cite{vazquez09, vazquez17}. It can be seen as an extension of the competitive accretion model, when accretion through inflowing gas streams driven by gravity replaces the Bondi-Hoyle accretion \citep{smith09}. 
In this scenario high-mass protostars would then be fed from the gas of their surrounding MDC/clumps, following the clump-fed scenario of  protostellar accretion, in contrast with the the core-fed scenario of low-mass protostellar accretion and of the \cite{MKT02} model.

In the global hierarchical collapse scenario, the high-mass equivalent of prestellar cores could therefore be low-mass cores within massive infalling MDCs/clumps. The inner low-mass core will first be prestellar then protostellar before it becomes a high-mass protostar. Structural studies alone would not be able differentiate massive MDCs which will form high-mass stars, from those forming a cluster of intermediate-mass stars. In contrast, kinematics studies can themselves give clues on the future gas mass accretion expected toward the inner low-mass cores. Ridges and hubs, defined in Sect.~\ref{s:ridge}, may constitute the parsec-scale gas reservoir/clumps, from which gas is accreted onto 0.02~pc-scale cores. The global hierarchical collapse theory invokes that ridges and cores simultaneously form and get denser. 

\begin{figure}[h]  
\vskip -0.5cm
\hskip -1.5cm \includegraphics[width=21.5cm, angle=0]{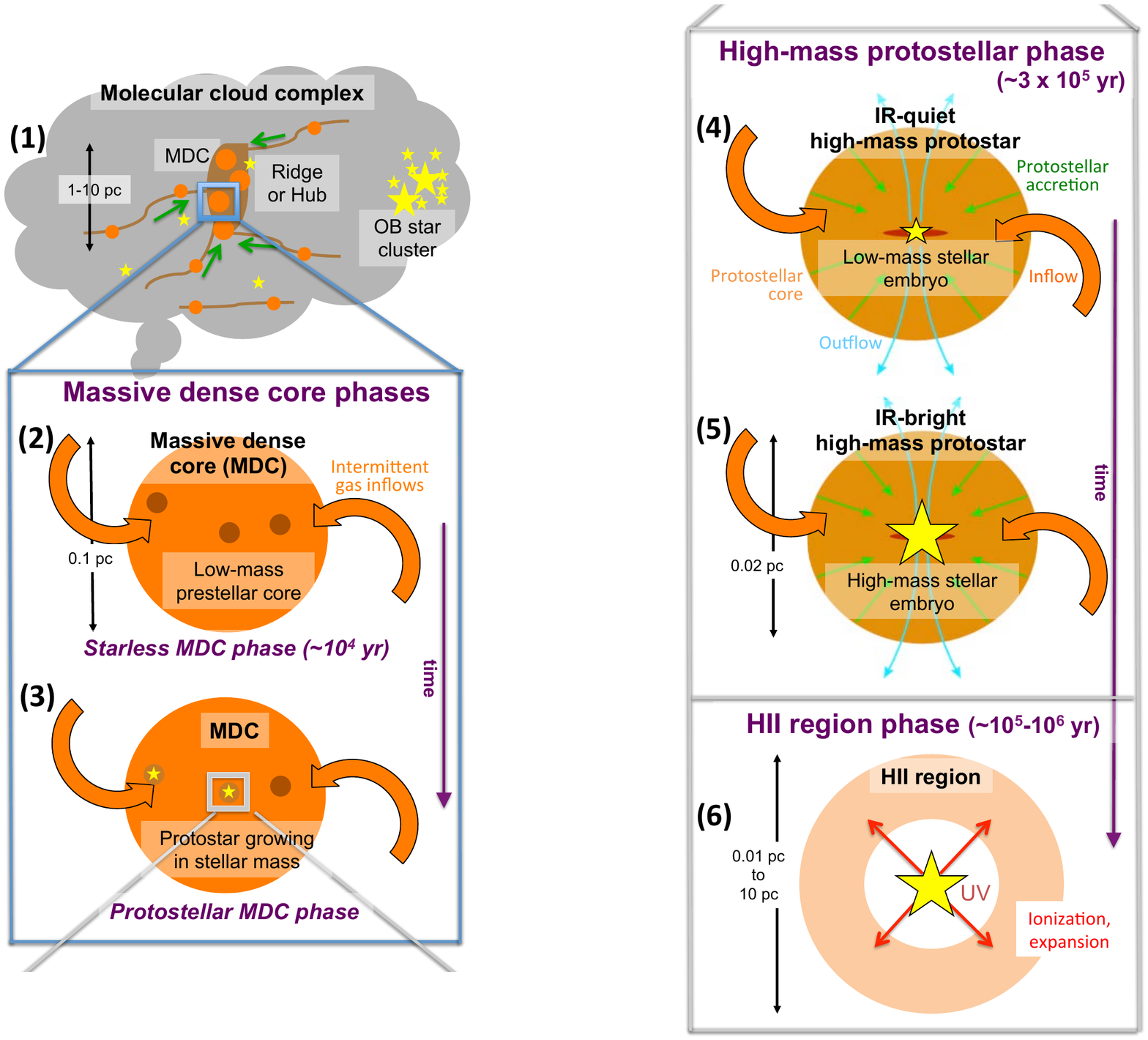}
\caption{Schematic evolutionary diagram proposed for the formation of high-mass stars. 
\textbf{(1)} Massive filaments and spherical clumps, called ridges and hubs, host massive dense cores (MDCs, 0.1~pc) forming high-mass stars.
\textbf{(2)} During their starless phase, MDCs only harbor low-mass prestellar cores.
\textbf{(3)} IR-quiet MDCs become protostellar when hosting a stellar embryo of low-mass. 
The local, 0.02~pc, protostellar collapse is accompanied by the global, $0.1-1$~pc, collapse of MDCs and ridges/hubs.
\textbf{(4)} Protostellar envelopes feed from these gravitationally-driven inflows, leading to the formation of high-mass protostars. The latter are IR-quiet as long as their stellar embryos remain low-mass.
\textbf{(5)} High-mass protostars become IR-bright for stellar embryos with mass larger than $8~\msun$.
\textbf{(6)} The main accretion phase terminates when the stellar UV field ionizes the protostellar envelope and an \hii region develops.
Adapted from \cite{tige17} with permission.}
\label{f:scheme}
\end{figure}

\textbf{Figure}~\ref{f:scheme} illustrates the evolutionary scheme we propose for the formation of high-mass stars. 
Based on observational constraints given in Sects.~\ref{s:lifetimeProt}, \ref{s:accretion}, \ref{s:lifetimePcore}, and  \ref{s:ridge}, it follows an empirical scenario qualitatively recalling the global hierarchical collapse and clump-fed accretion scenarios \citep{vazquez09, smith09}. Despite the large binary fraction of high-mass stars, present scenario cannot yet include their formation because observational constraints are lacking.
\begin{enumerate}
\item High-mass stars form in molecular complexes hosting massive clouds and often OB clusters. Parsec-scale massive clumps/clouds called ridges and hubs are the preferred, if not the only, sites for high-mass star formation. Their infall velocity and density structure suggest ridges/hubs undergo a global but controlled collapse.
\item At first, IR-quiet massive dense cores (MDCs) are 0.1~pc massive cloud fragments, which host low-mass prestellar cores. They represent the starless MDC phase lasting for about one free-fall time, $\sim$$10^5$~yr.
\item At the MDC center, low-mass prestellar cores become protostars with growing mass and \emph{not} high-mass prestellar cores. The global collapse of ridges/hubs generates gas flow streams, which simultaneously increase the mass of MDCs and, on 0.02~pc scales, that of their hosted protostar(s). Typically, in $\sim$$10^5$~yr, two high-mass protostars form in 0.1~pc MDCs.
\item When inflowing gas streams are efficient to reach and feed the low-mass protostellar cores, the latter become IR-quiet high-mass protostars. They have 0.02~pc sizes, super-Jeans masses, but still \emph{only} harbor low-mass, $<$$8~\msun$, stellar embryos. Their accretion rates are strong, they drive outflows and power hot cores.
\item When stellar embryos reach more than $8~\msun$, their luminosity sharply increases and high-mass protostars become IR-bright. Their hot cores grow in size and they soon develop HC\hii regions quenched by infalling gas or localized toward photo-evaporating disks.
\item Stellar embryos have increasing UV fields that develop \hii regions, which, along with other processes including outflows and winds, slow and later on eventually stop gas accretion toward the newborn star. This terminates the main accretion phase.
\end{enumerate}

\section{MASSIVE CLOUD AND MASSIVE CLUSTER FORMATION}
\label{s:cloud+cluster}

The scenario proposed in Sect.~\ref{s:EvolSeq} and shown in \textbf{Figure}~\ref{f:scheme} is consistent with a scenario proposing that high-mass star formation develops simultaneously and in tight link with the formation of massive clouds and massive clusters. In the following, we show that the structure and kinematics of massive clouds are extreme relative to those of low-mass star-forming regions (see Sect.~\ref{s:ridge}). As a consequence, massive clouds can sustain intense star formation activity, which impacts the content of their future stellar clusters (see Sect.~\ref{s:starburst}).

\subsection{High-density dynamical clumps quoted as ridges and hubs}
\label{s:ridge}

Because the formation of high-mass stars requires more mass than that of low-mass stars, high-mass star-forming sites have always been looked for among the most massive cloud structures. Massive molecular cloud complexes, defined as 100~pc cloud ensembles with $3\times 10^5-3\times 10^6~\msun$ masses, are quantitatively larger than those of Gould Belt clouds (see \textbf{Table}~\ref{tab:hobysC}). Mass is however not a sufficient parameter and several authors proposed that cloud gas density is the one allowing, or not allowing, the formation of high-mass stars \citep[e.g.,][see also \textbf{Figure}~\ref{f:mass-radius}]{MKT02,motte07}. 

\begin{figure}[h]
\vskip 0.5cm
\hskip 1cm \includegraphics[angle=0,width=8cm]{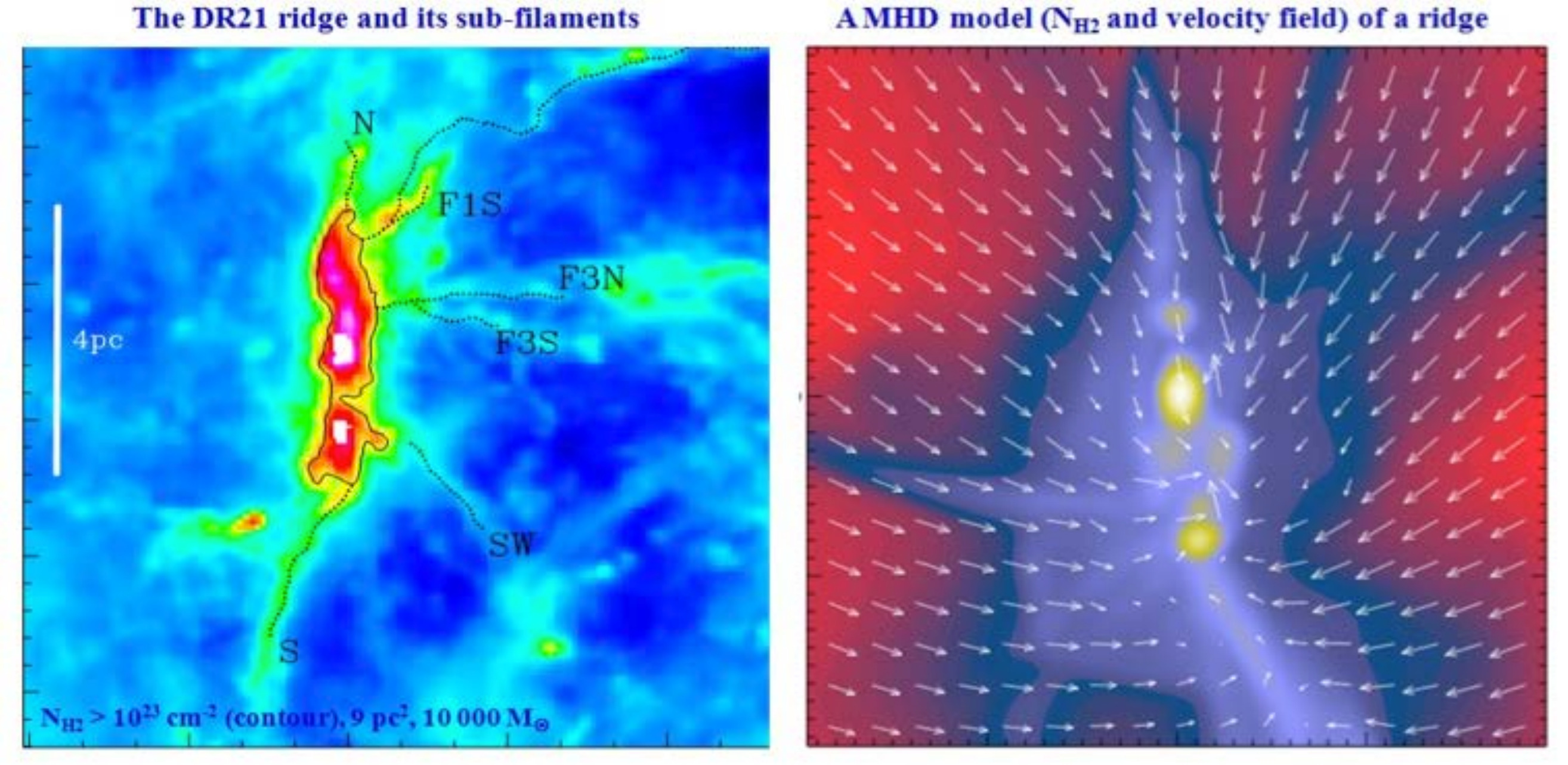}
\vskip 0.8cm
\caption{The DR21 ridge and its feeding sub-filaments network {\bf (Left)}, compared to numerical simulations of the collapse of a massive elongated clump {\bf (Right)}. 
 {\bf Left:} The \emph{Herschel} column density map is used to outline the DR21 ridge by the $\sim$$10^{23}$~cm$^{-2}$ contour and delineate sub-filaments found by \cite{schneider10} with dots.
 {\bf Right:} Simulated velocity streams (arrows) are overplotted on the modeled column density image.
Adapted from \cite{hennemann12} and \cite{schneider10} with permission.
}
\label{f:ridge}
\end{figure}

At the $1-10$~pc clump to cloud scales, some massive filaments and extreme IRDCs have been found close to the gravity center of molecular cloud complexes. One striking example is the prototypical DR21 ridge, located at the heart of the CygX-North cloud \citep[][see \textbf{Figure}~\ref{f:cygnus}-Left]{motte07, schneider10}. \emph{Herschel} allowed for quantitative studies of massive filaments through their column density and temperature images and direct comparison with their protostellar population (see \textbf{Figures}~\ref{f:ridge} and \ref{f:SKdiag} Left). Within the massive molecular complexes imaged by the HOBYS key program, high-density dominating clumps, are confirmed to be the preferred sites for forming massive stars  \citep[e.g.][]{hill11, nguyen11a, tige17}. 
The so-called ridges are high-density filaments, $>$$10^5$~cm$^{-3}$ over $\sim$5~pc$^3$, forming clusters of high-mass stars \citep[e.g.,][]{schneider10, nguyen13, hennemann12} while hubs are more spherical, smaller clumps forming at most a couple of high-mass stars \citep[e.g.,][see \textbf{Figure}~\ref{f:irdc}]{peretto13, rivera13, didelon15}. The most extreme IRDCs qualify as ridges or hubs and the densest ridges coincide with the precursors of young massive clusters  \citep[e.g.,][see also Sect.~\ref{s:starburst}]{nguyen11a,nguyen13,ginsburg12}. 
The existence of ridges/hubs is predicted by dynamical models of cloud formation such as colliding flow simulations \citep[e.g.,][]{HeHa08, federrath10} and some analytical theories of filament collapse \citep{myers09}. 

\begin{figure}[h]
\vskip -1cm
\hskip -0.5cm \includegraphics[angle=0,width=6.8cm]{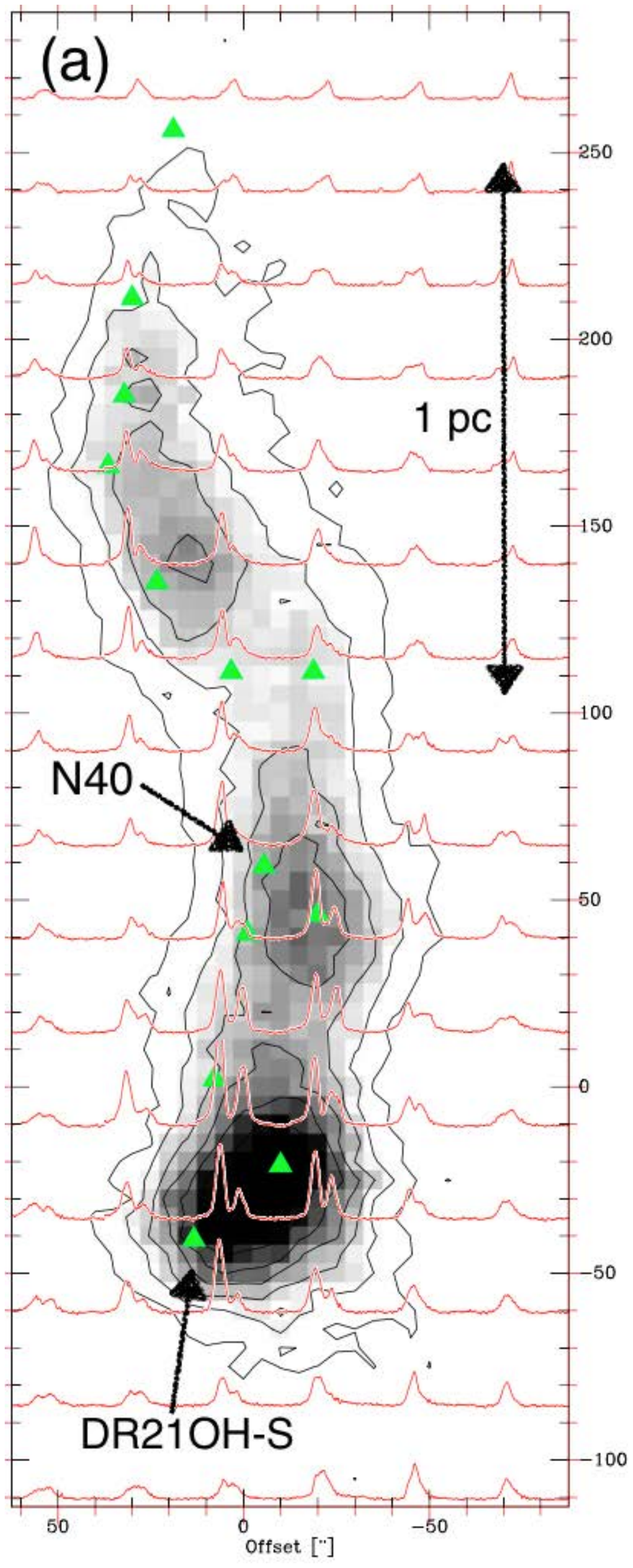}\vskip -14cm\hskip 6.3cm\includegraphics[angle=270,width=9.5cm]{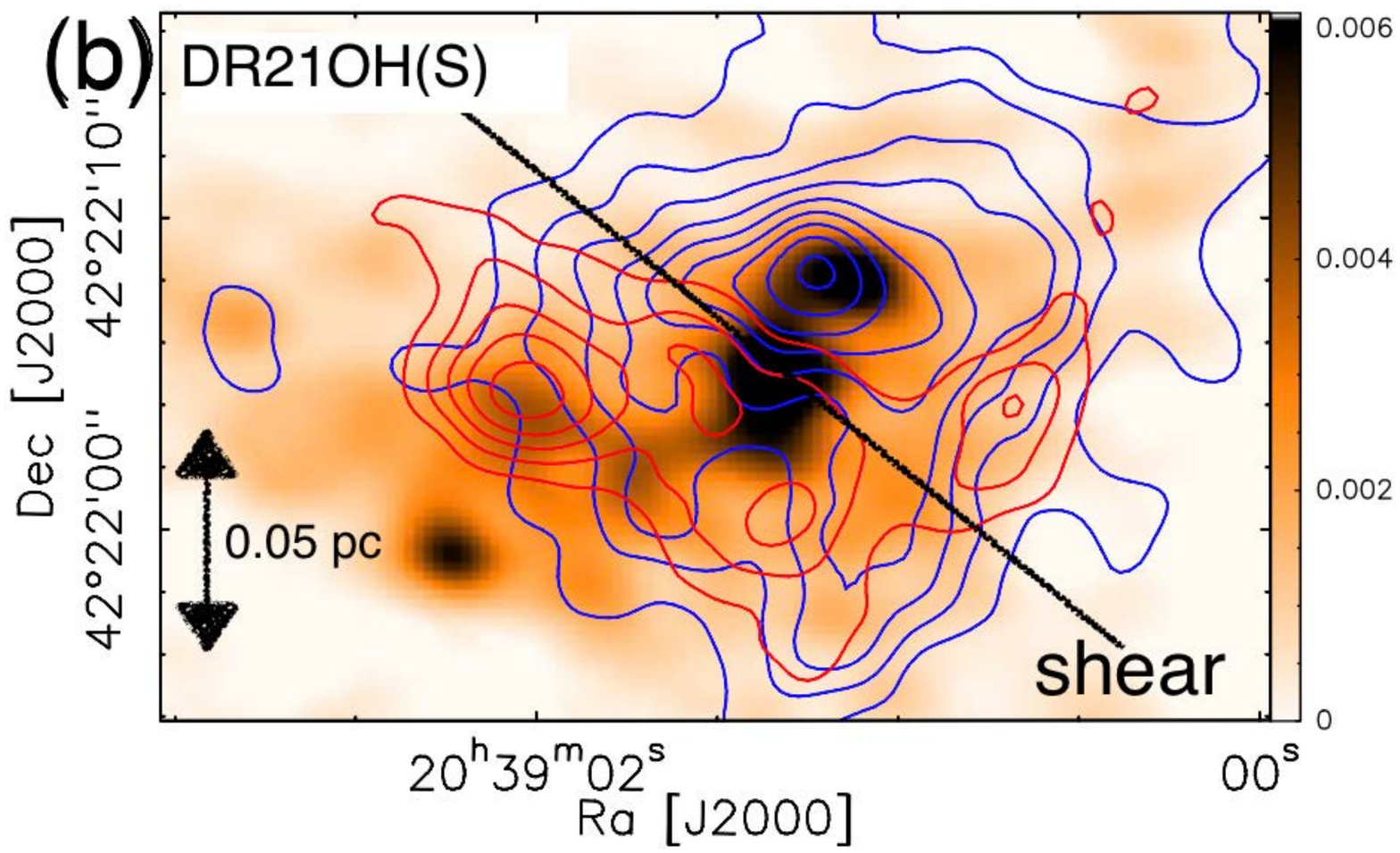}\vskip 2.5cm\hskip 5.8cm\includegraphics[angle=270,width=10cm]{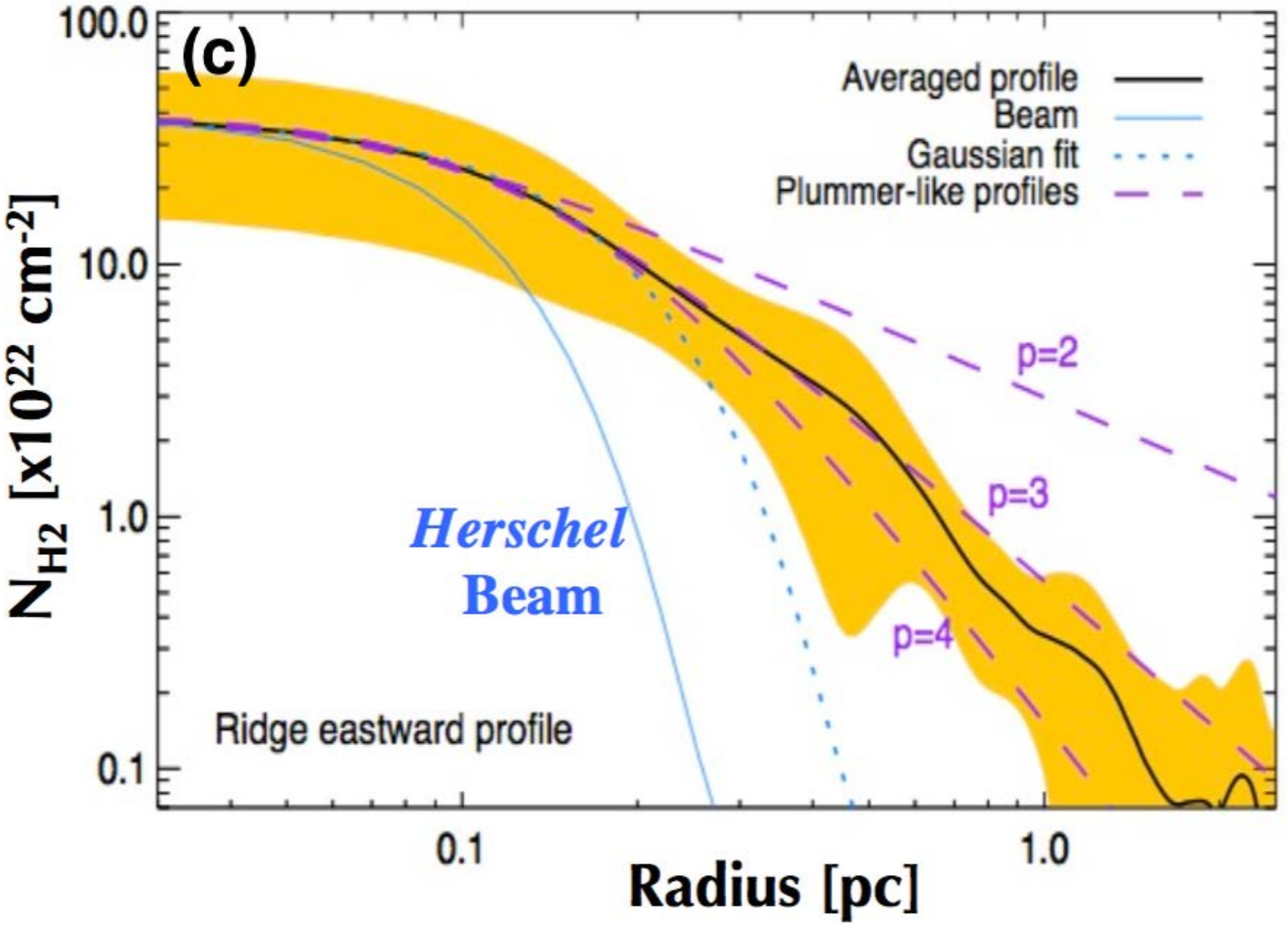}
\vskip 2cm
\caption{The dynamical environment of high-mass star forming regions illustrated by, in {\bf (a):} the global collapse of the DR21 ridge on pc scales, in {\bf (b):} local shears close to the location of 0.02~pc high-mass protostars in DR21OH-S (3~mm continuum in heat colors), and in {\bf (c):} the column density transverse profile of the DR21 ridge, which is steeper than the classical $\rho(r)\propto r^{-2}$ ($p=2$) law found for Gould Belt filaments.
{\bf (a):} Optically thick HCO$^+$ (1--0) lines (red spectra) suggest supersonic infalling velocities, $v_{\rm inflow} =0.5-1$~km\,s$^{-1}$. 
{\bf (b):} N$_2$H$^+$ (1--0) gas flow streams (red and blue contours) display $\sim$$\rm 2~km\,s^{-1}$ velocity jumps across the black line, assimilated as shears.
{\bf (c:)} This steeper transverse profile suggest that either adiabatic heating, rotation, or magnetic field slows down the ridge collapse. 
Adapted from \cite{schneider10}, \cite{csengeri11a}, and \cite{hennemann12} with permission.}
\label{f:collapse}
\end{figure}

When imaged with molecular lines, column density, or extinction, ridges/hubs seem to be the focus points of large amounts of gas structured in filaments (see \textbf{Figures}~\ref{f:irdc} and \ref{f:cygnus}). These massive filament networks have a more spherical/elliptical geometry (see \textbf{Figure}~\ref{f:ridge}a) than the prototypical Taurus filament, which is perpendicularly crossed by subcritical filaments called striations \citep[B211/3][]{palmeirim13}. Density-wise sub-filaments are reminiscent of the most massive low-mass star-forming filaments. It would take the merging of tens of them to form ridges, and that could then be considered as a second generation of supercritical (gravitationally bound) filamentary structures \citep{hennemann12}. Velocity drifts of one to a few $\rm km\,s^{-1}$ are observed along sub-filaments, which converge toward ridges/hubs, suggesting drifts feed subfilaments by funneling the surrounding gas \citep[e.g.,][]{schneider10, peretto13}. Ridges and hubs would therefore be large gravity potentials attracting filaments, sometimes following a fan-shaped structure. Numerical simulations of high-density collapsing clumps agree with such a picture \citep[][see \textbf{Figure}~\ref{f:ridge}b]{HaBu07, schneider10, GoVa14}.

Molecular line imaging, with the combination of optically thick and thin lines, have revealed that ridges and hubs undergo global collapse with supersonic inward velocities, $v_{\rm inflow} \sim \rm 1-2~km\,s^{-1}$ over $1-10$~pc$^2$ \citep[][see \textbf{Figure}~\ref{f:collapse}a]{schneider10, peretto13}. This strong result builds on initial studies by, e.g., \cite{rudolph90}, \cite{WuEv03}, and \cite{motte05} and agrees with simulated line profiles of collapsing clumps \citep{smith13}. 
The general structure of collapsing ridges may reflect gas compression. They indeed have steeper radial density profiles than the classical $\rho(r)\propto r^{-2}$ law \citep[see \textbf{Figure}~\ref{f:collapse}c][]{hennemann12, didelon15} and display long gravity tails and even secondary tails in their column density probability distribution functions \citep[PDFs,][]{hill11, russeil13, schneider15a, schneider15b}.
In some intermediate-mass clouds, similar but slower global collapses have been observed \citep{loren77,peretto06, kirk13} and suggested to result from cloud-cloud collision \citep{nakamura14}. Infall motions, sometimes called gravitational focusing, are expected in colliding flow models \citep[e.g.,][]{vazquez07, HaBu07} as well as in ionization compression models \citep[e.g.,][]{tremblin12}. In any case, whatever the origin of the additional pressure, arising from colliding flows or ionization, it supersedes the thermal and micro-turbulence pressure inducing ridges and hubs to collapse. 

Detailed analyses of the ridge inner structure suggest they are braids/bundles of filaments \citep{hennemann12, henshaw14}, like the main Taurus filament \citep[B213,][]{hacar13}. Kinematic imaging of ridges, with high-density molecular lines such as N$_2$H$^+$, indeed revealed multiple velocity components, interpreted as sub-filaments which sometimes cross each others \citep[e.g.,][]{galvan10, henshaw13, henshaw14, tackenberg14}. 
In the high-density, complex medium of ridges, these sub-filaments are difficult to recognize and disentangle from each others. The non-homogeneous structure of ridges and their potential rotation along the big axis could partly explain why their inflow rate would be smaller than free-fall \citep{wyrowski16}.
An indirect evidence of these filament bundles arises from the detection of shocks associated with gas shears created by this braiding. Large-scale SiO emission has been found along several ridges, suggesting shock velocities of $\rm 1-5~km\,s^{-1}$ \citep{jimenez10,nguyen13, sanhueza13, duarte14, louvet16}. A detailed shock modeling of the strong and extended, $\sim$5~pc, SiO emission found along the W43-MM1 ridge proved that the low-velocity shock, developing within high-density ridges can liberate SiO from the grains \citep{louvet16}. 

The complex structure and high dynamics of ridges may have direct consequences for the building phase of $\sim$0.1~pc MDCs and their $\sim$0.02~pc protostellar cores. In the global hierarchical collapse model of 
 \cite{vazquez09} \citep[see also][]{smith09}, the gas mass accretion rate onto MDCs and individual cores is determined by their tidal radii and initial ridge structure. In contrast, the isolated turbulent core model of  \cite{MKT02} postulates that the ridge kinematics does not impact much the protostellar collapse as global infall should somehow be stopped at MDC scales.
 A pioneer study has been performed within Cygnus~X and especially the DR21 ridge MDCs \citep{csengeri11a} \citep[see also][]{galvan09}. This kinematical study found high-density gas streams inflowing on $\sim$0.05~pc scales and developing $\sim$2~km\,s$^{-1}$ shears in the immediate proximity of Cygnus~X high-mass protostars \citep[with H$^{13}$CO$^+$ and CH$_3$CN lines,][see \textbf{Figure}~\ref{f:collapse}b]{csengeri11b, csengeri11a}. Investigating many more ridges, with a series of angular resolutions, is necessary to properly follow inflowing gas from the ridge to the MDC, and finally the protostellar scales. This is a challenging but mandatory step to better understand star formation in ridges. As for  now, the only secure conclusion one can derive is that ridges and hubs are highly dynamical medium, within which prestellar cores can probably not be long-lived objects, in agreement with statistical studies \citep[see Sects.~\ref{s:lifetimeProt} and \ref{s:scenario},][]{motte07, tige17}.

\subsection{Mini-starburst activity within ridges}
\label{s:starburst}

\begin{figure}[h]
\vskip -0.5cm
\hskip -3.5cm\includegraphics[angle=180,width=5cm]{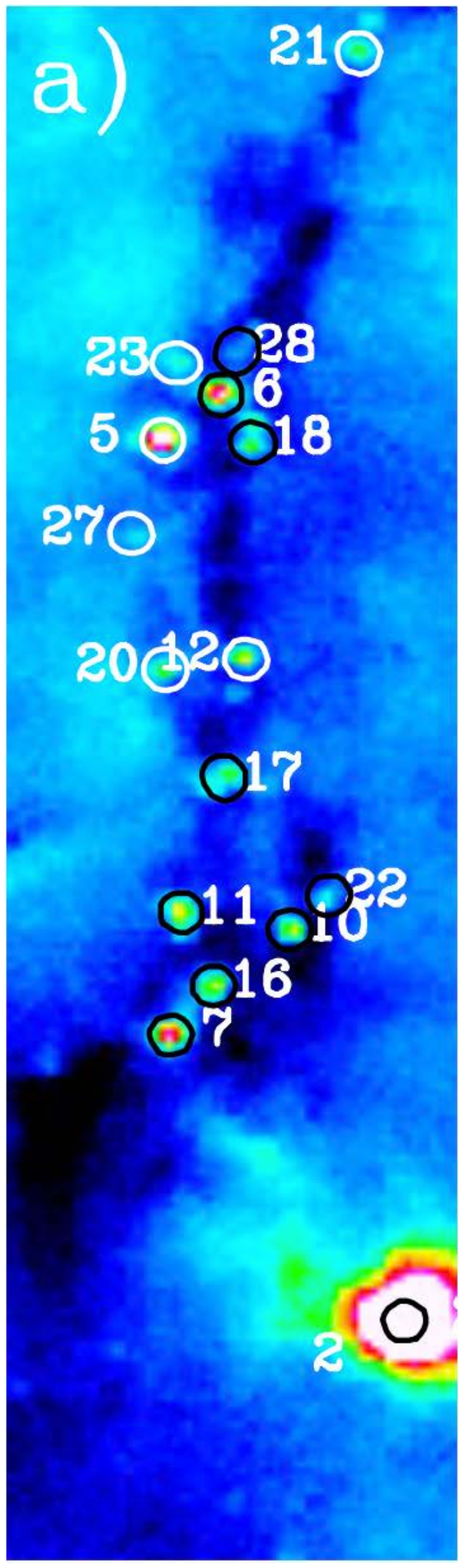}\vskip -11.5cm\hskip 3cm \includegraphics[angle=0,width=8cm]{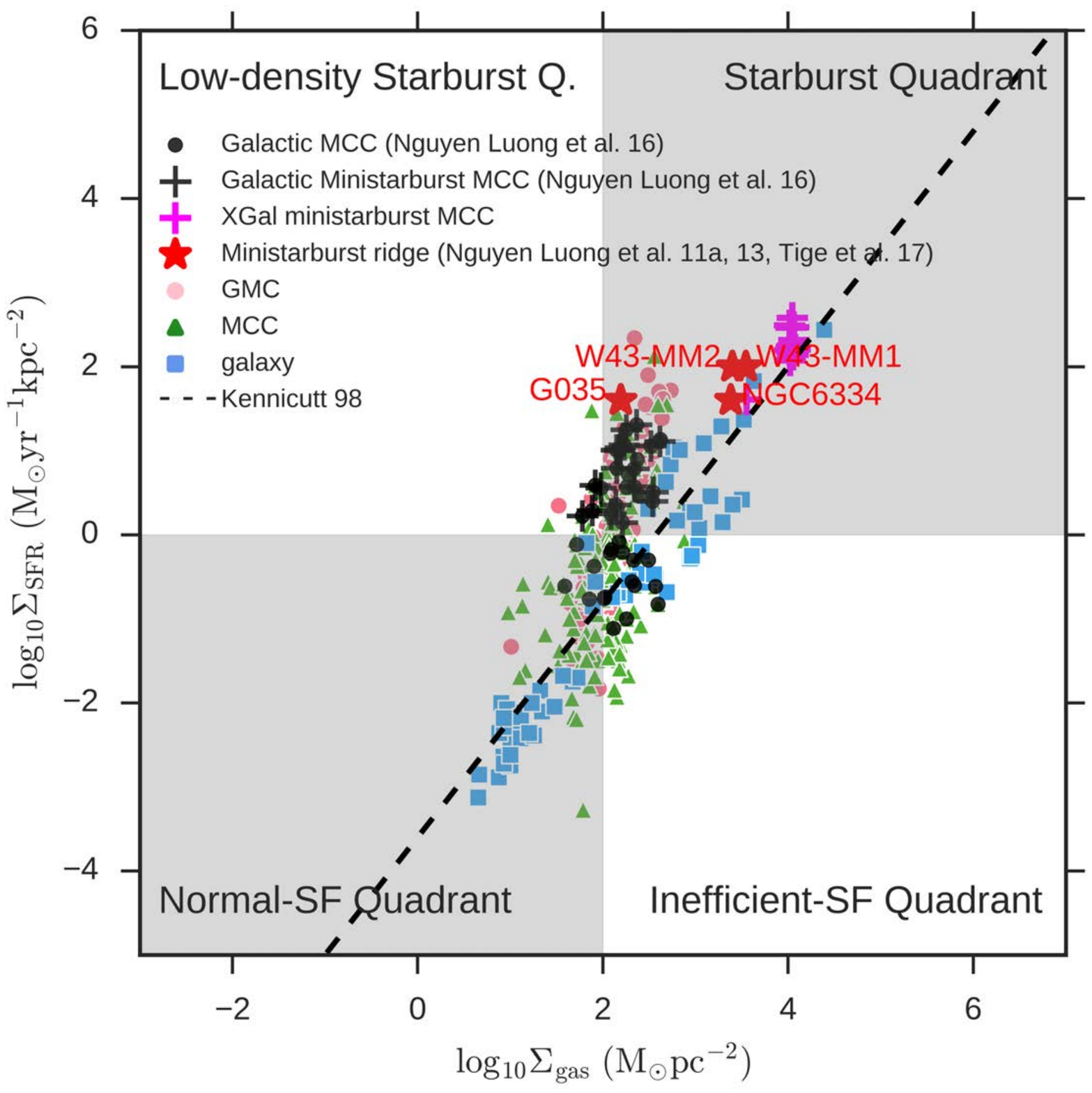}
\caption{With their high-density and intense star formation activity \textbf{(Left)}, ridges qualify as mini-starburst clumps \textbf{(Right)}. 
\textbf{Left:} Massive protostar cluster ($70~\mu$m emission and circles) within IRDC G035-00.33 ridge (absorption silhouette in blue colors). Protostar counting provides estimates of the current and instantaneous SFR.
\textbf{Right:} Schmidt-Kennicutt  diagram \citep{kennicutt98}, which plots SFR density as a function of the gas mass surface density. Ridges lie in the starburst quadrant (red star markers) like mini-starburst molecular cloud complexes (MCC). 
Adapted from \cite{nguyen11a, nguyen16} with permission.}
\label{f:SKdiag}
\end{figure}

The extreme characteristics of ridges and hubs, in terms of density and kinematics (see Sect.~\ref{s:ridge}), could lead to an atypical star formation activity. Star formation efficiencies (SFE) and rates (SFR) are indeed predicted to continuously increase with gas density \cite[e.g.,][]{HeCh11}. The clear accumulation of high-mass protostars observed along ridges (see, e.g., \textbf{Figure}~\ref{f:SKdiag} Left) in fact tends to suggest an intense star formation activity.

Among the numerous methods used to estimate SFRs, the two more direct ones, based on young star or protostar counting, are the most relevant for estimates in nearby clouds of the Milky Way \citep[see, e.g.,][]{vutisalchavakul16}.
Several authors have used near- to mid-IR imaging, such as those done with \emph{Spitzer}, to measure SFRs. They either counted pre-main sequence, T Tauri, stars or integrated the diffuse mid-IR polycyclic aromatic hydrocarbon (PAH) emission attributed to the luminosity impact of recently formed OB stars on the cloud. The first method is the only direct one. It was applied in nearby, $<$500~pc, low-mass star-forming regions \citep[e.g.,][]{heiderman10, dunham15}. The second method was used when the angular resolution was not sufficient for counting purposes, for example for Galactic molecular complexes (\citealt{nguyen11b}; see also \citealt{eden12} using \emph{Herschel} $70~\mu$m). If the star formation activity varies with time, these two methods, based on $\sim$$2\times10^6$ years-old T Tauri and $\sim$$10^6$ years-old OB stars, would measure past and integrated SFRs. 

In contrast, counting $\sim$$3\times 10^5$ years-old protostars permits evaluation of the current and instantaneous star formation activity, meaning the star formation developing for a few free-fall times of MDCs. This method was applied first on regions imaged by ground-based submillimeter radio-telescopes \citep[e.g.][]{motte03,maury11} and more recently on \emph{Herschel} images \citep[e.g.][]{nguyen11a,sadavoy14}.

On the ridge spatial scale, $\sim$1~pc, current and instantaneous SFRs have been estimated for a few ridges  \citep[e.g.,][]{motte03,nguyen11a, louvet14}. They have star formation rate densities, $\Sigma_{\rm SFR}~\sim 10-100~\msun$\,yr$^{-1}$\,kpc$^{-2}$, on $1-10$~pc$^2$ areas, worthy of starburst galaxies, usually defined by $\Sigma_{\rm SFR} > 1$ (see \textbf{Figure}~\ref{f:SKdiag} Right). For this reason, G035.39-00.33 and W43-MM1 were called mini-starburst ridges, i.e. miniature and instantaneous models of starburst galaxies. 

These mini-starburst events most probably follow the formation of the ridges, which was proposed to develop through gravitationally driven, and possibly colliding, flows (see Sect.~\ref{s:ridge}). Indeed, short mini-bursts of star formation are to be expected after a fast episode of cloud formation \cite[e.g.,][]{vazquez08} or, equivalently, for a cloud under compressive turbulent forcing \citep{FeKl12}. In this scenario the star formation should gradually settle within ridges. We should thus measure different SFR levels depending on the evolutionary status of ridges. To investigate this statement, one must calculate current and instantaneous SFRs, whose timescale, over which the SFR is integrated, remains much shorter than the ridge formation timescale, $\sim$$3\times10^5$~yr versus $\sim$$10^6$~yr.

 \cite{louvet14} measured instantaneous SFRs within regions of the W43-MM1 ridge and found a clear correlation of SFR with cloud density. This result recalls the correlation of the core formation efficiency (CFE, cloud concentration at high densities) with cloud density \cite[see][]{motte98,bontemps10, palau13}. Interestingly, SFR is smaller within the part of the ridge where the shocks associated with most recent cloud formation are the strongest \citep{louvet14, louvet16}. Therefore, the dynamical ridge formation may well be followed by a series of intense bursts of star formation.

\section{TOWARD GALAXY-WIDE SURVEYS}
\label{s:prospects}

For a long time, the only way to work on comprehensive samples of high-mass star-forming sites was to focus on the nearest, massive molecular complexes (see Sect.~\ref{s:HOBYSclouds}). Galaxy-wide surveys performed during the past decade are now entering the maturity phase and provide the well-characterized samples of high-mass star and massive cluster precursors, which are necessary to make progress (see Sect.~\ref{s:GPsurveys}). In parallel, the physical processes of high-mass star and cluster formation will soon be investigated for variations throughout the Milky Way (see Sect.~\ref{s:starburstclouds}), with the ultimate goal to extrapolate them to other galaxies.

\begin{table}[h]
\begin{center}
\caption{Massive molecular cloud complexes forming high-mass stars at less than 3~kpc and the reference Orion region.}
\begin{tabular}{lcccccl}
\hline \hline
Complex      &     d$_{\rm Sun}$   & Gas mass$^{\mathrm{a}}$   & Size & $<n_{\rm H2}>$ & Ref.$^{\mathrm{b}}$ & High-mass star-forming\\
name      &   (kpc)  & ($M_\odot$)       & (pc)      &   (cm$^{-3}$)   &               &  complex\\
\hline
Cygnus~X$^\star$       & 1.4      & $3.4\times 10^6$    & 200   &  7.9       & (1)     & Richest and most nearby\\
       &          &                   &           &           &              & \\
Rosette$^\star$         & 1.6      &  $3.1\times 10^5$  & 96     &  6.1                  &  (2)(3)          &  Relatively isolated\\
       &          &                   &           &           &              & \\
M16/M17$^\star$        & 1.7      & $8.6\times 10^5$  & 120   &  9.1          & (4)         & In the Sagittarius arm\\
      &          &                   &           &              &           & \\
NGC6334-6357$^\star$       & 1.7      & $6.6\times 10^5$  & 99  &  11.8          & (4)          &  In the Carina-Sag. arm\\
      &          &                   &           &               &          & \\
Vulpecula           & 2.0      & $7.7\times 10^5$    & 140  &    5.1 & (4)(5)          & Recently identified\\
       &          &                   &           &             &            & \\
G345           & 2.0      & $5.6\times 10^5$    & 116  &    6.1 & (4)          & Recently identified\\
      &          &                   &           &             &            & \\
W3/KR140$^\star$          & 2.2      & $7.4\times 10^5$    & 140      &  4.6             & (6)(7)     & In the Perseus arm\\
      &          &                   &           &             &            & \\
Carina           & 2.3      & $4.5\times 10^5$    & 110  &    6.3 & (4)(8)          & Formed a massive cluster\\     
       &          &                   &           &           &              & \\
NGC~7538$^\star$       & 2.8      & $3.2\times 10^5$    & 65    &  20             & (9)         & In the Perseus arm\\
      &          &                   &           &           &              & \\
W48$^\star$            & 3.0      & $1.6\times 10^6$    & 170  &   5.7  & (4)          & In the molecular ring or\\
             & 1.6      & $4.5\times 10^5$    & 90  &    10.6 & (4)          & at a much closer  distance \\
\hline
Orion           & 0.45      & $3.2\times 10^5$    & 100  &    5.5 & (10)          & Formed the ONC cluster\\
\hline
\end{tabular}
\label{tab:hobysC}
\end{center}
\begin{small}
\begin{list}{}{}
\item[$^\star$]
Molecular complexes imaged by the \emph{Herschel}/HOBYS survey \citep{motte10}.
\item[$^{\mathrm{a}}$]
The listed masses and average densities are, for homogeneity reasons, derived from 2MASS extinction maps. They are found to be similar to those derived using other methods such as CO surveys.
\item[$^{\mathrm{b}}$] References:
(1) \cite{schneider06};
(2) \cite{williams95};
(3) \cite{Heyer06};
(4) Bontemps et al. in prep.;
(5)  \cite{billot10};
(6) \cite{lada78};
(7) \cite{carpenter00};
(8)  \cite{preibisch12};
(9) \cite{ungerechts00};
(10) \cite{tatematsu98}.
\end{list}
\end{small}
\end{table}

\subsection{Most nearby, massive molecular cloud complexes}
\label{s:HOBYSclouds}

The most nearby cloud complexes are particularly interesting because they offer the opportunity to reach the smallest spatial scales and separate individual collapsing objects. To actually probe the accreting phase for high-mass star formation, it is however necessary to focus on the most massive such complexes. With a total mass of $3\times10^5~\msun$ \citep{tatematsu98}, Orion is itself not massive enough to contain more than a couple of high-mass protostars. Such low numbers are estimated assuming typical cloud lifetime and SFE, $10^7$~yr and $3\%$, a typical IMF with $\sim$10$\%$ of the stellar mass in high-mass stars, and a protostellar lifetime of $3\times10^5\,$yr (see \textbf{Table}~\ref{tab:lifetime}). If we restrict ourselves to the IR-quiet high-mass protostellar phase, whose lifetime is even shorter, less than one such object is expected in Orion and indeed no high-mass IR-quiet protostar is known in Orion.

About ten years ago, to prepare the \emph{Herschel} surveys \citep[such as \emph{Herschel}/HOBYS, ][]{motte10}, S. Bontemps derived a near-IR extinction image of the Milky Way. Partly shown in Figs.~4-6 of \cite{schneider11}, it was built from the stellar reddening measured by 2MASS and confirmed by CO surveys according to the method presented in \cite{schneider11}. It led to the identification of $\sim$100~pc molecular cloud complexes, which are massive enough, $\ge$$3\times10^5~\msun$, to contain high-mass protostars. Located at less than 3~kpc from the Sun, these cloud complexes can be imaged by \emph{Herschel} with a spatial resolution below 0.1~pc at 70~$\mu$m.
\textbf{Table}~\ref{tab:hobysC} lists the seven complexes selected for the \emph{Herschel}/HOBYS survey, Carina, and two recently identified complexes. They constitute a complete sample of molecular cloud complexes, which currently form high-mass stars at less than 3~kpc. The nearby cloud complexes of  \textbf{Table}~\ref{tab:hobysC} contain $\sim$30 times more mass than Orion and one/two of them is/are among the most massive molecular cloud complexes known to date, $\ge$$10^6~\msun$ (see \citealt{nguyen16}). The amount of molecular gas contained in these cloud complexes should provide about 80 OB star precursors and should statistically permit studying the precursors of stars with masses from 8 to $20~\msun$.
\begin{marginnote}[]
\entry{\emph{2MASS}}{The Two Micron All Sky Survey scanned 70\% of the sky in J, H, and K.}
\end{marginnote}

Because all Gould Belt clouds sum up to about $8\times 10^5~\msun$ but do not host high-mass star precursors \citep{andre10}, the combination of large mass and high density may be prerequisites for a cloud structure to be able to form high-mass stars. 
Orion indeed appears in \textbf{Table}~\ref{tab:hobysC} as those with both the lowest average density and lowest cloud mass. The Carina molecular complex is also interesting because it formed already several clusters hosting high-mass stars but may not form high-mass stars anymore \citep{gaczkowski13}. It remains to be investigated if the strong feedback effects of Carina \hii regions do prevent further generation of high-mass stars to form in this rather massive cloud complex.

The most nearby, massive molecular cloud complexes are rich star-forming sites, whose study allowed researchers to make definite progress in the understanding of high-mass star formation, massive cloud and cluster formation (see Sects.~\ref{s:HMSF}--\ref{s:cloud+cluster}). However, located up to $\sim$3~kpc, these complexes can only offer limited statistics to study the shortest, and therefore rarest, phases of the formation of high-mass stars. Therefore, Galaxy-wide surveys are required to fully investigate each step of the high-mass star formation process.

\begin{figure}[h]
\vskip -1.5cm
\hskip 1.5cm\includegraphics[angle=0,width=5.5cm]{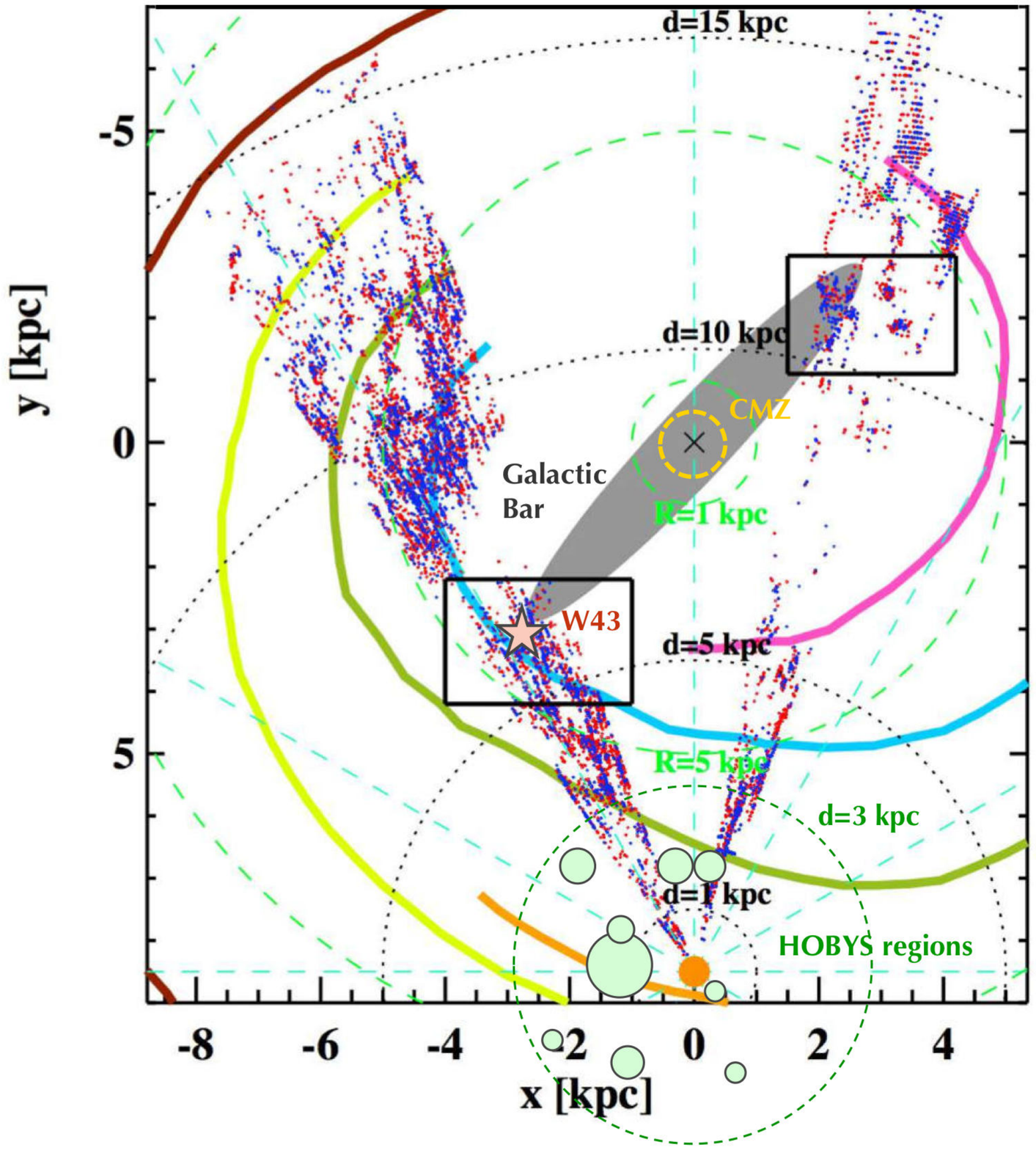}\vskip -8cm \hskip 7.5cm\includegraphics[angle=270,width=9cm]{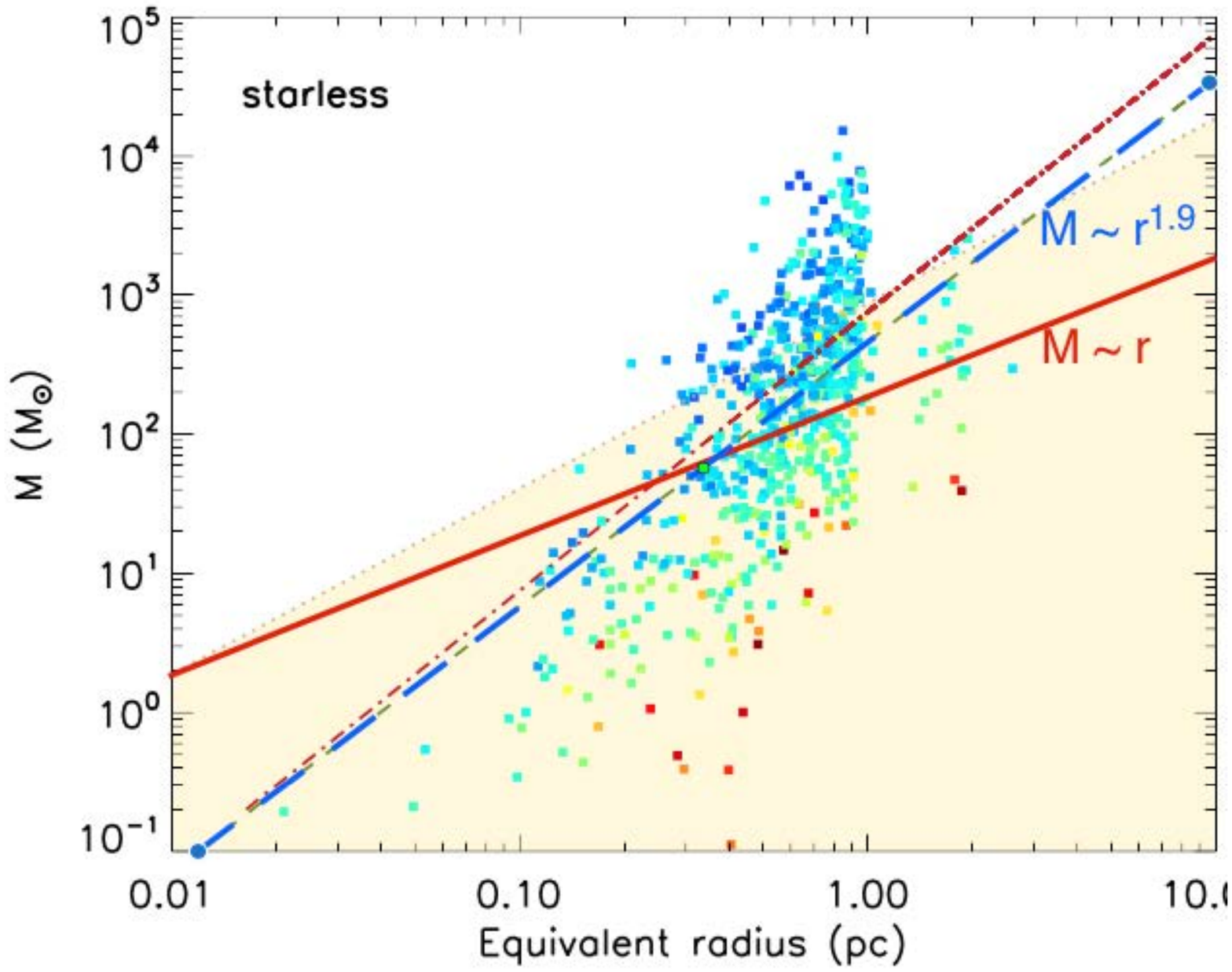}
\vskip 2.5cm
\caption{First analyses of clumps, which were identified and characterized (distance to the Sun, mass, luminosity) by the \emph{Herschel}/Hi-GAL survey \citep{molinari16b, elia17}.
{\bf Left:} Galactic distribution of clumps in two cones covering the tips of the long bar and the W43 cloud complex. The locations of the nearby cloud complexes selected for the \emph{Herschel}/HOBYS survey are shown as light green circles with areas proportional to the total masses of the individual complexes (from \textbf{Table}~\ref{tab:hobysC}). The colored thick curves trace the empirical Galactic spiral arms and the orange circle represents the Sun.
{\bf Right:} Mass versus Radius diagram of the starless clumps at $15-55^{\rm o}$ longitude (color-coded for their SED fit temperature around $\sim$15~K).  The empirical thresholds of \cite{KaPi10} and \cite{urquhart14} (dotted and dashed-dotted red lines) suggest that 171 of the most massive clumps could form high-mass stars. 
Their distribution is however much steeper than the $M(<r)\propto r$ relation of gravity-dominated clumps (red line) and closer to the $M(<r)\propto r^2$ relation found for turbulence-dominated structures. According to Sect.~\ref{s:gasConcent}, it thus remains unclear if present massive starless clumps could host high-mass prestellar cores. 
Adapted from \cite{veneziani17} and \cite{traficante15} with permission.}
\label{f:higal}
\end{figure}

\subsection{Combination of Galaxy-wide surveys and detailed images with ALMA}
\label{s:GPsurveys}

It is the convergence of large far-IR to millimeter imaging surveys of the Galactic plane, proper distance derivation analyses, and high-spatial resolution follow-up observations which has recently opened a new window on high-mass star formation. Large imaging surveys are required to get unbiased lists of high-mass star precursors. To recognize individual high-mass protostars and prestellar cores, Galactic plane studies need precise distance derivations and high-spatial resolution follow-ups. High-mass star-forming sites provided by Galactic plane surveys indeed spread over large, typically $1-15$~kpc, distance ranges, and most of them are expected to lie further than 5~kpc from the Sun (see, e.g., \textbf{Figure}~\ref{f:higal}a). We summarize below the results and prospects of the most relevant programs in this context.

\begin{marginnote}[]
\entry{\emph{GLIMPSE}}{The Galactic Legacy IR Mid-Plane Survey Extraordinaire covered the Galactic plane at $4-8~$$\mu$m.}
\entry{\emph{MIPSGAL}}{The MIPS GALactic plane survey covered the Galactic plane at 24~$\mu$m (70~$\mu$m band almost useless).}
\end{marginnote}
The advent of large-format millimeter cameras on ground-based telescopes and of two space missions, \emph{Spitzer} and \emph{Herschel}, have provided a whole set of complete and sensitive surveys of the inner Galactic plane. The most sensitive and complete surveys are \emph{Spitzer}/GLIMPSE and \emph{Spitzer}/MIPSGAL from 4 to 24~$\mu$m  \citep{benjamin03}, \emph{Herschel}/Hi-GAL from 70 to 500~$\mu$m \citep{molinari10}, APEX/ATLASGAL at  870~$\mu$m \citep{schuller09}, and the CSO/BGPS at 1.1~mm \citep{aguirre11, ginsburg13b}.
The \emph{Spitzer}/GLIMPSE survey has been particularly successful to recognize and build complete samples of IRDCs \citep{simon06a,PeFu09}. The extremely green objects also appeared as useful probes of the earliest phases of (high-mass) star formation \cite[EGOs,][]{cyganowski11}. 
\begin{marginnote}[]
\entry{\emph{ATLASGAL}}{The APEX Telescope Large Area Survey of the Galaxy covered the whole inner Galactic plane at 870~$\mu$m.}
\entry{\emph{BGPS}}{The BOLOCAM Galactic Plane Survey imaged it at 1.1~mm from the Northern hemisphere.}
\end{marginnote}

\emph{Herschel}/Hi-GAL becomes a reference survey for the earliest phases of high-mass star-forming sites in the Galactic plane \citep{molinari16b}.
With its five far-IR bands covering the SED peak of stellar precursors, it has the potential to trace both the column density and temperature of dusty cloud fragments. However, its highest resolution at the shortest wavelengths, $7''-25''$ at $70-250~\mu$m, is barely sufficient to resolve clumps hosting protoclusters of $\sim$0.5~pc typical sizes when located at $5$~kpc from the Sun \citep[see, e.g., Fig.~3 of][]{beltran13}. Interferometric follow-ups at (sub)millimeter wavelengths, with e.g. ALMA, will thus be mandatory to probe individual protostellar or prestellar cores forming high-mass stars in these clumps.
Until now, the Hi-GAL survey started the census of clumps throughout the Milky Way and discussed their evolutionary sequence from their earliest phases to the development of UC\hii regions \citep[e.g.,][]{elia17}. Among the large numbers of high-mass star-forming sites, Hi-GAL identified hundreds of starless clump candidates, whose gas concentration resemble that of starless MDCs \citep[e.g.,][compare \textbf{Figures}~\ref{f:mass-radius} and \ref{f:higal}b]{traficante15}. The distribution, mass, and luminosity of clumps in the Galaxy was also used to estimate SFRs of some specific areas such as the tips of the Galactic long bar \citep[][see \textbf{Figure}~\ref{f:higal}a]{veneziani17}. 

\begin{figure}[htbp]
\hskip 2.4cm\includegraphics[angle=0,width=9.7cm]{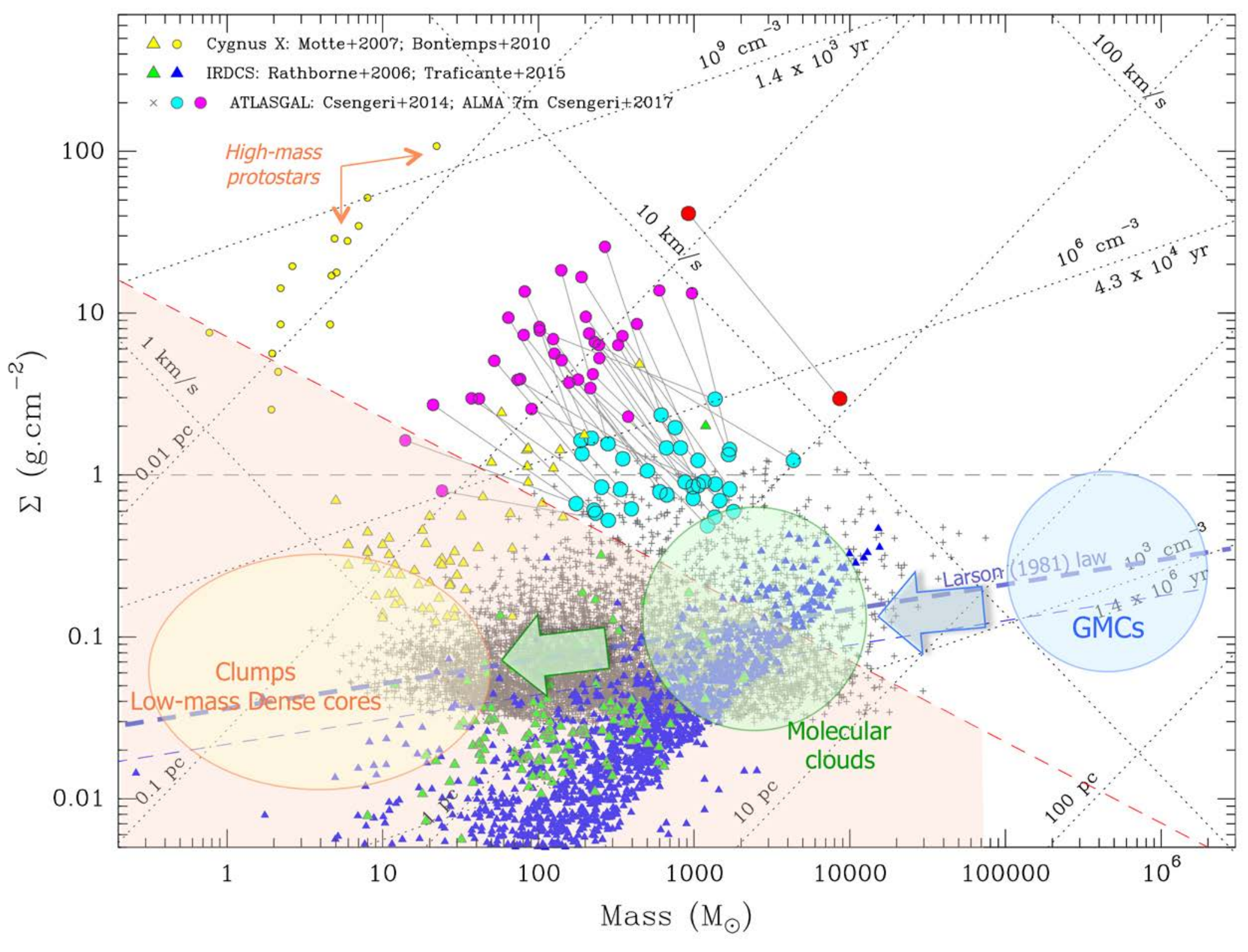}
\vskip 0.5cm
\caption{Surface density versus mass diagram of the most massive IR-quiet clumps of our Galaxy (cyan circles), which generally fragment into single dominating MDCs (pink circles)  \citep{csengeri14, csengeri17}. These clumps and MDCs should be able to form high-mass stars, according to their location above, e.g., the empirical threshold by \cite{KaPi10} (red dashed line). In average, the most massive ATLASGAL clumps and hosted MDCs are as dense as Cygnus~X MDCs and protostars \citep[yellow markers above the red dashed line,][]{motte07,bontemps10} and the W43-MM1 clump and MDC \citep[red circles,][]{motte03,louvet14}. In contrast, they are much denser than IRDC clumps \citep[green and blue triangles,][]{rathborne06, traficante15} and clumps of the whole ATLASGAL sample \citep[black crosses,][]{csengeri14}. Three grids of dotted lines represent gas mass concentration at constant sizes, constant densities (or free-fall times), and constant escape velocities \citep{tan14}. The mass concentration in the 35 most massive IR-quiet clumps of ATLASGAL, which are located at less than 4.5~kpc, generally is consistent with the $M(<r) \propto r$ relation, parallel to constant escape velocity lines. It recalls gravity-dominated cloud structures with $\rho(r)\propto r^{-2}$ densities (see also \textbf{Figure}~\ref{f:mass-radius}) and clearly departs from the Larson's law, emphasized here with large arrows linking GMCs/cloud complexes, molecular clouds, and dense cores \citep[large shaded ellipses,][]{BeTa07}. 
Adapted from \cite{csengeri17} with permission.}
\label{f:surfdens}
\end{figure}

The ATLASGAL survey covered the whole inner Galactic plane at 870~$\mu$m, with a $19''$ resolution \citep{schuller09}. This uniquely complete survey of the submillimeter range revealed several thousands of dense clumps, mostly located at 2 to 8~kpc distances \citep[e.g.,][see \textbf{Figure}~\ref{f:surfdens}]{contreras13, csengeri14}. Huge efforts have been dedicated to distance determination, using NH$_3$ line detections of the most massive ATLASGAL clumps and innovative methods to resolve the distance ambiguity \citep{wienen15}. Recent VLBI determinations of maser parallaxes have shown that kinematic distances can still be inaccurate when there are large deviations from the mean rotation curve of the Milky Way. \cite{wienen15} incorporated the recent results from VLBI distance determinations to improve ATLASGAL clump distances. In the very near future, the release of \emph{GAIA} catalogs will greatly improve the distance determination of Galactic young clusters (some massive stars in most, even distant, clusters are optically visible and will be part of the  \emph{GAIA} catalogs) and thus of their associated star formation sites. As for now, virtually all massive clumps detected by the ATLASGAL survey have reasonably well-determined distances. 
\begin{marginnote}[]
\entry{\emph{VLBI}}{The Very Long Baseline Interferometry technique derives parallax distances for strong maser sources.}
\entry{\emph{GAIA}}{The Gaia mission aims to chart a three-dimensional map of the Milky Way.}
\end{marginnote}

ATLASGAL and Hi-GAL are the perfect surveys to identify the targets to observe at high-resolution when the goal is to recognize and statistically study high-mass protostars and prestellar cores. Indeed,  ATLASGAL and Hi-GAL catalogs \citep{csengeri14,elia17} include $\sim$0.1~pc MDCs and $\sim$1~pc massive clumps that need to be investigated down to the 0.02~pc scale of protostars. While several surveys already identified a few cloud structures able to form high-mass stars \citep[e.g.,][see Sect.~\ref{s:survey}]{motte07}, only Galaxy-wide surveys like ATLASGAL and Hi-GAL can reveal a statistically-significant number of clumps, which sit in the high-mass star formation regime \cite[][see \textbf{Figures}~\ref{f:surfdens} and \ref{f:higal}b]{csengeri14, traficante15}. \textbf{Figure}~\ref{f:surfdens} shows the distribution of ATLASGAL clumps in a surface density versus mass diagram adapted from \cite{tan14}. As part of a first ALMA imaging survey, as many as 35 of these ATLASGAL clumps were observed down to a $\sim$0.01~pc spatial resolution with the ALMA 7~m and 12~m antennas. They are the most massive ATLASGAL IR-quiet clumps at less than 4.5 kpc and thus are candidate precursors of the richest, highest-mass clusters of the Milky Way. The first results on the fragmentation properties of these clumps with the ALMA 7~m compact array is displayed in \textbf{Figure}~\ref{f:surfdens} \citep{csengeri17}. It shows that most of these clumps are much more centrally concentrated than cloud structures following the Larson's law \citep{larson81}. Their gas mass concentration close to $M(<r)\propto r$ is an indication that they are gravity-dominated and could have formed dynamically, in agreement with their short lifetime \citep[$7.5\times10^4$~yr,][]{csengeri14}. The most massive ATLASGAL clumps also are overdense by one to two orders of magnitude with respect to the density-size Larson's relation shared by typical cloud complexes and low-mass dense cores (sizes from 100 pc to 0.1 pc). Massive ATLASGAL clumps may thus originate from gas concentration loci associated with large-scale collapses rather than typical turbulent fluctuations \citep{csengeri17}.

This ALMA survey will investigate if MDCs further concentrate down to the individual core scale of 0.02~pc, as observed for Cygnus~X and W43-MM1 protostellar MDCs \citep[][see \textbf{Figure}~\ref{f:surfdens}]{bontemps10, louvet14}. This survey, along with other similar ALMA projects, will also have the potential to finally determine if high-mass prestellar cores exist or not.

\subsection{Extreme molecular cloud complexes in the Milky Way and starburst clusters}
\label{s:starburstclouds}

The Milky Way offers a wide variety of star formation and cluster formation environments, from the Galactic disk to the bar and central regions which can be probed for environ,mental effects on the process of star and cluster formation.  One can first investigate variations of the star formation activity across the Milky Way and test star formation models up to their limit by studying the extreme molecular cloud complexes of our Galaxy. They host very massive ridges or hubs, called mini-starburst clumps, precursors of young massive clusters, or even precursors of super star clusters  \citep[e.g.,][]{motte03,bressert12}. Their study is just starting as presented in the recent review by \cite{longmore14}.

W43 probably is the best-studied of these mini-starburst regions. The molecular complex was identified from a combination of its CO and H~{\small I} line cubes and ATLASGAL 870~$\mu$m emission \citep{nguyen11b}. It has
a 290~pc diameter and $7\times 10^6~\msun$ mass, characteristics quantitatively larger than those of nearby massive molecular complexes (see \textbf{Table}~\ref{tab:hobysC}). 
W43 is located at the junction of the Scutum arm and the Galactic bar \citep{nguyen11b, carlhoff13}. Its star formation activity estimated from cloud concentration at 0.1~pc scales and from its  \emph{Spitzer} $8~\mu$m emission suggests it qualifies as a mini-starburst \citep{motte03, nguyen11b}. W43 host two ridges, whose cloud structure and star formation content are as extreme as in mini-starbursts \citep[][see Sect.~\ref{s:starburst} and \textbf{Figure}~\ref{f:SKdiag} Right]{nguyen13, louvet14}. The exceptional star formation activity measured in W43 is probably related to cloud-cloud agglomeration or collision events suggested from few hundreds of pc \citep{motte14, renaud15} down to $\sim$1~pc \citep{nguyen13, louvet16}.
ALMA imaging of the W43-MM1 ridge, which is the most concentrated pc-scale cloud at less than 6~kpc, revealed a rich protocluster. It is currently being investigated to look for high-mass prestellar cores and to constrain the origin of the IMF in extreme clouds (Nony et al. in prep.; Motte et al. subm.).

Other well-known mini-starburst regions of the Milky Way galactic disk are W49 \citep[e.g.,][]{galvan13}, W51 \citep[e.g.,][]{ginsburg15}, and Sgr~B2 \citep[e.g.,][]{schmiedeke16}. A catalog of mini-starburst complexes of the Milky Way has been built using CO and centimeter free-free surveys and aims to help go beyond these very few examples \citep{nguyen16}.
 
In contrast to these mini-starburst regions of the Galactic disk, the Central Molecular Zone contains high-density molecular clouds with low star formation activity \citep{longmore13}, even when investigating the current SFR through protostellar core counting in ALMA images \citep{henshaw17}. An analytical model proposed that the low star formation efficiency of this cloud is because of the strong shearing effects developing in the central Galactic regions \citep{kruijssen14}.

\section{CONCLUSIONS AND PERSPECTIVES}
\label{s:conc}
The fifteen past years have seen an increasing interest in approaching the issue of the formation of high-mass stars and massive clusters, from both the theoretical and observational sides. Here we reviewed the progress that was made from observations, especially with submillimeter radiotelescopes, the \emph{Herschel} far-IR observatory, and submillimeter interferometers.

Current knowledge of high-mass star formation is mainly based on statistical studies of distance-limited samples of molecular complexes (see \textbf{Table}~\ref{tab:hobysC}). High-mass star formation scenarios currently undergo a change of paradigm, in which this process is no longer quasi-static but simultaneously evolves with both cloud and cluster formation. 
The lifetime of high-mass protostars and the lack of high-mass, $\sim$0.02~pc-scale, prestellar cores presented in Sect.~\ref{s:HMSF} are consistent with the large dynamics of their hosted ridges, hubs, and MDCs on $\sim$1~pc to $\sim$0.1~pc scales (see Sect.~\ref{s:cloud+cluster}). As a consequence, we propose an evolutionary scenario, inside which the high-mass analogs of prestellar cores are replaced by large-scale, $\sim$$0.1-1$~pc, gas reservoirs, called starless MDCs or starless clumps. \emph{During their protostellar phase, these mass reservoirs would concentrate their mass into high-mass cores at the same time as they accrete stellar embryos, skipping the high-mass prestellar core phase}. Figure~\ref{f:scheme} illustrates this evolutionary scheme.

Although star cluster properties, among them the IMF \citep[see, e.g.,][]{kroupa01}, seem universal, our review suggests that massive stellar clusters form in extreme clouds called ridges. Unlike the case of low-mass stars which accrete their final mass from well-defined pre-stellar cores, with a mass distribution mimicking the IMF \citep{motte98, konyves15}, (high-mass) stars within ridges should have a much more complex accretion history (see Sect.~\ref{s:EvolSeq}). One should therefore investigate the detailed properties of mini-starburst protoclusters to constrain the outcome characteristics of massive clusters.

Because Galactic-scale surveys are starting to provide well-constrained samples of high-mass star-forming sites and because we are entering the ALMA era, we are at the dawn of 1/ definitively stating the evolutionary scenario for both high-mass star and massive cluster formation and 2/ following their evolution throughout the Milky Way and beyond.

\section*{DISCLOSURE STATEMENT}
The authors are not aware of any affiliations, memberships, funding, or financial holdings that might be perceived as affecting the objectivity of this review. 

\section*{ACKNOWLEDGMENTS}
We wish to thank Ewine van Dishoeck, Quang Nguy$\tilde{\hat{\rm e}}$n~Lu{\hskip-0.65mm\small'{}\hskip-0.5mm}o{\hskip-0.65mm\small'{}\hskip-0.5mm}ng, Pierre Didelon, and ARAA reviewers for helpful comments. We also thank all the authors who provided figures for this review: Henrik Beuther, Paolo Cortes, Timea Csengeri, Ana Duarte Cabral, Benjamin Gaczkowski, Martin Hennemann, Quang Nguy$\tilde{\hat{\rm e}}$n~Lu{\hskip-0.65mm\small'{}\hskip-0.5mm}o{\hskip-0.65mm\small'{}\hskip-0.5mm}ng, Nicolas Peretto, Nicola Schneider, J\'er\'emy Tig\'e,  Alessio Traficante, Marcella Veneziani, Ke Wang, Qizhou Zhang.
During the writing of this review we were supported by CNRS, CNES, and Programme National de Physique Stellaire (PNPS) and program Physique et Chime du Milieu Interstellaire (PCMI) of CNRS/INSU, France. This project has received funding from the European Union?s Horizon 2020 research and innovation programme under grant agreement No 687528.

\bibliographystyle{ar-style2.bst}


\newpage

\setcounter{figure}{0}
\begin{figure}[h]
\vskip -2cm
\centerline{\includegraphics[angle=0,width=2.8cm]{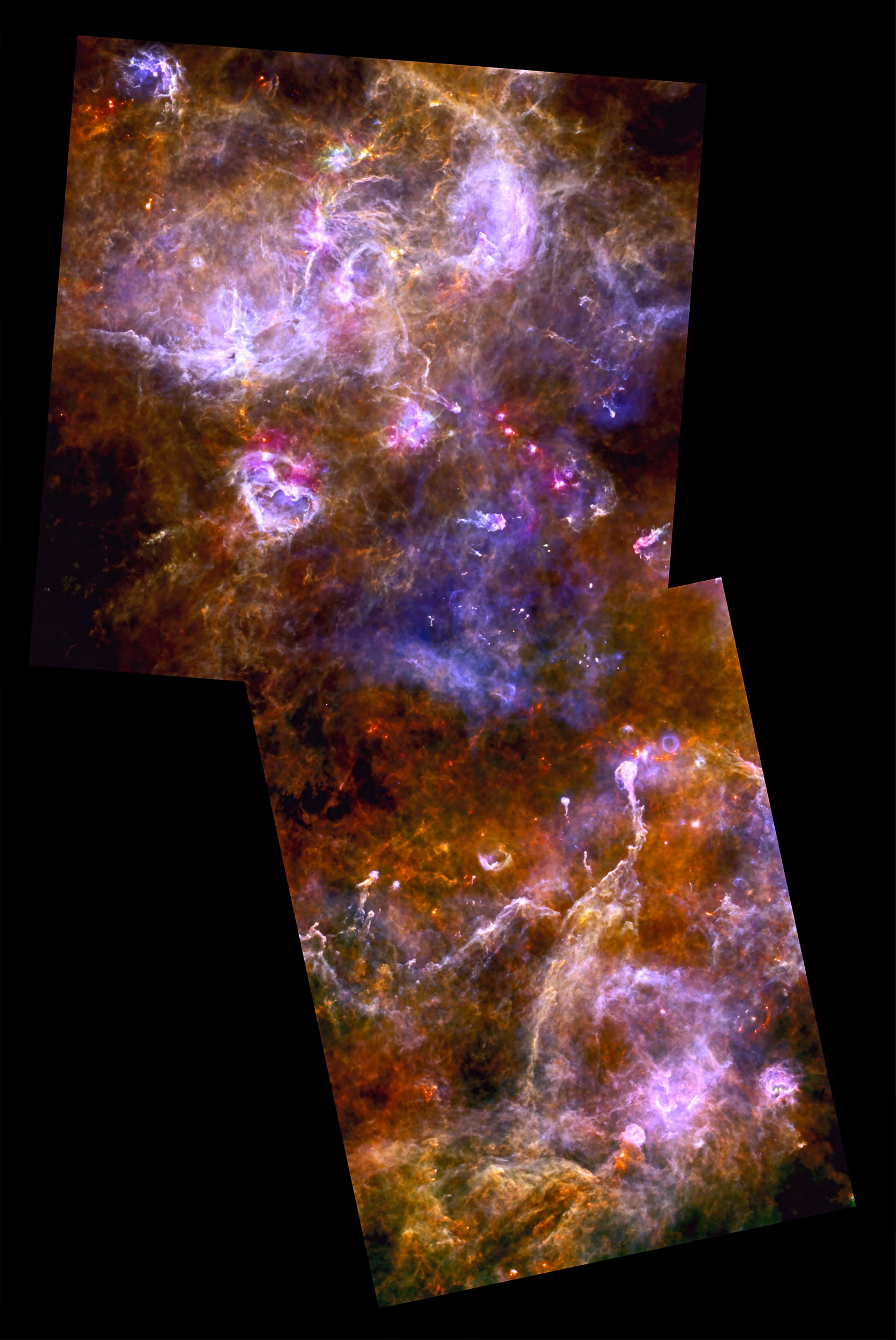}}
\vskip 0.5cm
  \caption{Among the 10 most massive molecular cloud complexes forming high-mass stars at less than 3~kpc (see \textbf{Table}~\ref{tab:hobysC}), Cygnus~X was imaged by the \emph{Herschel}/HOBYS key program \citep{motte10}.
Composite three-color \emph{Herschel} image with red\,=\,250\,$\mu$m, green\,=\,160\,$\mu$m, and blue\,=\,70\,$\mu$m. Blue diffuse emission corresponds to photo-dissociation regions around massive stars or clusters. Earlier stage star-forming sites are themselves seen as red filaments and orange MDCs. Abbreviation: MDC, massive dense core. Adapted from \cite{hennemann12} and \cite{schneider16b} with permission. }
\end{figure}

\begin{figure}[h]
\vskip -2cm
\hskip 3cm \includegraphics[angle=0,width=9.5cm]{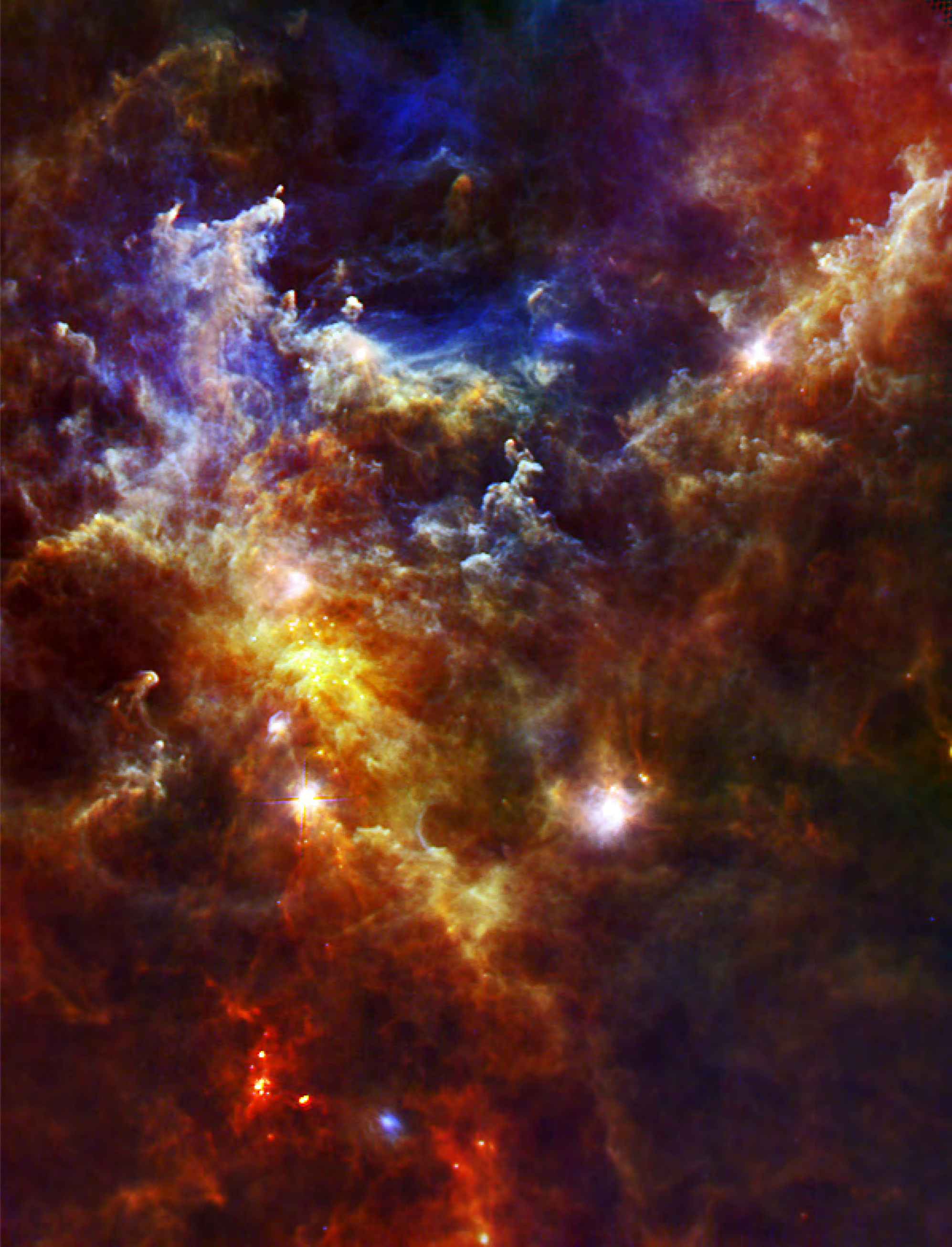}
\vskip 0.5cm
  \caption{Among the 10 most massive molecular cloud complexes forming high-mass stars at less than 3~kpc (see \textbf{Table}~\ref{tab:hobysC}), Rosette was imaged by the \emph{Herschel}/HOBYS key program \citep{motte10}.
Composite three-color \emph{Herschel} image with red\,=\,250\,$\mu$m, green\,=\,160\,$\mu$m, and blue\,=\,70\,$\mu$m. Blue diffuse emission corresponds to photo-dissociation regions around massive stars or clusters. Earlier stage star-forming sites are themselves seen as red filaments and orange MDCs. Abbreviation: MDC, massive dense core.
Adapted from \cite{motte10} and \cite{schneider10} with permission.
 }
\end{figure}

\begin{figure}[h]
\vskip -2cm
\hskip 3.5cm\includegraphics[angle=0,width=3.5cm]{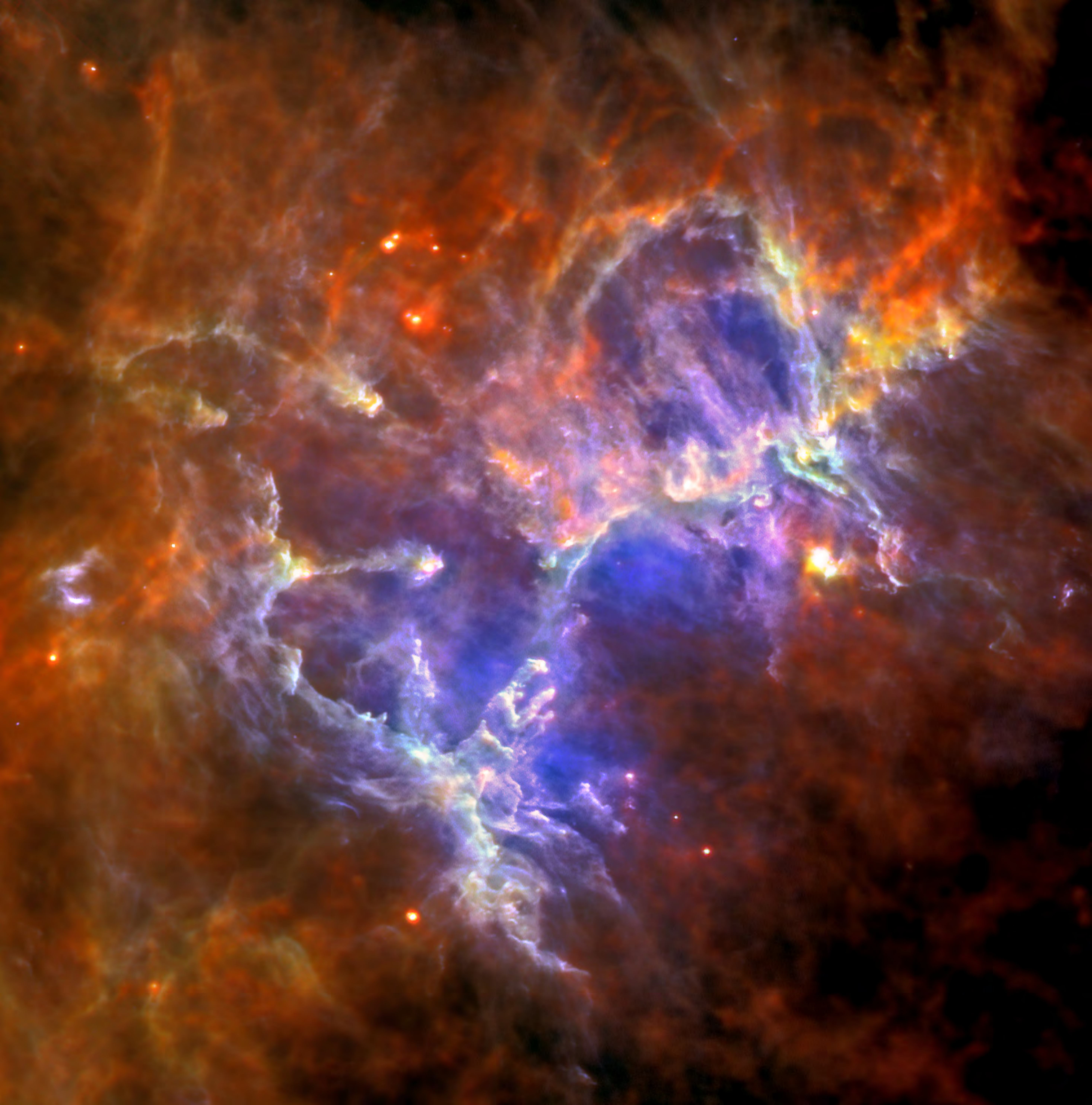}
\vskip 0.5cm
  \caption{Among the 10 most massive molecular cloud complexes forming high-mass stars at less than 3~kpc (see \textbf{Table}~\ref{tab:hobysC}), M16/M17 was imaged by the \emph{Herschel}/HOBYS key program \citep{motte10}. 
Composite three-color \emph{Herschel} image with red\,=\,250\,$\mu$m, green\,=\,160\,$\mu$m, and blue\,=\,70\,$\mu$m. Blue diffuse emission corresponds to photo-dissociation regions around massive stars or clusters. Earlier stage star-forming sites are themselves seen as red filaments and orange MDCs. Abbreviation: MDC, massive dense core.
Adapted from \cite{hill12} with permission.
 }
\end{figure}

\begin{figure}[h]
\vskip -4cm
\hskip 2.5cm \includegraphics[angle=0,width=10.5cm]{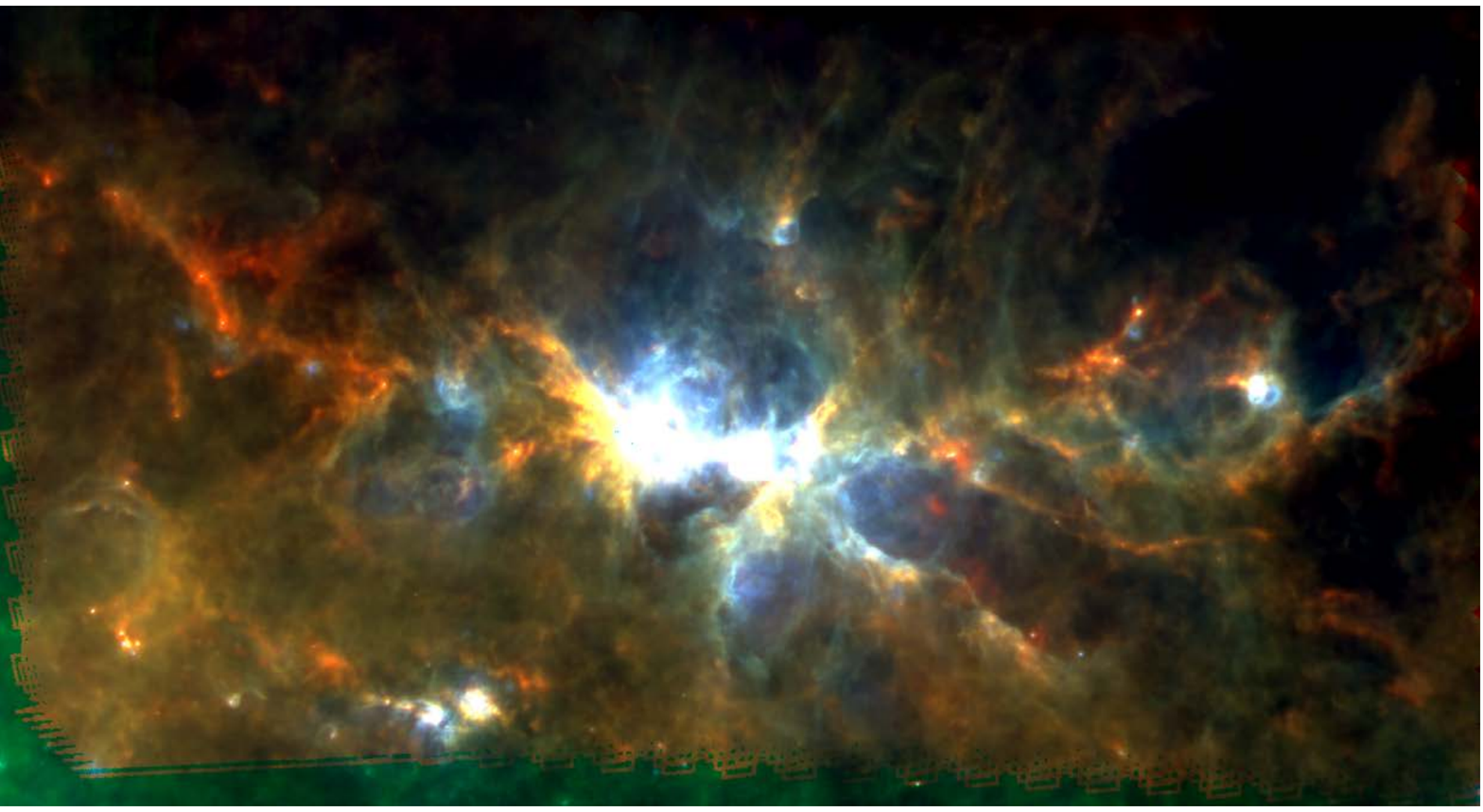}
\vskip -6.5cm
  \caption{Among the 10 most massive molecular cloud complexes forming high-mass stars at less than 3~kpc (see \textbf{Table}~\ref{tab:hobysC}), NGC~6334-6357 was imaged by the \emph{Herschel}/HOBYS key program \citep{motte10}. 
Composite three-color \emph{Herschel} image with red\,=\,250\,$\mu$m, green\,=\,160\,$\mu$m, and blue\,=\,70\,$\mu$m. Blue diffuse emission corresponds to photo-dissociation regions around massive stars or clusters. Earlier stage star-forming sites are themselves seen as red filaments and orange MDCs. Abbreviation: MDC, massive dense core.
Adapted from \cite{russeil13} with permission.
 }
\end{figure}

\begin{figure}[h]
\hskip 1cm\includegraphics[angle=0,width=9cm]{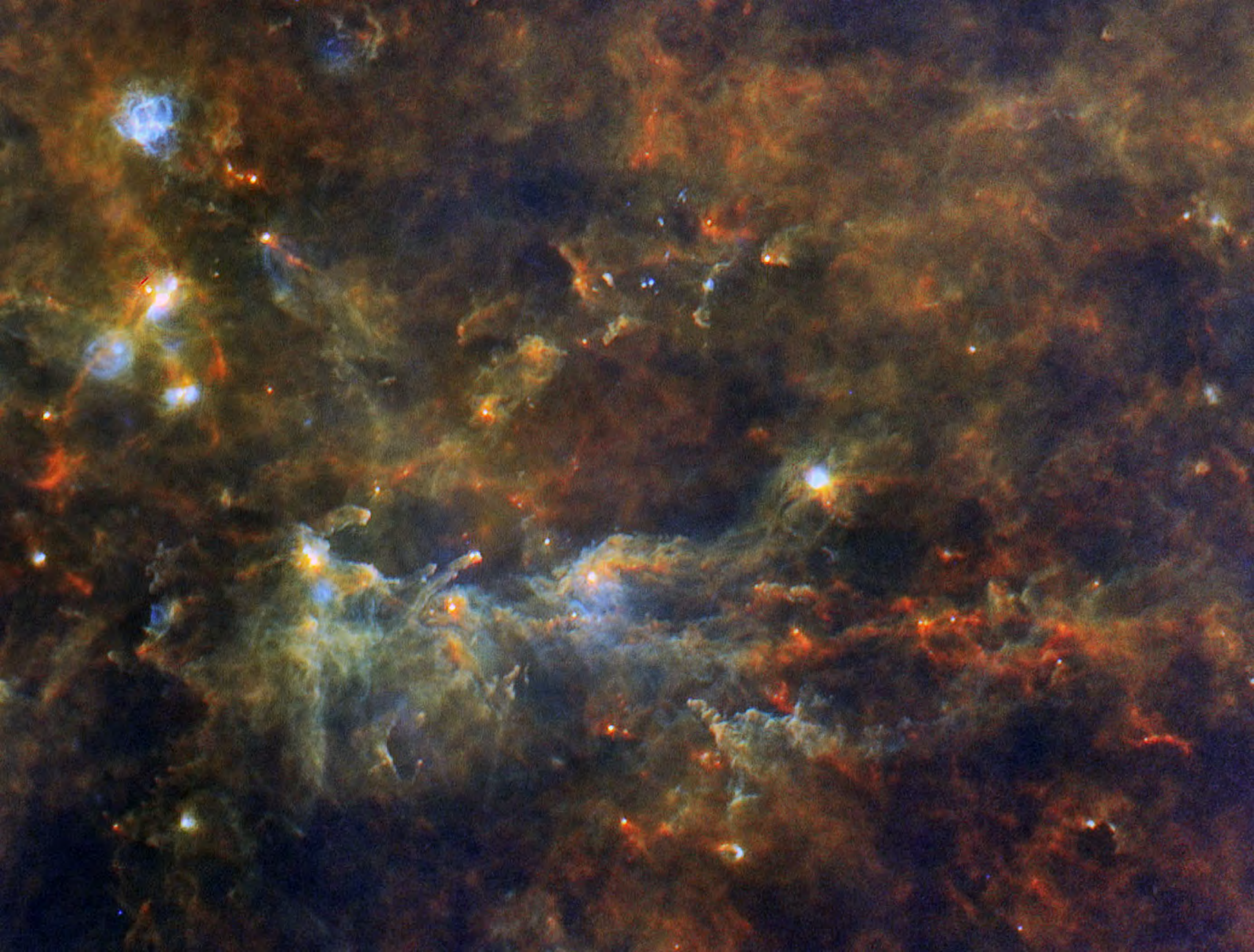}
\vskip 0.5cm
  \caption{Among the 10 most massive molecular cloud complexes forming high-mass stars at less than 3~kpc (see \textbf{Table}~\ref{tab:hobysC}), Vulpecula was imaged  the \emph{Herschel}/Hi-GAL key program \citep{molinari10}.
Composite three-color \emph{Herschel} image with red\,=\,250\,$\mu$m, green\,=\,160\,$\mu$m, and blue\,=\,70\,$\mu$m. Blue diffuse emission corresponds to photo-dissociation regions around massive stars or clusters. Earlier stage star-forming sites are themselves seen as red filaments and orange MDCs. Abbreviation: MDC, massive dense core.
Adapted from \cite{billot10} with permission.
 }
\end{figure}

\begin{figure}[h]
\hskip 7cm \includegraphics[angle=0,width=1.8cm]{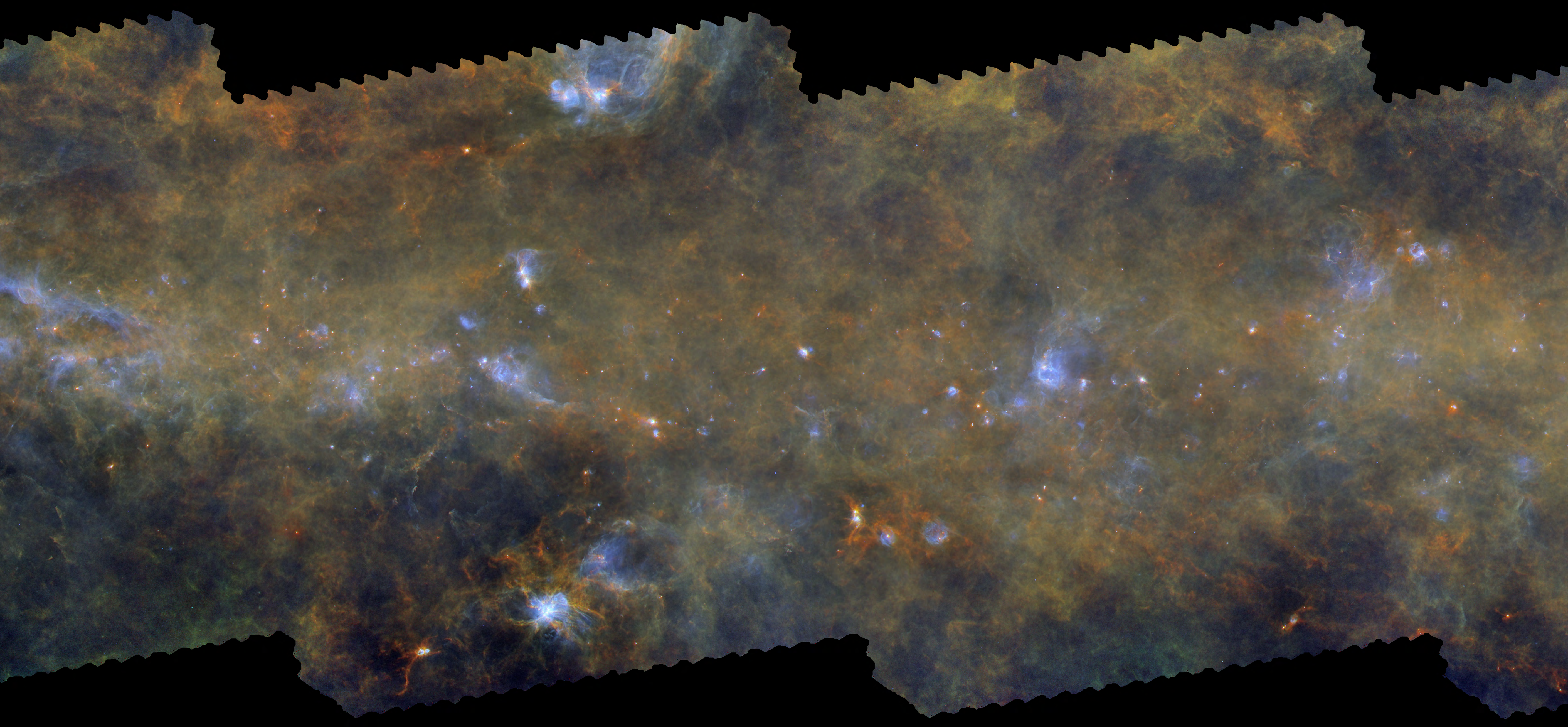}
\vskip 0.5cm
  \caption{Among the 10 most massive molecular cloud complexes forming high-mass stars at less than 3~kpc (see \textbf{Table}~\ref{tab:hobysC}), G345 was imaged  the \emph{Herschel}/Hi-GAL key program \citep{molinari10}. 
Composite three-color \emph{Herschel} image with red\,=\,250\,$\mu$m, green\,=\,160\,$\mu$m, and blue\,=\,70\,$\mu$m. Blue diffuse emission corresponds to photo-dissociation regions around massive stars or clusters. Earlier stage star-forming sites are themselves seen as red filaments and orange MDCs. Abbreviation: MDC, massive dense core.
Adapted from \cite{molinari16b} with permission.
 }
\end{figure}

\begin{figure}[h]
\vskip -2cm
\hskip 4cm\includegraphics[angle=0,width=2cm]{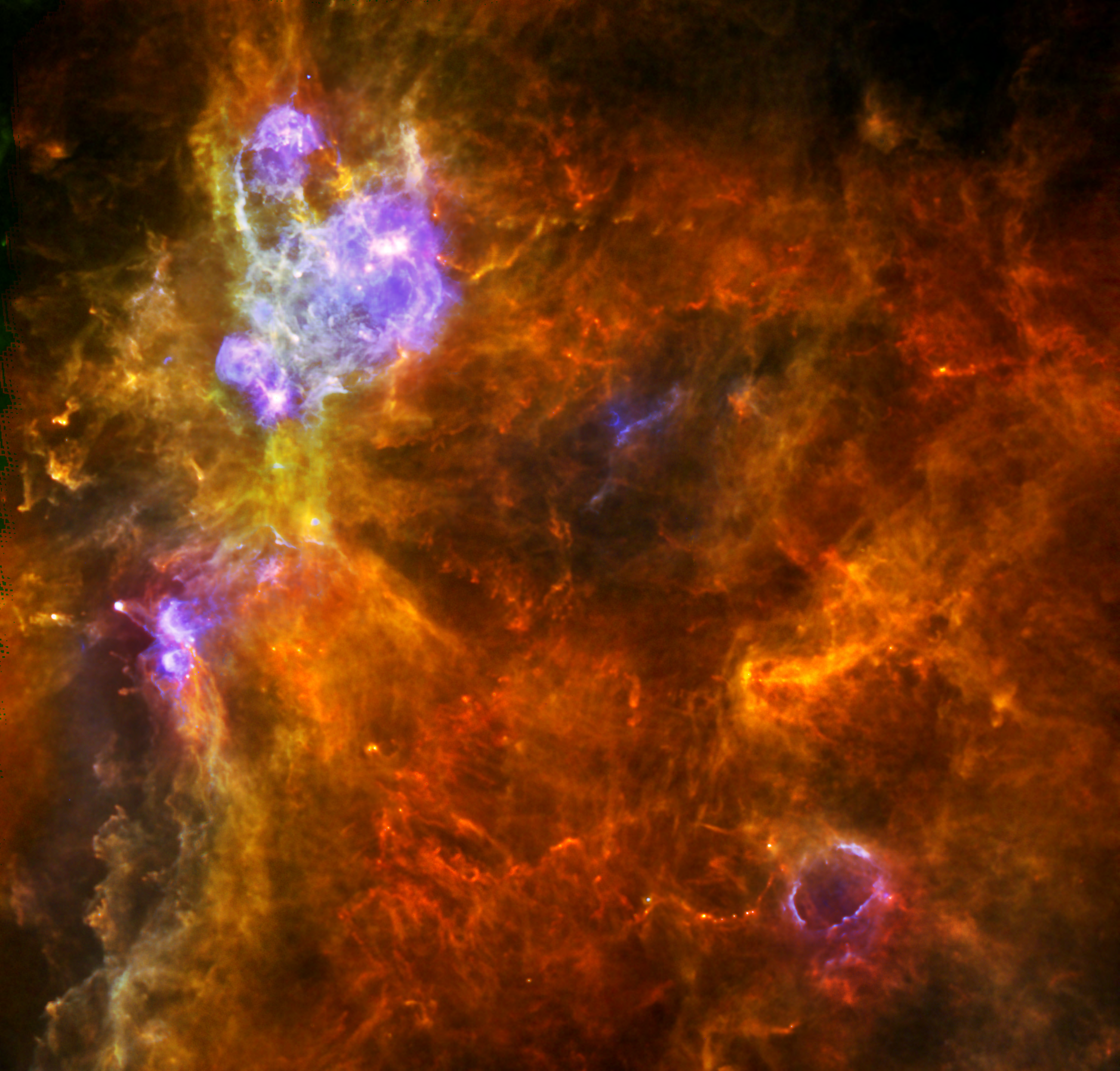}
\vskip 0.5cm
  \caption{Among the 10 most massive molecular cloud complexes forming high-mass stars at less than 3~kpc (see \textbf{Table}~\ref{tab:hobysC}), W3/KR140 was imaged by the \emph{Herschel}/HOBYS key program \citep{motte10}. 
Composite three-color \emph{Herschel} image with red\,=\,250\,$\mu$m, green\,=\,160\,$\mu$m, and blue\,=\,70\,$\mu$m. Blue diffuse emission corresponds to photo-dissociation regions around massive stars or clusters. Earlier stage star-forming sites are themselves seen as red filaments and orange MDCs. Abbreviation: MDC, massive dense core.
Adapted from \cite{rivera13} with permission.
 }
\end{figure}

\begin{figure}[h]
\vskip -2cm
\hskip 7cm \includegraphics[angle=0,width=2.2cm]{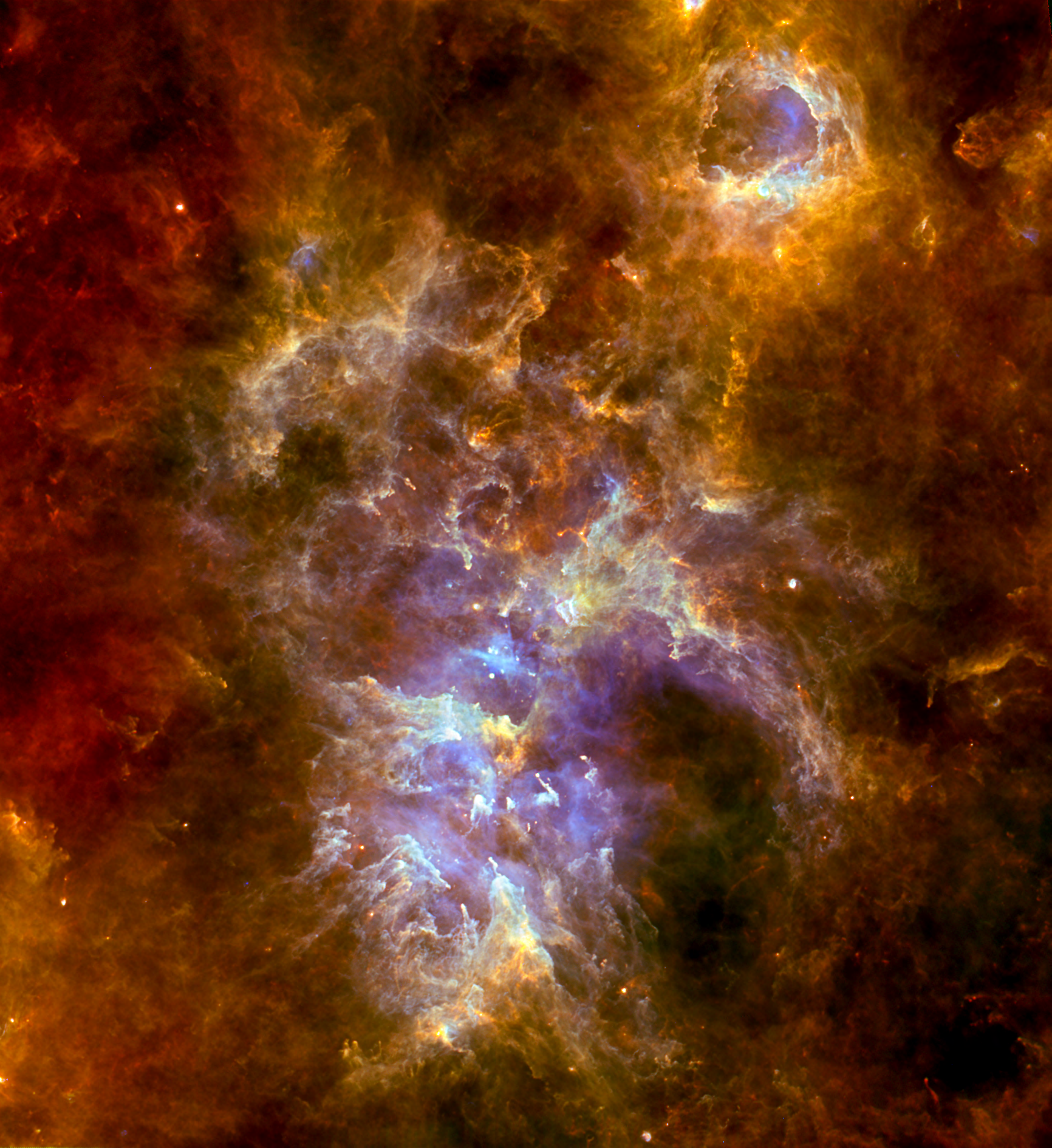}
\vskip 0.5cm
  \caption{Among the 10 most massive molecular cloud complexes forming high-mass stars at less than 3~kpc (see \textbf{Table}~\ref{tab:hobysC}), Carina was imaged by \emph{Herschel}.
Composite three-color \emph{Herschel} image with red\,=\,250\,$\mu$m, green\,=\,160\,$\mu$m, and blue\,=\,70\,$\mu$m. Blue diffuse emission corresponds to photo-dissociation regions around massive stars or clusters. Earlier stage star-forming sites are themselves seen as red filaments and orange MDCs. Abbreviation: MDC, massive dense core.
Adapted from \cite{preibisch12} and \cite{gaczkowski13} with permission.
 }
\end{figure}

\begin{figure}[h]
\vskip -5cm
\hskip -0cm \includegraphics[angle=0,width=11cm]{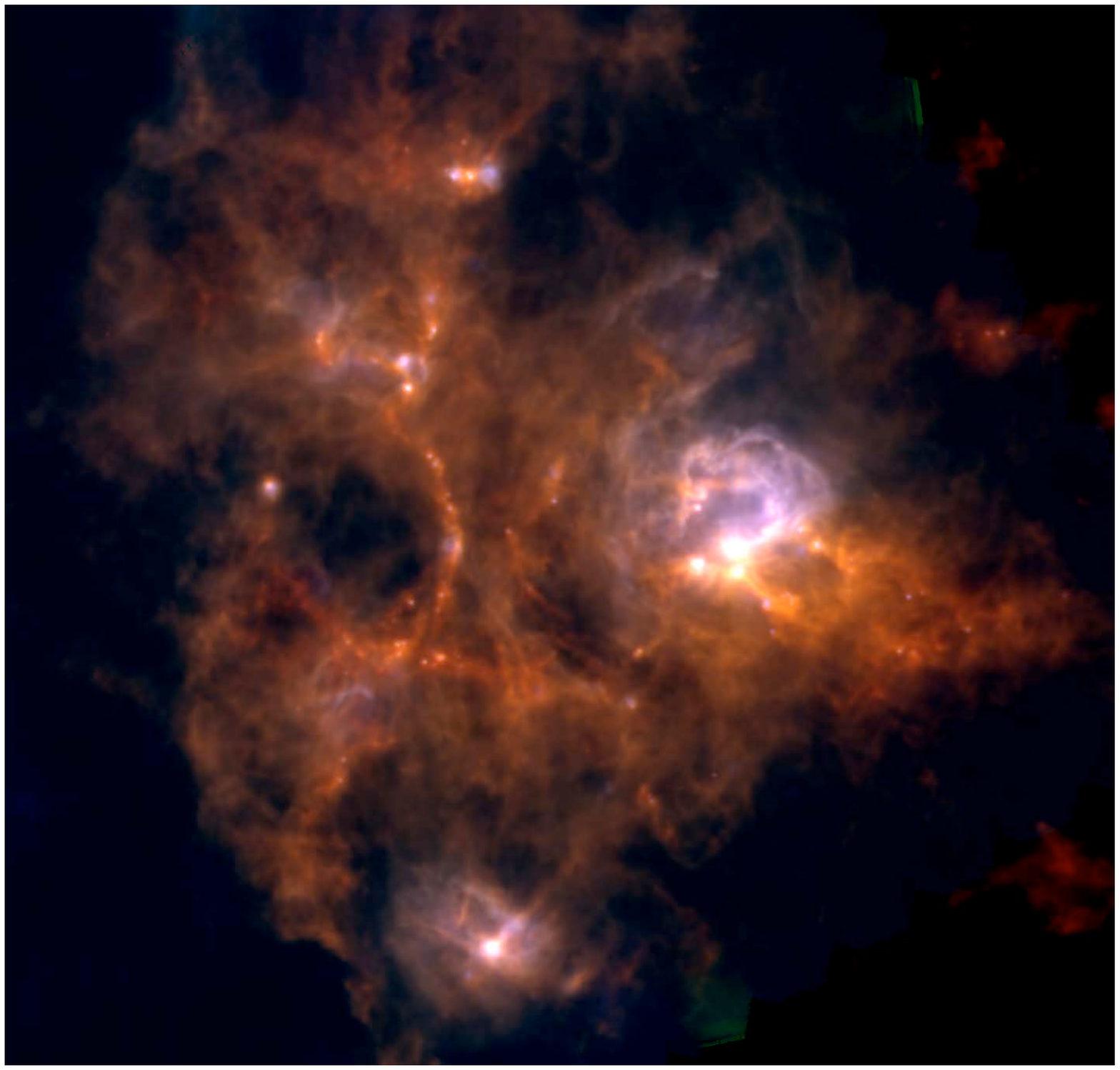}
\vskip -3cm
  \caption{Among the 10 most massive molecular cloud complexes forming high-mass stars at less than 3~kpc (see \textbf{Table}~\ref{tab:hobysC}), NGC~7538 was imaged by the \emph{Herschel}/HOBYS key program \citep{motte10}. 
Composite three-color \emph{Herschel} image with red\,=\,250\,$\mu$m, green\,=\,160\,$\mu$m, and blue\,=\,70\,$\mu$m. Blue diffuse emission corresponds to photo-dissociation regions around massive stars or clusters. Earlier stage star-forming sites are themselves seen as red filaments and orange MDCs. Abbreviation: MDC, massive dense core.
Adapted from \cite{fallscheer13} with permission.
 }
\end{figure}

\begin{figure}[h]
\hskip 7cm\includegraphics[angle=0,width=2.5cm]{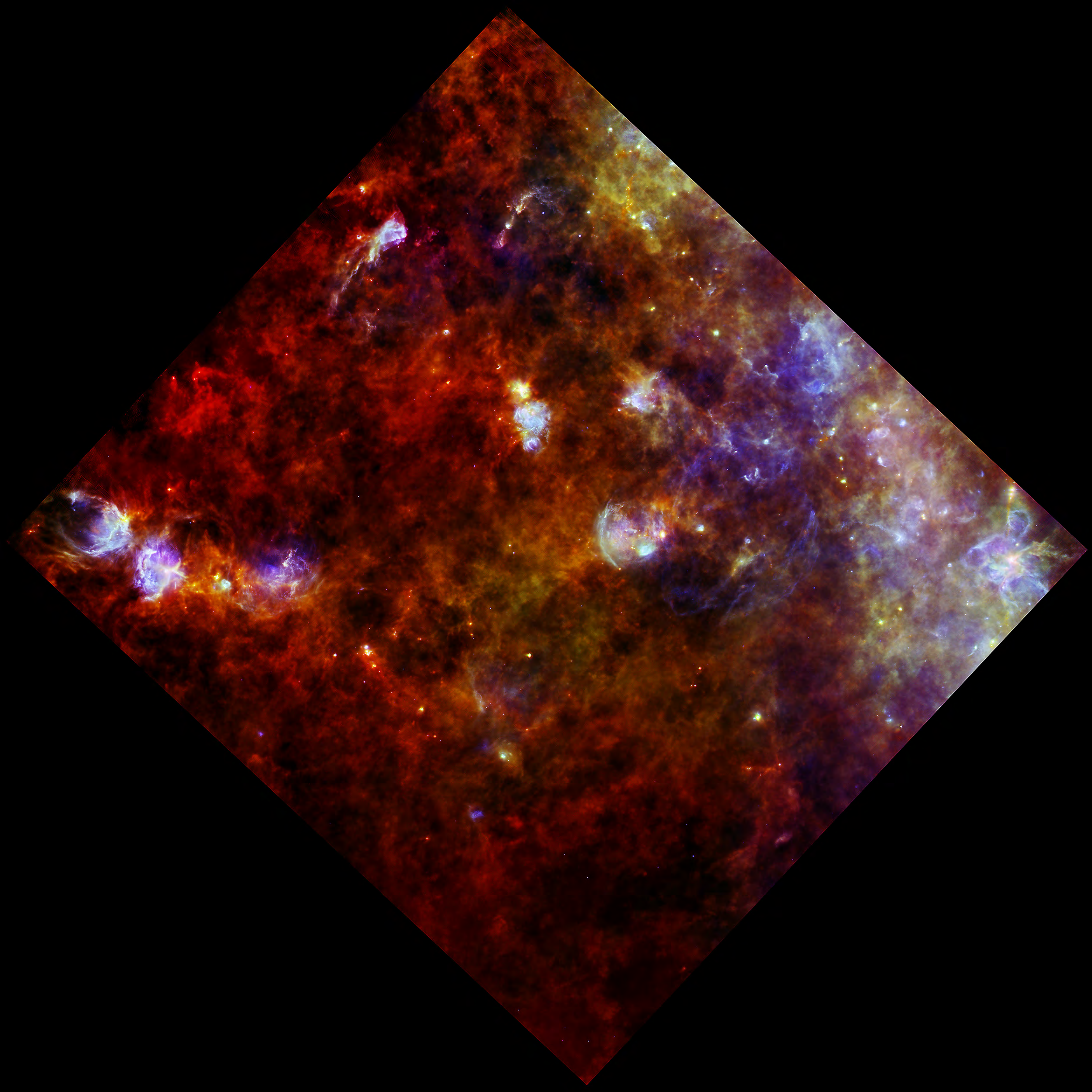}
\vskip 0.5cm
  \caption{Among the 10 most massive molecular cloud complexes forming high-mass stars at less than 3~kpc (see \textbf{Table}~\ref{tab:hobysC}), W48 was imaged by the \emph{Herschel}/HOBYS key program \citep{motte10}. 
Composite three-color \emph{Herschel} image with red\,=\,250\,$\mu$m, green\,=\,160\,$\mu$m, and blue\,=\,70\,$\mu$m. Blue diffuse emission corresponds to photo-dissociation regions around massive stars or clusters. Earlier stage star-forming sites are themselves seen as red filaments and orange MDCs. Abbreviation: MDC, massive dense core.
Adapted from \cite{nguyen11a} and \cite{rygl14} with permission.
 }
\end{figure}

\begin{figure}[h]
\vskip -2cm
\centerline{\includegraphics[angle=0,width=3.8cm]{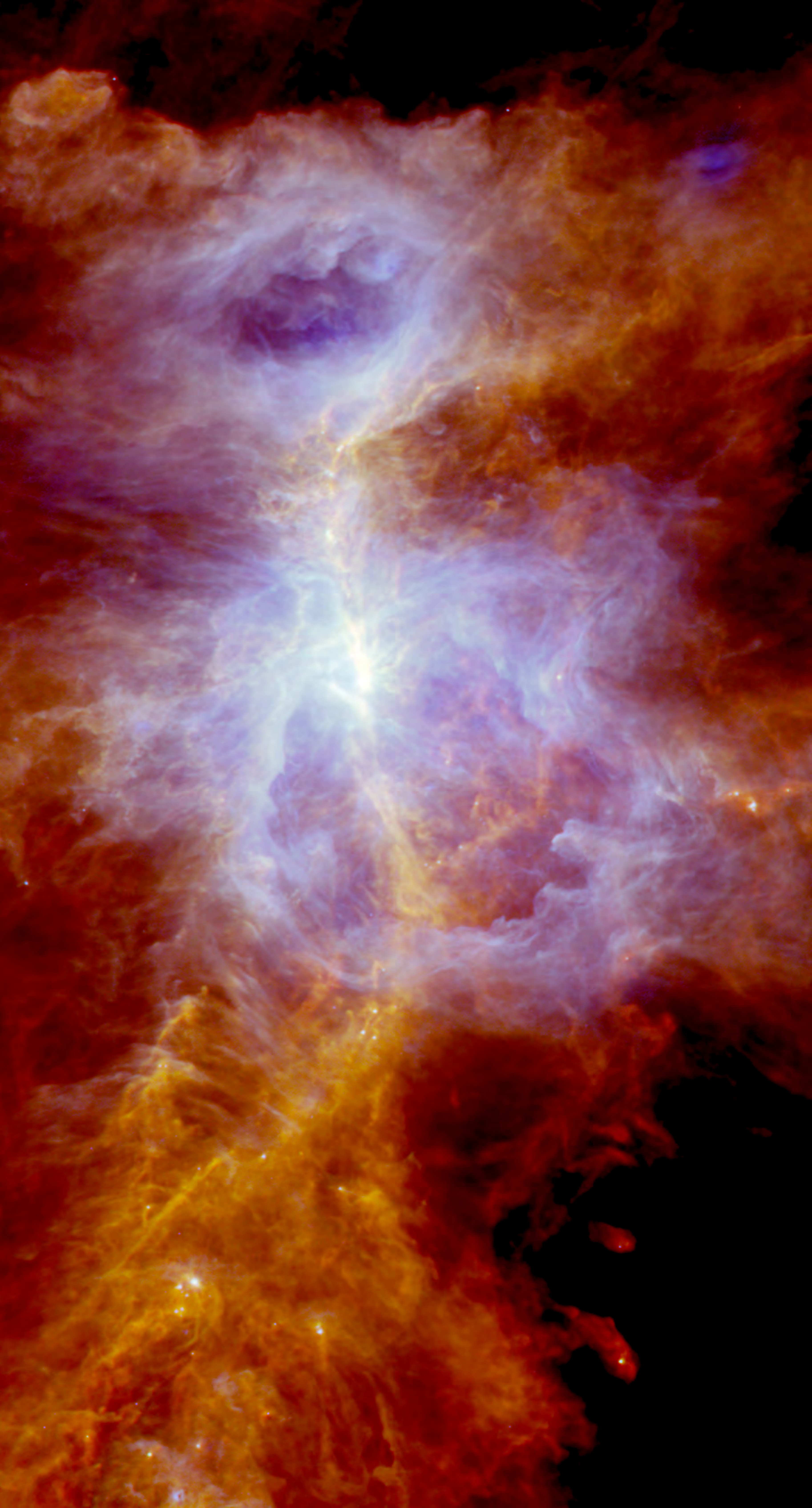}}
\vskip 0.5cm
  \caption{The reference molecular cloud complex Orion was imaged by the \emph{Herschel}/HGBS key program \citep{andre10}. Composite three-color \emph{Herschel} image with red\,=\,250\,$\mu$m, green\,=\,160\,$\mu$m, and blue\,=\,70\,$\mu$m. 
Adapted from \cite{polychroni13} with permission.
 }
\end{figure}

\end{document}